\documentclass[3p]{elsarticle}
\usepackage{hyperref}

\usepackage{fleqn}

\usepackage{graphicx}
\usepackage{amssymb,latexsym,amsthm,amsmath,dsfont,mathtools}
\usepackage{color}
\usepackage[utf8x]{inputenc}
\usepackage[T1]{fontenc}
\usepackage[polutonikogreek,english]{babel}
\usepackage{mathdots}
\usepackage{multirow}
\usepackage{hyperref}
\usepackage{array}

\usepackage{bm} 
\newcommand{\be}{\begin{equation}}
\newcommand{\ee}{\end{equation}}
\newcommand{\ba}{\begin{eqnarray}}
\newcommand{\ea}{\end{eqnarray}}
\def\bea{\begin{eqnarray}}
\def\eea{\end{eqnarray}}

\journal{Physics Reports}

\begin{document}

\begin{frontmatter}

\title{Combining complex networks and data mining: why and how}

\author[MZ,MZ2]{M. Zanin\corref{corrauthor}}
\cortext[corrauthor]{Corresponding author}
\ead{mzanin@innaxis.org}
\address[MZ]{Innaxis Foundation \& Research Institute,  Jos\'e Ortega y Gasset 20, 28006 Madrid, Spain}
\address[MZ2]{Departamento de Engenharia Electrot\'ecnica, Faculdade de Ci\^ecias e Tecnologia, Universidade Nova de Lisboa, 2829-516 Caparica, Portugal}

\author[DP]{D. Papo}
\address[DP]{Center for Biomedical Technology, Universidad
  Polit\'ecnica de Madrid,  28223 Pozuelo de Alarc\'on, Madrid, Spain}

\author[MZ2]{P. A. Sousa}

\author[DP]{E. Menasalvas}

\author[AN]{A. Nicchi}
\address[AN]{Università degli Studi Guglielmo Marconi, Centro Studi in Scienze della Vita, Via Plinio 44, 00193 Roma, Italy}

\author[EK]{E. Kubik}
\address[EK]{Columbia University, School of International and Public Affairs, 420 W 118th St \#1411, New York, NY 10027, United States}

\author[SB]{S. Boccaletti}
\address[SB]{CNR- Institute of Complex Systems, Via Madonna del Piano,
  10, 50019 Sesto Fiorentino, Florence, Italy}

\begin{abstract}
The increasing power of computer technology does not dispense with the need to extract meaningful information out of data sets of ever growing size, and indeed typically exacerbates the complexity of this task. To tackle this general problem, two methods have emerged, at chronologically different times, that are now commonly used in the scientific community: data mining and complex network theory. Not only do complex network analysis and data mining share the same general goal, that of extracting information from complex systems to ultimately create a new compact quantifiable representation, but they also often address similar problems too. In the face of that, a surprisingly low number of researchers turn out to resort to both methodologies. One may then be tempted to conclude that these two fields are either largely redundant or totally antithetic. The starting point of this review is that this state of affairs should be put down to contingent rather than conceptual differences, and that these two fields can in fact advantageously be used in a synergistic manner. An overview of both fields is first provided, some fundamental concepts of which are illustrated. A variety of contexts in which complex network theory and data mining have been used in a synergistic manner are then presented. Contexts in which the appropriate integration of complex network metrics can lead to improved classification rates with respect to classical data mining algorithms and, conversely, contexts in which data mining can be used to tackle important issues in complex network theory applications are illustrated. Finally, ways to achieve a tighter integration between complex networks and data mining, and open lines of research are discussed.
\end{abstract}

\begin{keyword}
Complex networks; data mining; Big Data
\MSC[2010] 05C82 \sep  62-07 \sep 92C42
\end{keyword}

\end{frontmatter}
\tableofcontents


\section{Introduction}

As much as they may sometimes seem chaotic or random, large-scale natural and man-made phenomena show characteristic complex structure. Parts or components are related to each other in a non-random way, with identifiable regularities and recurring patterns, ultimately making for internally coherent entities. The ability to extract regularities of all sorts from the environment is already functional within the first few months of humans’ life. In fact, the ability to recognise and use patterns is likely the gateway to some important function acquisition such as that of language, and appears to be key to the adaptive capabilities of living organisms \cite{huettel2002perceiving, zeki2008encoding}.

Scientific knowledge can be seen as an endeavour to extract and quantify reproducible regularities of phenomena. Up until relatively recently, save for a few notable {\it Gedankenexperiment} examples in physics, science generally progressed through similar inductive rule extraction cycles of carefully sampled phenomena, then reordered in a compact, {\it e.g.} mathematical, formulation. However powerful and general, the classical theory-based approach encounters limits more often than scientists were initially ready to admit. Some systems simply turn out to be too complex to be modelled in a synthetic form. Furthermore, the relevant mechanisms may be hidden. In some other cases, the seemingly intuitive notion of system may be of little help in identifying the underlying structure of a constellation of separate entities.

While methods to extract patterns from data have a long history –  Bayes' theorem (1700s) and regression analysis (1800s) arguably belong to this tradition – over the past few decades, the increasing power of computer technology has dramatically increased data collection, storage, and manipulation ability. While uncovering more information, ever more precise instruments with constantly improved sampling capabilities do not yield compact representations of the entities that are being studied, and therefore do not dispense with the need to extract and model regularities, patterns or rules from the data that these instruments generate. Often, quite the contrary, in fact, as the multiplication of samples makes up for overwhelming and intricate jumbles of always increasing size.

As data sets have grown in size and complexity, direct data analysis has started been augmented with indirect, automated data processing methods including neural networks, cluster analysis, genetic algorithms (1950s), decision trees and decision rules (1960s) and, more recently support vector machines (1990s) \cite{friedman2001elements, vapnik2013nature}. {\it Data mining} was born: an interdisciplinary subfield of computer science applying these methods with the intention of uncovering hidden patterns in large data sets, and transforming them into understandable structures for further use. Data mining bridges the gap from applied statistics and artificial intelligence to database management, to execute learning and discovery algorithms more efficiently, allowing such methods to be applied to ever larger data sets. When data sets are so complex that managing their size is a challenge in itself, data mining is also called {\it Big Data} \cite{mayer2013big}.

Over the last two decades, {\it complex network theory} has emerged as a new way to understand the structure of relationships characterising complex systems. A statistical physics understanding of graph theory (a much older branch of pure mathematics), complex networks have been used by scientists from very different fields to describe systems ranging from power grids to social interactions and the brain \cite{albert2002statistical, newman2003structure, boccaletti2006complex, costa2011analyzing, boccaletti2014structure}.

Complex network analysis and data mining have similar goals: given some data, usually representing a complex system, the aim is to extract (or synthesise) some information from them, and create a new representation (either a complex network, or a data mining model) that can be used in successive steps of the analysis. Likewise, many of the problems encountered in both fields are similar, although bearing different names: for instance, selecting the most relevant nodes in a network is mirrored by the feature selection problem; evaluating differences between groups of networks in classification tasks; and so forth. At the same time, complex network theory and data mining also present some important differences. Networks aim at {\it describing} the relations between the elements of a single (complex) system, by means of patterns that have a statistical nature and that are free from any topological space. On the other hand, data mining characterises the relations between the observables associated to different instances (not of a single system), by means of patterns pinned to a set of well-identified elements/features that are used to {\it explain} reality.

Given the similarity in the general purpose and sometimes even in procedures, it may be natural to think that both approaches can be used interchangeably. Nevertheless, the differences between network theory and data mining may, in some situations, provide an added value when both of them are used in combination. The question is then: why and when would one want to resort to both data mining and complex networks? In what kind of problems do the differences between both of them result in a synergic effect? Note that the primary goal of this review is not to make network-minded individuals fluent in the language of data mining or vice versa, but to convince either side of this field divide that there is reason enough to resort to both.

Some examples of such synergy are easy to identify. Probably the clearest ones can be drawn from biology and biomedicine, two fields in which it is simple to identify different conditions ({\it e.g.} control subjects and patients suffering from some disease) that have to be diagnosed and prognosed. 
Data mining is well suited for constructing models allowing the evaluation of the health condition of a person, for instance by detecting a pattern between two or more genes that is symptomatic of a disease; yet, it requires the target elements and the corresponding relations to be regular across instances - in other words, the same pattern, but involving other genes, would not be correctly identified.
Complex network theory may solve the problem, by identifying target structures even if they vary across subjects (provided they are topological invariant). Similar situations can be found in neuroscience, where complex networks can be used to describe the structure of relationships between different regions of the brain, and data mining can be used to create diagnostic models. This is not limited to biology: many interdisciplinary fields face similar problems, as for instance econophysics, where networks may represent the economical structure of countries, which can then be modelled through data mining techniques.

This review aims at creating bridges between two communities that have so far had little contact and cross-fertilisation, mainly due to contingent and background differences: Section \ref{sec:literature} thus starts by providing an overview of both fields, laying down some fundamental concepts, and providing references to allow the interested reader to dig deeper into them. Next, in Section \ref{sec:classification} we focus on one of the most important tasks in data mining, {\it i.e.} classification, to understand how complex networks can be used to provide a richer representation of some complex systems. Section \ref{sec:improve} changes the point of view, by dealing with how data mining can be used to tackle one of the most important problems in the application of complex network theory, that is, the reconstruction of meaningful functional networks. This latter section is complemented by Section \ref{sec:featureSel}, in which we discuss the potential applications of {\it feature selection} in network analysis. Finally, Section \ref{sec:other} discusses some additional problems and scenarios in which complex network theory and data mining have been (or could be) jointly used. The review is closed by Section \ref{sec:conclusions}, in which conclusions are drawn about two important topics: what are the problems to overcome for a tighter integration of complex networks and data mining, and open lines of research.
Additionally, hands-on sections will illustrate with worked examples the main points discussed in each chapter.


\section{Some basic concepts}
\label{sec:literature}

This Section is intended to be a primer of several concepts that will be used throughout this review: what complex networks are, how to quantify their structure, up to a first definition of the most important data mining algorithms and concepts. Over the past twenty years, both complex network theory and data mining have undergone changes that are far too dramatic and extensive to be discussed in detail here.
What we propose is a first introduction to both topics: some general ideas and concepts of data mining, for the complex network practitioner that has never worked with them; and conversely some basic notions of complex networks for data mining specialists. Throughout this review, we refer the interested reader to excellent reviews and books in both fields. Readers familiar with these concepts can skip the present section, and go straight to Section \ref{sec:classification}.

\subsection{Complex networks}
\label{sec:CNDef}

Network theory is a statistical-mechanics understanding of a past branch of pure mathematics: graph theory. In a network representation of a system, all unnecessary details are removed, to extract only its constituent parts and the interactions between them. 
Graph theory allows characterising a system once its {\it boundaries}, {\it constituent parts} and {\it relationships} are defined, but is indifferent as to what should be treated as a system and how to isolate its constituent parts; a challenge that will be addressed in Section \ref{sec:improve}.
The structure created by such interactions is called the network topology.
Most social, biological, and technological networks (including the brain) display substantial non-trivial topological properties, {\it i.e.} patterns of connections between their elements are neither purely regular nor purely random \cite{costa2007characterization}. These properties can be thought of as features describing the network's structure.

\begin{figure*}[!tb]
	\centering
		\includegraphics[width=0.99\textwidth]{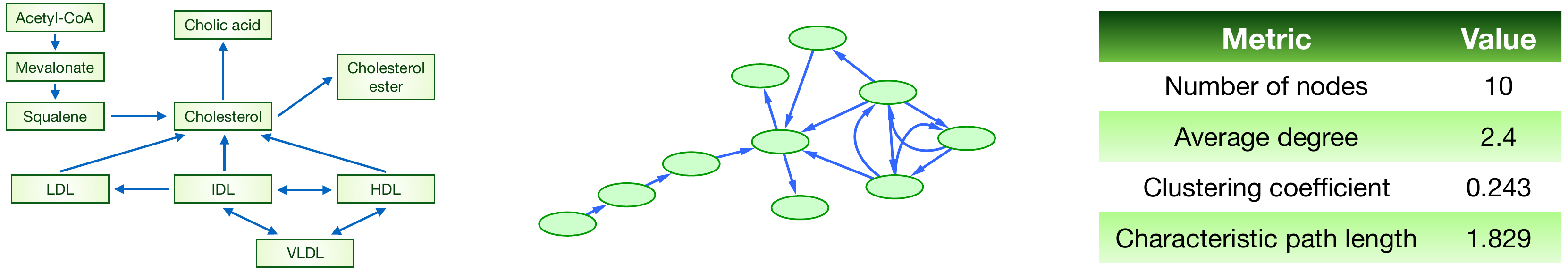}
	\caption{Creating a complex network representation. (Left) Simplified cholesterol pathway. (Center) Network representation of the same pathway. (Right) Some topological metrics of the network.}
	\label{fig:NetworkExample}
\end{figure*}

The process of evolving from a complex system to a complex network is depicted in Fig. \ref{fig:NetworkExample}. The left panel represents a simplified version of the {\it statin pathway} focused around cholesterol. This shows the set of interactions occurring within a human cell that is responsible for many cardiovascular diseases \cite{chen2000direct}. When the nature of the elements (molecules, metabolites, {\it etc.}) and of the interactions is disregarded, the result is a network, as shown in the central panel. Finally, the topological structure of the network can be represented as a set of {\it topological metrics}, as reported in Fig. \ref{fig:NetworkExample} Right.

Roughly speaking, complex network measures can be divided into three classes, according to the relevant {\it scale}: micro- (single nodes or links), meso- (groups of few nodes) and macro-scale (the network as a whole). See Table \ref{tab:cnscales} for further examination. In the last decade, tens of different metrics have been proposed for each of these classes and attempts to describe different aspects of the structure. However, some overlap and present different characteristics, like different computational costs and sensitivity to noise. 

A few of these metrics will be used throughout this text and are reported in Table \ref{tab:cnmetrics}. Although a complete review of existing complex network metrics is outside the scope of this review, the interested reader can refer to some of the excellent analyses provided in the available literature, such as: Refs. \cite{newman2001scientific, boccaletti2006complex, costa2007characterization, boccaletti2014structure}.

For the sake of completeness, we will describe three important topics in complex network analysis, which repeatedly appear throughout in this review: {\it i}) the existence of different {\it classes} of networks, {\it ii}) the difference between {\it structural} and {\it functional} networks, and {\it iii}) some recent trends in network analysis.

\begin{table}[!t]
\caption{The scales of complex networks}\label{tab:cnscales}
\begin{center}
\begin{tabular}{ | m{0.15\textwidth} | m{0.75\textwidth} | }
  \hline
  Micro-scale & This scale focuses on the properties of a single link or single node. When the whole network is analysed these metrics are usually averaged over each node comprised within the network, in order to create a global picture. {\it Example: Number of connections a person has in a social network.} \\
  \hline
  Meso-scale & This is the intermediate level between studying elements of a system separately (micro-) and considering the system as a whole (macro-scale). We may define the meso-scale as any regular structure that affects the connectivity of groups of nodes in the network \cite{almendral2011introduction}. {\it Example: Communities of friends present within a social network.} \\
  \hline
  Macro-scale & The macro-scale depicts the system as a whole. It accounts for the overall structure of the network, addressing the movement of information throughout. {\it Example: The average path length to reach a person in a social network.} \\
  \hline
\end{tabular}
\end{center}
 \end{table}

\begin{table}[!t]
\caption{Some basic complex network metrics}\label{tab:cnmetrics}
\begin{center}
\begin{tabular}{ | m{0.15\textwidth} | m{0.12\textwidth} | m{0.60\textwidth} | }
  \hline
  {\bf Metric} & {\bf Scale} & {\bf Description} \\
  \hline
  Degree & Micro-scale & The number of connections a node has, {\it i.e.} the number of neighbours. The degree $k$ of node $i$ is calculated from the adjacency matrix as $k_i = \sum _{j} a_{ij}$. \\
  \hline
  Link density & Micro-scale & The number of links $l$ in the network, divided by the maximum number of links that could be present. For a network composed of $n$ nodes, the link density is thus $\frac{l}{n (n - 1)} = \frac{1}{n (n - 1)}\sum _{i,j} a_{ij}$. \\
  \hline
  \hline
  Clustering coefficient & Meso-scale & The clustering coefficient $C_i$ of a node $i$ is the fraction of pairs of its neighbours that are directly connected, and is thus a count of the number of triangles in the network. Such metric can be averaged over all nodes, yielding the network clustering coefficient $C = \frac{1}{n} \sum _{i} C_i$. \\
  \hline
  Modularity & Meso-scale & Assesses the presence of a community structure in the network, {\it i.e.} of groups of nodes densely connected between them \cite{newman2006modularity, fortunato2010community}. \\
  \hline
  Motifs & Meso-scale & Subgraphs (usually composed of three or four nodes) that appear more frequently than what could be statistically expected \cite{milo2002network}. \\
  \hline
  \hline
  Geodesic distance & Macro-scale & The geodesic distance $d_{i,j}$ between nodes $i$ and $j$ is the minimum number of steps needed to move between them, {\it i.e.} the length of the shortest path connecting them. The value can be averaged over all possible pairs of nodes, yielding the average shortest path length of the network. \\
  \hline
  Efficiency & Macro-scale & Mean value of the inverse of the geodesic distance between pairs of nodes, {\it i.e.} $E = \frac{1}{n(n-1)} \sum _{i, j \neq i} \frac{1}{d_{i,j}}$. It assesses the ease of information flow in the network, and has the advantage of being defined for disconnected networks \cite{latora2001efficient, crucitti2003efficiency}. \\
  \hline
\end{tabular}
\end{center}
 \end{table}

\subsubsection*{Classes of networks}

In the last decade the analysis of a large number of real-world networks has revealed that some structures, or networks topologies, are ubiquitous across many natural and man-made systems \cite{costa2011analyzing}. Such classes are of relevance, as they constitute the basis of the classification of real systems.

From a historical point of view, the start of classifications was the random graphs, also known as Erd\"os-R\'enyi graphs after the mathematician who discovered them, Paul Erd\"os.
Given a set of $n$ disconnected vertices, links between all pairs of nodes are created with a probability $p$. Many theoretical results were obtained in random graphs, as for instance, the expected size of the largest component (groups of nodes connected between them), or the critical value of $p$ for which the graph was connected \cite{renyi1959random}. A comprehensive review of all results obtained in random graph analysis can be found in Ref. \cite{bollobas1998random}.
If random graphs are characterised by a complete absence of structure, the other extreme is represented by regular graphs, {\it i.e.} networks where all nodes have the same number of connections. 
While these two extrema have limited interest for modelling real-world systems, which seldom present such simple structures, the importance reside in their use as {\it null models}. For instance, the frequency of appearance of a motif is usually compared with the frequency expected in equivalent (same number of nodes and links) random graphs, in order to assess its statistical significance \cite{milo2002network}.

In 1998, Watts and Strogatz realised that real networks were neither regular nor completely random graphs; that real systems lie somewhere between these two extremes. Specifically, random graphs are characterised by both a low mean geodesic distance and clustering coefficient. On the other hand, regular graphs show both high mean geodesic distance and clustering. By analysing social and biological networks, they discovered that most of them are characterised by a low mean geodesic distance, but also by a high clustering coefficient: in this case, networks are said to have the {\it small-world property} \cite{watts1998collective}. A third class of networks was born: the small-world ones. Simple models can be designed to synthetically recover such effect \cite{watts1998collective}, and the property can be numerically estimated through the {\it small-worldness} metric \cite{humphries2008network, zanin2015alternative}, which has been particularly popular in neuroscience, though some important caveats to its applications have recently been pointed out \cite{papo2016beware}.

Until 1998, three classes were thus known: random, regular, and small-world ones. A fourth class of network topologies emerged when it was first noticed that real-world networks are characterised by heterogeneous nodes, some of them (the {\it hubs}) of high connectivity. A clear example can be found in transportation networks, where few airports create the backbone of the network \cite{zanin2013modelling}. Mathematically, these networks are called {\it scale-free}, as their degree distribution follows a power law, and thus have no characteristic scale \cite{barabasi2009scale}. Many models for obtaining scale-free topologies have been proposed \cite{barabasi2001deterministic, dorogovtsev2001size, holme2002growing, caldarelli2002scale, klemm2002highly, klemm2002growing, valverde2002scale, jung2002geometric, saramaki2004scale, catanzaro2005generation}, the most praised is the Barab\'asi-Albert one \cite{barabasi1999emergence}. It refers to a {\it preferential attachment} process, in which new nodes are added to the network, each connected to existing nodes with a probability proportional to their degree.

\subsubsection*{Structural {\it vs.} functional networks}
\label{sec:functional}

When analysing networks, it is useful to make a distinction between two types: {\it structural} (also called {\it physical}) and {\it functional} networks.

Physical networks are described by explicit interactions; reconstructing them is tantamount to mapping the connections into the corresponding links, as seen in the real system. For instance, in a transportation system nodes can be airports, bus stops, train stations, {\it etc}, while scheduling tables provide information on the corresponding links. However, some real-world systems lack explicit information about links. One may nevertheless suppose that each element's dynamics is a function of its neighbours. The first step then requires quantifying these functions (hence the name {\it functional networks}). The resulting network's topological features can finally be extracted as in the previous case.

The study of brain structure and dynamics provides a clear example of this difference. On the one hand, different brain regions are connected by fibres and are organised in a {\it physical structure}, called the {\it connectome} \cite{sporns2005human, sporns2011human}, which represents the substrate on which brain dynamics takes place. However, on the other hand, one may be interested in understanding how the brain executes a specific task, and thus on how the connectome is used to transfer information between different regions. This can be achieved by reconstructing the functional network associated to the task, by connecting pairs of nodes (brain regions) when some common dynamics (or {\it synchronisation}) is observed \cite{bullmore2009complex, papo2014reconstructing}.

\begin{figure*}[!tb]
	\centering
		\includegraphics[width=0.7\textwidth]{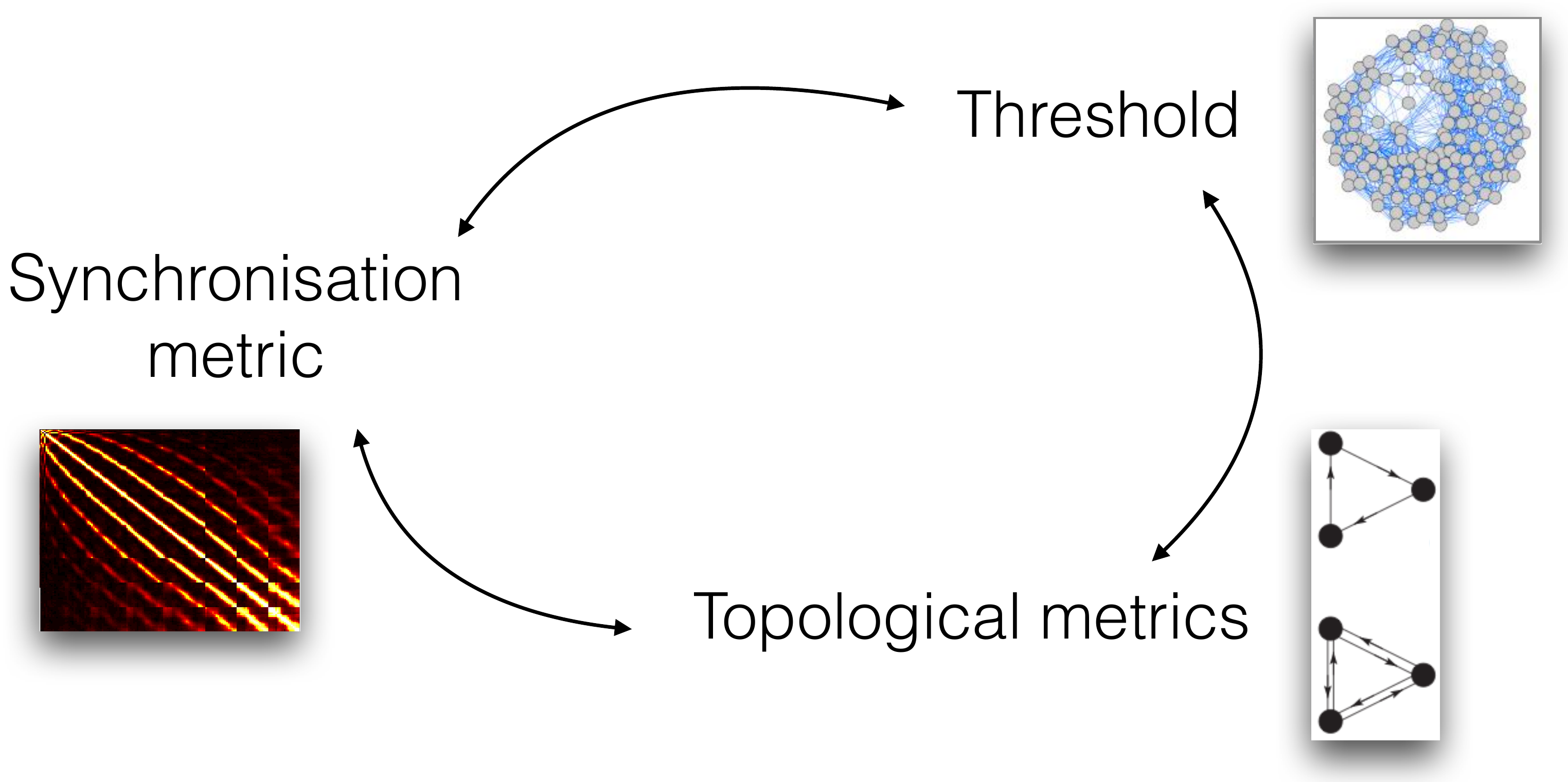}
	\caption{Interactions between different aspects of functional network reconstruction: selection of the synchronisation metric, of the threshold for its binarisation, and of the topological metrics. See the main text for details.}
	\label{fig:NetworkOptimisation}
\end{figure*}

In the most general case, reconstructing functional networks involves choosing (or tuning) three different aspects of the analysis, namely: 
\begin{enumerate}
	\item {\it The connectivity metric}. This is the measure assessing the presence of some relationship between the dynamics of two elements of the system.
	\item {\it A threshold}. While not essential, it is customary to eliminate those links that have a low statistical significance, and to consider the binarised version of the network, {\it i.e.} to discard weights.
	\item {\it The topological metrics}. These metrics describe the structure of the resulting networks and eventually compare different groups of subjects.
\end{enumerate}

In the absence of a set of rules, these three aspects bring in a degree of subjectiveness. For instance, researchers can be tempted to arbitrarily choose a threshold only based on their experience.
More importantly, these three steps are strongly interconnected, as depicted in Fig. \ref{fig:NetworkOptimisation}. Each connectivity metric is designed to detect different aspects of dynamics. In the case of functional brain networks, this may include local {\it vs.} global dynamics, linear and non-linear dependencies, correlations {\it vs.} causalities, and so forth. Each one of these metrics, in turns, changes the resulting topology, thus requiring different thresholds and topological metrics.
Changing the threshold has important implications on the topological metrics that can be assessed. For instance, in dense networks the average geodesic distance always tends to one, while in sparse networks it is difficult to observe complex motifs.
Closing the circle, if one wants to observe a given topological metric, it is necessary to choose a suitable synchronisation metric and threshold. For example, motifs containing bi-directional connections are mostly excluded by using a causality metric.
These problems will be tackled in Section \ref{sec:improve}, which is devoted to the use of data mining methods for improving network representations.

While the analysis of physical networks is usually straightforward, it may still require some pre-processing.  This pre-processing can include filtering out spurious links (links that are the result of noisy measurements, and that therefore distort the real structure of the network) or reconstructing links in unexplored regions of the system space. These two topics are respectively discussed in Sections \ref{sec:featureSel} and \ref{sec:LinkPrediction}.

\subsubsection*{Recent trends in network theory}
\label{sec:multilayer}

Until now we have concisely presented the main elements of complex network theory that are used in the analysis of real systems. Due to the inability of this standard framework to describe scenarios observed in complex real-world scenarios, network theory has been extended within the last years to include concepts such as {\it temporal} and {\it multi-layer} networks.

{\it Temporal networks} are composed of edges that are not continuously active. As an example, in networks of communication via e-mail, text messages, or phone calls, edges represent sequences of instantaneous or practically instantaneous contacts. Sometimes, edges are active for non-negligible periods of time. This is for instance the case of inpatients proximity patterns within a hospital, where individuals are pairwise connected while they are at the same ward. It is clear that the temporal structure of edge activations can affect the dynamics of the elements interacting through the network, from disease contagion in the patient network to information diffusion over an e-mail network. See \cite{holme2012temporal, holme2013temporal} for reviews on the topic.

The idea of {\it multi-layer networks} emerged by observing that connections between the elements constituting real-world networks are seldom of a single type. Let us consider the following three examples, starting with social networks. Social networks as people (or groups of people) have some pattern of contacts or interactions between them \cite{vega2007complex}. Social interactions seldom develop on a single channel and more than one relationship can bind pairs of people, as was initially proposed by Erving Goffman in 1974, along with his theory of frame analysis \cite{goffman1974frame}.
Second, consider a transportation network, as for instance the Air Transportation Network (ATN) \cite{zanin2013modelling}. Flights connecting pairs of cities are not homogeneous, as airlines have to be considered. Furthermore, passengers cannot easily connect two flights operated by different airlines, at least connect airlines belonging to different alliances \cite{cardillo2013emergence, cardillo2013modeling}. 
Finally, biology provides a third example as scientists try to understand the role of specific components in a biological system. The full mapping of the {\it Caenorhabditis elegans} (or {\it C. elegans}) neural network is now known, comprising $281$ neurons along with more than two thousand connections \cite{white1986structure}. Since neurons can be connected by chemical and electrical (ionic) links, this results in two different dynamics, and ultimately does not form a single network.

These three examples explain the efforts for generalising the traditional network theory by developing a novel framework for the study of multi-layer networks, {\it i.e.} graphs where several different layers of connections are taken into account \cite{de2013mathematical, lee2014multiplex, boccaletti2014structure}.
Multi-layer networks explicitly incorporate multiple channels of connectivity and therefore constitute the natural environment to describe systems interconnected through different categories of connections. A layer represents each channel (relationship, activity, category) and the same node may have different kinds of interactions (different set of neighbours in each layer). For example, in social networks, one can consider several types of relationships: friendship, vicinity, kinship, membership of the same cultural society, partnership or coworker-ship, {\it etc}.

\subsubsection*{Software for network analysis}
\label{sec:softwareNets}

\begin{description}

	\item \emph{NetworkX} \cite{schult2008exploring}. Python library used for the creation, manipulation, and study of the structure, dynamics, and functions of complex networks. This allows the creation of networks with different algorithms, evaluation of a large set of standard metrics, and finally display the results in an easily understood way. Freeware. Available at \url{https://networkx.github.io}.

	\item \emph{Cytoscape} \cite{shannon2003cytoscape}. Software specialised on the representation of networks, with some additional tools for the integration of biological data. It also provides some basic network analysis capabilities. Freeware. Available at \url{http://www.cytoscape.org}

	\item \emph{Gephi} \cite{bastian2009gephi}. Interactive visualisation and exploration platform. Freeware. Available at \url{https://gephi.github.io}

	\item \emph{Pajek} \cite{batagelj1998pajek}. Software for representing complex networks, with some basic analysis capabilities. Freeware. Available at \url{http://mrvar.fdv.uni-lj.si/pajek/}

	\item \emph{VisANT} \cite{hu2004visant}. Software for the visual study of metabolic networks and pathways. Freeware. Available at \url{http://visant.bu.edu}

	\item \emph{IBM \circledR~i2 Analyst's Notebook}. Software for the integration of social data and network analysis. Commercial. Information at \url{http://www-03.ibm.com/software/products/en/analysts-notebook}
	
	\item \emph{SAS \circledR~Social Network Analysis}. Software for the analysis of social networks. Commercial. Information at \url{http://support.sas.com/software/products/sna/index.html}
\end{description}

\subsection{Data Mining}
\label{sec:dataMining}

The proliferation, ubiquity and increasing power of computer technology have dramatically enhanced our ability for data collection, storage, and manipulation. This has created a new need for automatic data analysis, classification, and understanding. {\it Data mining}, with its vision of automatically extracting knowledge from data bases, has attracted growing attention since late 80's. In the current information age, data generated and stored by modern organisations increase in an extraordinary way, and data mining tasks \cite{fayyad1996advances} become a necessary and fundamental technology for sustainability and improvement.

The term has also evolved in \cite{fayyad1996advances}. A distinction was made between KDD (\textit{Knowledge Discovery in Databases}), which refers to the overall process of discovering useful knowledge from data; and data mining, which refers to a particular step in such process, in which specific algorithms are applied to extract patterns from data. 
Nowadays, this distinction is almost lost, and ``data mining'' is used to refer to the overall process performed by combining methodologies and techniques from different fields, such as statistics, databases, machine learning and visualisation. The size of the data increases and becomes challenging and data mining evolves into \textit{Big Data}; a topic that will be further discussed in Section \ref{sec:BigData}. 

Modern data mining faces two major challenges. The first is being able to share the obtained knowledge. In this sense, the Predictive Model Markup Language (PMML) \cite{pechter2009s} has become one of the most widely adopted data mining standards used today. The second challenge is defining a methodology to guide the process of  discovery. In 1997, an industry group called the \textit{Cross-Industry Standard Process for Data Mining} (CRISP-DM) \cite{wirth2000crisp} proposed a methodology for organising the KDD process in six standard steps:

\begin{description}

	\item \emph{Business (or Problem) Understanding}: initial phase that focuses on understanding the project objectives and requirements from a business perspective. This knowledge is then converted into a data mining problem, and a preliminary plan is designed to achieve the project objectives.

	\item \emph{Data Understanding}: starts with an initial data collection and proceeds with several activities, all aimed at: familiarise with the data; identify data quality problems; discover initial insights about the data; and detect interesting data subsets to form hypotheses about hidden information.

	\item \emph{Data preparation}: arguably the most important of the whole process, as the success of the final analysis strongly depends on it, and may consume up to the $90\%$ of time and resources. This covers all of the activities required to construct the final dataset, such as identifying data that will be fed into the modelling tools from the initial raw data along with data cleaning and preprocessing. Data preparation tasks are likely to be performed multiple times, and not in any prescribed order. They include selection and transformation of tables, records, attributes, and data cleaning. Ref. \cite{Cooley99datapreparation} presents several data preparation techniques and algorithms, which can be used for preprocessing data in web mining applications. Additionally, Ref. \cite{Zhang03datapreparation} presents a more general review of techniques, along with a deep discussion on the motivations for data preparation. 
		It is worth noting that the importance of this step is reduced when one is in the Big Data field; a concept that will be further discussed in Section \ref{sec:BigData}. Due to the complexity and volume of data, cleaning and preparation may not be feasible and it could lead to an unwanted loss of useful information. In these cases, it is preferable to work with ``data in the wild'', and sacrifice some accuracy.

	\item \emph{Modelling}: phase in which data mining algorithms are applied and parameters are calibrated to optimal values. Some of these techniques will be reviewed in detail in the following subsections. Typically, different techniques can be chosen to solve the same modelling problem, each having specific requirements on the format of input data and hypothesis on the patterns to be detected. In these situations, models are optimised and compared; the models reaching a higher performance are passed to the next phase for a final evaluation.

	\item \emph{Evaluation}. Once the models have been evaluated from a data mining perspective, it is necessary to review the output patterns obtained considering the business requirements identified in the first phase. Only when all relevant questions have been addressed, can one then move to the deployment of the extracted knowledge.

	\item \emph{Deployment}. When all of the information about the business problems has been gathered, the information and knowledge then has to be organised and presented.

\end{description}

\subsubsection*{Data Mining tasks and models}

Data mining tasks, or the tasks performed in the modelling phase of the KDD process, can be classified into \textit{predictive} and \textit{descriptive} tasks. Predictive analytics includes all data analysis processes that perform inference in order to extract patterns to be used to make predictions. On the other hand, descriptive tasks group all the processes characterising the general properties of the data. For any of these tasks, a tremendous amount of techniques and algorithms have been described in the literature, the most important of which are reviewed in the next subsections. Delving deeper into such concepts, both families can be defined as follows:

\begin{description}
	
	\item \emph{Descriptive Modelling}, also called \textit{Exploratory Analysis}, has the main purpose of describing, through patterns, the information encoded in the data set being studied. These techniques have an exploratory nature, in the sense that allows a better characterisation of the existing data, without providing any forecast. They are oriented towards data interpretation, which focuses on understanding the underlying data relations, such as finding correlations, trends, groups, clusters and anomalies. Descriptive models mostly rely on {\it Clustering} and {\it Association Rules} techniques and algorithms.

	\item \emph{Predictive Modelling}: its main goal is to find a model, constructed over the information already labelled in the data set, which can be used in the future to predict information. This model aims at predicting the value of a particular attribute, such as the target or dependent variable, based on the values of other attributes (variables), assuming a set of labelled data (training set) is given. The underlying assumption of the inductive approach is that the data used for training are representative of the whole universe, meaning of all the possible unknown data that may be encountered, and therefore the trained model is able to accurately predict the values of future unseen instances.

\end{description}

It is important to mention that in the case of predictive and descriptive tasks, the methods used to extract knowledge are based on inductive learning, where a model is constructed by generalising a set of training records. In both cases, though, knowledge description comes in different forms. For classification problems, this may be a set of characteristic or discriminant rules, or a decision tree or a neural network with fixed sets of weights. In case of association analyses, it may be a set of associations, or association rules. For cluster analysis, it consists of a set of clusters, each of which has its own description and cluster name (class name). 

In the following we will review the main data mining tasks, with a special emphasis will be placed on classification, due to its relevance in problems involving complex networks.

\begin{description}

\item \emph{Clustering}. Also called \textit{unsupervised learning}, it involves the division of data into groups (clusters) that contain similar records (according to some similarity measures), and those subsequent, separate dissimilar records organised into different clusters. In Ref. \cite{kaufman2009finding} clustering is defined as: \textit{partition a given data set in groups, called clusters, so that the points belonging to a cluster are more similar to each other than the rest of the items belonging to other clusters}. In Ref. \cite{Jain1999} a taxonomy of clustering techniques is presented, with an overview of its fundamental concepts and methods. Moreover, it describes several successful applications of clustering, such as image segmentation or object and character recognition. Classifying clustering algorithms is not an easy task, as the categories very often overlap. According to the survey found in Ref. \cite{Berkhin02surveyof}, one can establish the following classification:
  
\begin{itemize}

	\item \emph{Hierarchical-Based}. The hierarchical clustering combines instances of the dataset, to form successive clusters, resulting in a tree form called {\it dendrogram}. In the lower level of the tree, only one instance is associated to each cluster, and upper levels of the tree are aggregations of the nodes below. Agglomerative and divisive clustering can be distinguished, according to the criteria to group nodes. 

	\item \emph{Partitions-Based}. The clustering methods based on partitions divide the data set into different disjoint subsets. The operation involves assigning points to different clusters, whose number is initially fixed, and iteratively improving the division, until a previously defined heuristic finds the optimal solution. The praised {\it K-means} \cite{kmeans1979} algorithm belongs to this class.

	\item \emph{Density-Based}. In the previously described algorithms, such as the K-means algorithm, the similarity of the points (instances) for cluster assignment is assessed through a specific distance measure. In density-based algorithms, however, clusters are based on density measures. The {\it DBSCAN} \cite{Ester96adensity} algorithm belongs to this kind of clustering techniques.

\end{itemize}

\item \emph{Association Rule Mining}. This explores the relations between attributes that exist in data, thus detecting attribute-value conditions that occur frequently together. It has been popularised by Agrawal \cite{agrawal1993mining}, who proposed {\it Apriori}, the best-known algorithm to mine association rules. Association rules can be used as the basis for decisions about marketing activities, as their best known application conducts analysis of market baskets. In Ref. \cite{zaki1997new, agrawal1996fast} the authors present two new algorithms for solving basket market analysis. The empirical evaluation presented in the aforementioned paper shows that these algorithms outperform the existing methods for both small and large problems. {\it Apriori Hybrid} is also shown as a combination of the best features of the two proposed algorithms. A more recent survey of association rule mining can be found in Ref. \cite{Kotsiantis_associationrules}.

\item \emph{Prediction}. Also called {\it supervised learning methods}, its goal is to learn from a set of training data, with the aim of predicting the class of new unlabelled records. More specifically, let $\mathcal{X}$ be the feature space and its possible values, and $\mathcal{Y}$ be the space of possible labels of a target variable.  The underlying assumption is that within exists a function $f(\mathcal{X}) \rightarrow \mathcal{Y}$ that assigns a record to a class depending on the values of its describing features. The function $f$ thus aims to predict values of a class attribute from other attributes based on training data. The classification algorithm will try to minimise the distance between the real and the predicted value ({\it i.e.}, minimise the expected error). Consequently, the classifiers are usually evaluated by assessing their predictive accuracy on a test set. The most frequent measures to validate the models are: ({\it i}) accuracy, defined as the proportion of correct predictions obtained; ({\it ii}) precision, the proportion of correct positive forecast; ({\it iii}) recall, the fraction of relevant instances that have been retrieved; and the ({\it iv}) F-measure, or $F1$ measure, the harmonic mean of precision and recall.
 
Depending on the type of variables to predict, we can distinguish between: 
\begin{itemize}
	\item Numeric Value Prediction or regression: which attempts to predict a numeric attribute value by modelling a continuous-valued function.
	\item Classification: which aims to predict a discrete class label. 
\end{itemize}

\end{description}

\begin{figure*}[!tb]
	\centering
		\includegraphics[width=0.6\textwidth]{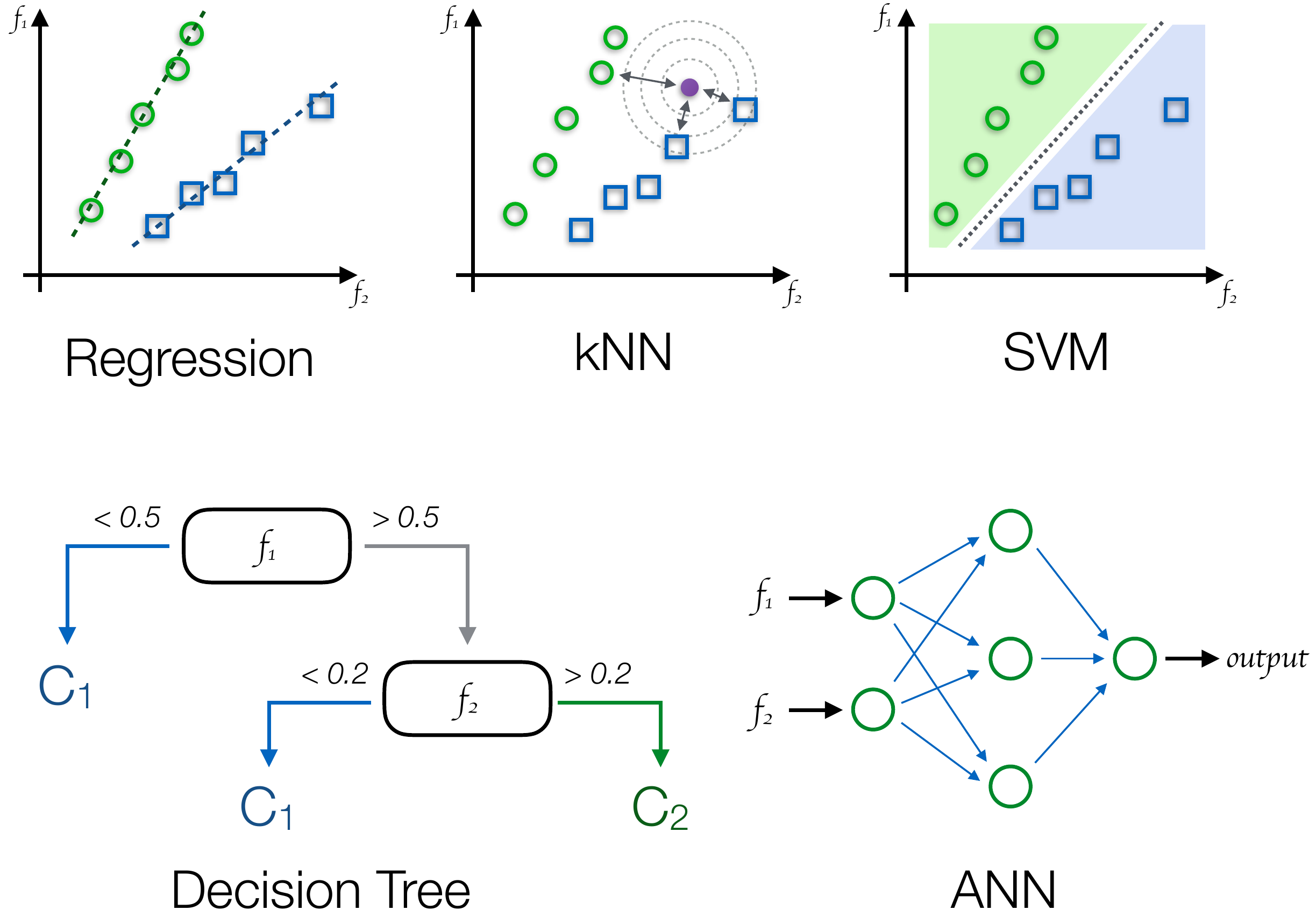}
	\caption{Example of five selected data mining classification algorithms. Green circles and blue squares represent two classes, {\it e.g.} control subjects and patients. The three algorithms in the top row respectively represent simple linear regression, kNN (in which new instances are classified according to the class of their nearest neighbours), and SVM (which divides the feature space is two non-overlapping regions). In the bottom row, Decision Tree, in which numerical conditions are sequentially tested of the available features; and Artificial Neural Network, which optimises a mathematical function on the input features. Additional graphical examples can be found in Figs. \ref{fig:GeneticExample} and \ref{fig:FigureHO02}.}
	\label{fig:DMAlg}
\end{figure*}

\subsubsection*{Algorithms and models}

Many classification and prediction algorithms have been proposed in the last few decades, each bringing different advantages and disadvantages, along with their own requirements on the format of the data. Below are the most successful and well-known techniques.
 
\begin{description}

\item \emph{Na\"ive Bayes classifiers}. This refers to a family of simple probabilistic classifiers based on applying Bayes' theorem with the assumption of independence between the features. Consider an instance, described by a vector of features $X$, which may belong to several classes $C_K$ (being $K$ the number of classes). The probability of that instance to belong to a class $i$ is then given by:
\begin{equation}
	p(C_i | X) = \frac{p(C_i) p(X | C_i)}{p(X)}.
\end{equation}
In other words, the {\it posterior} probability of the instance to belong to $C_i$ is proportional to the {\it prior} probability, to the likelihood, and inversely proportional to the evidence.

Na\"ive Bayes classifiers are highly scalable, requiring adjustments of parameters that grows linearly with the number of variables in the problem, and thus leading to a linear computation cost. More information can be found in Ref. \cite{Rish}.

\item \emph{Regression}. Regression analysis is the most basic and commonly used predictive tool, especially in practical applications \cite{Freedman91}, and is based on a statistical process for estimating the relationships among variables. 
As it  can be seen in Fig. \ref{fig:DMAlg} Top Left, the goal of regression analysis is fitting a linear function through a scatter plot. In the simplest case of univariate regression analysis ({\it i.e.} with one dependent and one independent variable), this goal can be defined as:

\begin{equation}
h_{\theta}(x)= \theta_{0}+\theta_{1}(x).
\end{equation}

The training process thus entails finding a set of $\theta_{i}$ that fit the target population; in other words, this implies minimising the following cost function through, for instance, a gradient descent algorithm or a least squares approach:

\begin{equation}
	J(\theta_0,\theta_1 ) =  \frac{1}{2m}  \sum^{m}_{i=1}( h_{\theta_{x}}^{i} - y^{i} ) ^ {2}.
\end{equation}

Real-world problems are usually not restricted to one independent variable, thus requiring the use of a multivariate regression. This, nevertheless, leads to the problem of overfitting: when too many features are present, the learned hypothesis may fit the training set very well, but fail to generalise to new examples.

If reducing the number of features is not an option, the solution is to resort to \textit{regularisation}: all features are kept, but the parameter values are reduced. In regularised linear regression, we choose  $\theta$ to minimise:

\begin{equation}
	J(\theta )=  \frac{1}{2m}  \sum^{m}_{i=1} (h_{\theta_{x}}^{i} - y ^{i}))^{2}+ \Lambda \sum^{n}_{i=1} \theta^{2}_{j}.
\end{equation}

\item \emph{Logistic regression}.
There are many situations in which we need to predict the value of a dependent variable as a function of other independent variables, as in regression, but in which the former is discrete, and specifically binary ({\it e.g.} the patient improves after a treatment, the credit is paid, the email is  spam, {\it etc.}). The value associated to each observation $i$ is thus either $y_i=0$ or  $y_i=1$. 
Mathematically, $y_i$ is a realisation of a random variable $Y_i$ that can take  values one and zero with probabilities $p$ and $1- p$, respectively. 
Assume that $Pr(Yi_ = 1|X = x) = p(x;\theta)$, for some function $p$ parameterised by
$\theta$, and  assume that observations are independent of each other. The conditional likelihood function is
\begin{equation}
	\prod^{n}_{i}  Pr{Y = y_i | =x_i} = \prod^{n}_{i} p(x_i; \theta )^{y_i} (1 - p(x_i;\theta)^{1-y_i}).
\end{equation}

In this case it is important to see that the mean and variance depend on the underlying probability. As a consequence, any factor affecting the probability will also alter both the mean and the variance of the observations. This suggests that a linear model that allows the predictors to affect the mean but assumes that the variance is constant will not be adequate for the analysis of binary data.
 
In this kind of models, the constraint, $\pi = p(x_i ;\theta)$, tells us that $\pi$ must be the same whenever $x_i$ is the same, and if $\pi$ is a continuous function, then similar values of $x_i$ must lead to similar values of $\pi$ . Assuming $p$ is known (up to parameters), the likelihood is a function of $\theta$, then we can estimate $\theta$ by maximising its likelihood.

Summing up, we have a binary output variable $Y$, and want to model the conditional probability $Pr(Y = 1|X = x)$ as a function of $x$; any unknown parameters in the function are to be estimated by maximum likelihood. How can we use linear regression to solve this problem? Logistic regression uses a logistic (or {\it logit}) transformation of  $\log p$, by  $\log \frac{p}{ 1-p} $. This can be transformed into a linear function of $x$ without fear of nonsensical result.
Formally, the model logistic regression model is:

\begin{equation}
	\log \frac{p(x)}{1 - p(x)}= \beta_0+ x\beta 
\end{equation}

The final classification model is then given by the solution of the equation $\beta_0 + x\beta=0$, which represents the line that separates the two classes. 
The way to minimise the misclassification error is to predict $Y = 1$ when $p\geq 0.5$
and $Y = 0 $ when $p < 0.5$. This is to say that non-negative values of  $\beta_0 + x\beta$ predict value 1, and 0 otherwise.

\item \emph{Bayesian networks}. Probabilistic graphical model that represents a set of random variables and their conditional dependencies through a directed acyclic graph \cite {friedman1997bayesian}. Edges represent conditional dependencies, while nodes that are not connected represent variables that are conditionally independent of each other. Each node is associated with a probability function that takes a particular set of values for the node's parent variables as the input, and gives the probability (or probability distribution, if applicable) of the variable represented by the node as the output.

\item \emph{Decision trees}. Involves a set of techniques aimed at generating comprehensive tree structures that classify records by sorting them based on attribute values. Each node in a decision tree represents an attribute in a record to be classified, while each branch represents a value that the attribute can take - see Fig \ref{fig:DMAlg} Bottom Left for a graphical example. Decision trees where the target variable can take on continuous values are called regression trees. Let $D$ be the set of training records that reach a node. The general procedure to build the tree is as follows:

\begin{itemize}
	\item If $D$ contains records that belong to the same class, then this is a leaf node.
	\item If $D$ contains records that belong to more than one class, use an attribute test to split the data into smaller subsets. Apply the procedure recursively on the obtained subset.
\end{itemize}

Depending on the criteria (information gain, Gini index, {\it etc}.) chosen to decide the splitting point, different algorithms have been described in the following literature. Gini index is used in CART \cite{cart84}, SLIQ \cite{mehta1996sliq}, SPRINT \cite{shafer1996sprint};  information gain is used in ID3 \cite{quinlan1986induction} and in the well known $C.45$ \cite{quinlan2014c4}.
The main advantage of decision trees is that they are simple to understand and requires little data preparation, being able to handle both numerical and categorical variables. Its ability to perform well with large datasets has made decision trees one of most used techniques, even though they present the drawback of being based on heuristics where locally-optimal decisions are made at each node.

\item \emph{Random forests}.
When problems are characterised by a large number of variables, each one of them encoding very little information, the best strategy is to grow an ensemble of trees and letting them vote for the most popular class. Following this idea, random forests are a combination of tree predictors such that each tree depends on the values of a random vector sampled independently and with the same distribution for all trees in the forest \cite{breiman01}. Each tree in random forest is grown as follows:

\begin{enumerate}
	\item Sample with replacement a given number of cases from the training set at random. This sample will be the training set for growing the tree.
	\item Given $M$ input variables, select randomly at each node  $m<<M$ variables, and choose the best one to split the node. 
	\item Grow the tree with no pruning.
\end{enumerate}

Random forests have two important advantages: first, they do not suffer from overfitting, and can thus be use in small data sets. Second, they have been shown to outperform most known algorithms, in terms of accuracy.

\item \emph{Markov Random Field}.
Markov Random Fields (MRF) are the outcome of a branch of probability introduced in the 80s \cite{kindermann80mrf}. Specifically, an MRF is similar to a Bayesian network, in that its aim is to represent dependencies, but extends the latter by introducing concepts such as undirected and cyclic dependencies. This flexibility makes them very appropriate for classification tasks in vision problems, {\it i.e.} when the focus is on the relationship between a pixel and its neighbours.

\item \emph{Hidden conditional random fields}.
Hidden conditional random fields (HCRF) are discriminative latent variable models for classification problems in structured domains, where inputs can be represented by a graph of local observations \cite{quattoni2007hidden}. An HCRF framework learns a set of latent variables conditioned on local features. They have been shown to successfully learn the hidden structure of a given classification problem, provided an appropriate validation of the number of hidden states is provided. HCRFs learn not only the hidden states that discriminate one class (label) from all of the others, but also the structure that is shared among labels. The main limitation of HCRFs is that finding the optimal number of hidden states for a given classification problem is not always intuitive, and consequently learning the correct number of states often involves a trial-and-error process with cross-validation, which can be computationally very expensive.

\item \emph{Artificial Neural Networks(ANN)}. The structural aspects of biological neural networks inspired this method. ANNs are represented by a set of connected nodes in which each connection has a weight associated with it, and the network learns the classification function adjusting the node weights.

Here we describe the simplest possible neural network, one which comprises a single neuron. Let us imagine that the inputs of this neuron are  $x_1, x_2, x_3, \ldots x_n$, plus a constant term $1$. The output is defined by: 
\begin{equation}
	f(W^T, x)=\sum^{n}_{1}{W_ix_i+b},
\end{equation}

where $f$ is the activation function. If one chooses $f$ to be the sigmoid function
\begin{equation}
f(x)= \frac{1}{1 + \exp(-x)}	,
\end{equation}

then the neuron represents a logistic regression model.
A neural network is a set of interconnected neurons, where the neuron's output can represent the input for some other neuron in a different layer.

The simplest kind of neural networks, the single layer perceptron \cite{rosenblatt1958perceptron}, has two important drawbacks: ({\it i}) perceptron-like methods are binary, and therefore require splitting multi-class problems into multiple binary sub-problems; and ({\it ii}) single layer perceptrons are only capable of learning linearly separable functions, and thus are not suitable for the kind of problems usually found in real KDD applications. The back-propagation algorithm \cite{Werbos74}, used in conjunction with an optimisation method such as gradient descents, allows the avoidance of those problems. The method calculates the gradient of a loss function with respect to all of the weights in the network. The gradient is fed to the optimisation method, which in turn uses it to update the weights of the nodes in the network, trying to minimise the loss function - see Fig \ref{fig:DMAlg} Bottom Right. All of the basics to build neural networks can be found in Refs. \cite{hagan1996neural, Zurada92a}.

\item \emph{Instance based algorithms}, also called {\it memory-based learning} \cite{daelemans2005memory}. Consists of a family of learning algorithms that, instead of performing explicit generalisation, compares new problem instances with instances seen in training, which have been stored in memory. They are called instance-based because they construct hypotheses directly from the training instances themselves. One advantage they provide over other methods is their ability to adapt the model to previously unseen data. This type of algorithms requires less computation time during the training phase compared to other algorithms, as no model should be created beforehand. {\it kNN} \cite{altman1992introduction} is one of the most well known examples of ``lazy-learning algorithms''; it classifies a record using the most frequent class in its k-nearest records, by means of some distance metric - see Fig \ref{fig:DMAlg} Top Centre. kNN uses a function that is only locally approximated, and all computation is deferred until classification. The kNN algorithm is among the simplest of all machine-learning algorithms that can be used for classification and regression.

\item \emph{Support Vector Machines(SVM)} \cite{cortes1995support}. Binary linear classifiers that model concepts by creating hyperplanes in a multidimensional space, which can be used for both classification and regression. A good separation is achieved by the hyperplane that has the largest distance to the nearest training-data point of any class, as this minimises the error. The features available in the data set indicate the axes of this space, whose values should always have a numerical form. Records are mapped into this space, and the best linear separation between them is then calculated - see Fig \ref{fig:DMAlg} Top Right.

\item \emph{Multiple Kernel Learning} \cite{lanckriet2004learning}. Multiple Kernel Learning (MKL) methods expand the concept of kernel methods to which SVMs belong, by using a predefined set of kernels, and by learning an optimal linear (or non-linear) combination of kernels as part of the algorithm \cite{gonen2011multiple}. MKL models are able to select an optimal kernel and the corresponding parameters from a larger set of kernels, reducing the bias introduced in the kernel selection. 
These different kernels may correspond to using different notions of similarity, but may also respond to the use of information coming from multiple sources (different representations or different feature subsets). By using this technique, it is possible to generate a kernel for multiple data sources, by combining kernels already established for each individual source. For this reason, one of their applications is biomedical data fusion. MKL algorithms have been developed for supervised, semi-supervised, as well as unsupervised learning.

\item \emph{Rule induction}. Since regularities hidden in data are frequently expressed in terms of rules, rule induction is one of the tools of machine learning used for classification tasks. Rules are normally expressed in the form: {\it If} $a_1 = v_1$ {\it and} $a_2 = v_2, \ldots$ {\it and} $a_n= v_n, $ {\it then} $b = y$, where $a_i$ and $v_i$ are respectively the condition (independent) variables and their values, and $b, y$ are the decision (dependent) variable and its value. The algorithms aim at creating the smallest rule-set that is consistent with the training data. Refs. \cite{grzymala1997new} and \cite{Cohen95fasteffective} respectively present the LERS and RIPPER induction systems, which have been shown to be competitive with C4.5.

\end{description}

\subsubsection*{Feature selection}
\label{sec:featureSelection}

Feature selection methods became popular in the late 90's. They consist of algorithms used for selecting a subset of features, from the original data set, for subsequent analysis; discarding all other features which are expected to be irrelevant for the problem at hand. Data sets in the 90’s still had limited sizes and relatively few variables. Feature selection methods were developed to help in data understanding, reducing training times, improving prediction performance, and helping to deal with the problem known as \textit{curse of dimensionality} \cite{verleysen2005curse}. Data analytics tasks on domains such as gene and protein expression, chemistry or text classification have made feature selection a need not only for the high number of variables, but also for the shortage of records - {\it e.g.}, in biophysics, the reduced number of patients usually available. Refs. \cite{blum1997selection, guyon2003introduction} extensively review methods for feature selection. 

Broadly speaking, feature selection methods can be classified in three different families:

\begin{itemize}

	\item \textit{Filters} select subsets of variables as a pre-processing step, independently of the chosen predictor. In some cases they compete with wrappers as being the most efficient approach. One of the most relevant examples of feature selection filters is the {\it Recursive Feature Elimination} (RFE), which is based on repeatedly constructing a model (using some suitable classifier) and removing features with low weights ({\it i.e.} of low relevance). When features are added, instead of being eliminated, the result is a {\it forward} strategy.

	\item \textit{Wrappers}. The wrapper methods, popularised by Ref. \cite{kohavi1997wrappers}, assess subsets of variables according to their usefulness to a given predictor, using a search algorithm to search through the space of possible features, and evaluating each subset by running a model on it. Only when the number of variables is reduced, all variable combinations can be exhaustively evaluated, thus avoiding the search heuristic. The machine-learning algorithm is taken as a black box, to score subsets of variable according to their predictive power. Wrappers can be computationally expensive and have a risk of overfitting in the model, in which case coarse search strategies may be applied \cite{reunanen2003overfitting}.

	\item \textit{Embedded techniques} implement the same idea of the wrapper methods, but proceed directly to the optimisation of a two-part objective function, with a goodness-of-fit term and a penalty for a large number of variables. These methods perform variable selection in the process of training, and are usually specific to some learning algorithms. Embedded methods that incorporate variable selection as part of the training process may be more efficient in several aspects. First, they make better use of the available data by not needing to split the data into training and validation sets. Additionally, they reach a solution faster by avoiding retraining a predictor from scratch for every variable subset investigated.
\end{itemize}

\subsubsection*{Validation}
\label{sec:validation}

Validation is a broad term, encompassing two different families of metrics:

\begin{itemize}
	\item \emph{Metrics for Performance Evaluation}. The aim is to evaluate the performance of a model in a reliable way, {\it i.e.} in a way as general as possible.
	\item \emph{Methods for Model Comparison}. In this case the problem is how to compare the relative performance among competing models, especially in those cases in which the size of the datasets can make the difference in accuracy not statistically significant. A confidence interval should be established for comparison issues. While such problem will not be tackled in this review, more information can be found in Refs. \cite{moreno1998intrinsic, lahiri2001model, berger2001objective}.
\end{itemize}

Here we focus on the first problem, concerning the evaluation of the predictive capability of the model generated through a data mining task.

As previously discussed, the performance of a model is inversely proportional to the number of errors it produces, and can therefore be measured through the {\it precision} metric ({\it i.e.} the proportion of correct positive forecasts). However, this is seldom enough to assess the real performance, and an additional distinction should be implemented between the {\it training} and the {\it generalisation} errors. The former quantifies the error rate yielded by the model when processing the training set. Clearly, a good classification model should not only fit the training set, but it should also perform correctly against records never seen before. This second part is called the generalisation error, and it is meant to measure the model's performance in a real environment.

Note that the generalisation error can be significantly larger than the training error, and this usually occurs when the training is performed over a small data set, or when this data set does not correctly represent the characteristics of the whole universe. This situation is called {\it overfitting}, and its probability increases with the complexity of the model (as more training data are required to optimise a complex model). For this, the simplest model is chosen when two models are available and both display the same generalisation error. This is done by applying the {\it Occam's razor} \cite{Rasmussen01occam'srazor} principle.

The generalisation error of a data mining model is in principle easy to evaluate. For example, one just needs to consider a new data set, {\it i.e.} one not used in the training phase, and measure the error rate of the model within it. Since the amount of available data may often be limited, it is necessary to resort to the {\it hold out} method, in which the original dataset is split in training and validation subsets. The former (typically, two thirds of the total available records) is then used to build the model, while the latter (the remaining third) is used to test the model. The global accuracy is defined as the rate of correctly classified instances by the total number of predictions made, while the error rate is defined as the number of errors in the prediction divided by the total number of predictions. The hold out method can be repeated several times, to improve the accuracy of the estimation of the model performance.
 
A more complete and reliable method is the {\it cross-validation} \cite{golub79, kohavi1995study, efron1995cross}. It involves the estimation of the extra-sample error by applying the trained model to several validation sets, and averaging the error rates obtained. In the {\it k-fold cross-validation}, the original sample is randomly partitioned into $k$ equal sized subsamples. Of these $k$ subsamples, a single subsample is retained as the validation data for testing the model, and the remaining $k - 1$ subsamples are used as training data. The cross-validation process is then repeated $k$ times, with each of the $k$ subsamples used exactly once as the validation data. The average value of the error obtained in the $k$ executions will be used for estimating the error. In the special case when $K = N$, being $N$ the number of records in the data set, the method is called {\it Leave-One-Out Cross Validation} ({\it LOOCV} in short).

\subsubsection*{Data mining software}

\begin{description}

	\item \emph{Weka} \cite{holmes1994weka, frank2004data, hall2009weka}. Probably the best known and most used software for scientific data mining, Weka is a collection of machine learning algorithms written in Java. All algorithms can be executed on data files, or called from within any Java program. Freeware. Available at \url{http://www.cs.waikato.ac.nz/ml/weka/}.

	\item \emph{KNIME} \cite{berthold2008knime}. Based on the Weka libraries, KNIME is a data analytics platform which allows executing any data mining task (from data access, data transformation, predictive analytics and visualisation) through a graphical workbench. Freeware. Available at \url{https://www.knime.org}.

	\item \emph{R} \cite{ihaka1996r}. Free software environment for statistical computing and graphics. Available at \url{https://www.r-project.org}.

	\item \emph{SAS \circledR}. Software suite developed by the SAS Institute for advanced analytics, multivariate analyses, business intelligence, data management, and predictive analytics. Commercial. Information at \url{https://www.sas.com}.

\end{description}

Beyond these four software packages, the interested reader can refer to \url{http://www.kdnuggets.com/software/suites.html} for a more comprehensive guide.

\subsection{What is ``Big Data''?}
\label{sec:BigData}

The challenge to extract patterns from large volume of data was highlighted in the original definition of the Knowledge Discovery in Databases (KDD) process (see Section \ref{sec:dataMining}) \cite{fayyad1996advances}. This definition dates back to 1997, where ``large'' may have implied just tens or hundreds of megabits of information. Since then, the data targeted for real-world analyses have substantially increased in volume, velocity and variety, increasing the complexity of extracting value. This data explosion is behind the {\it Big Data} concept. 

Commonly found in any new field of research, there is still no generally accepted definition of Big Data and different contexts have given birth to different descriptions and characterisations. The origins of the name seem to date back to 1998, with two specific references: Ref. \cite{weiss1998predictive} in computer science and Ref. \cite{diebold2003big} in statistics/econometrics \footnote{While beyond the scope of this review, further information on the historical analysis of terms like ``data mining'', ``data science'' and ``Big Data'' can be found at: \url{http://www.forbes.com/sites/gilpress/2013/05/28/a-very-short-history-of-data-science/}}. For the sake of completeness, we will first review some of the most commonly used definitions and then describe the important goals of Big Data.

One of the first comprehensive definitions of Big Data was provided by McKinsey \& Company and states: ``Big Data refers to data sets that cannot be acquired, stored, processed and managed by classic database software within a tolerable time’’ \cite{manyika2011big}. This definition highlights the importance of the volume of data, which may grow faster than the capacity of storage and computation technologies. In the same year, the International Data Corporation (IDC) stated that ``big data technologies describe a new generation of technologies and architectures, designed to economically extract value from large volumes of a wide variety of data, by enabling high-velocity capture, discovery, and/or analysis'' \cite{gantz2011extracting}. This can be seen as new elements were introduced, beyond the data size: variety and velocity of data, plus the value of the analyses that can be performed on such data. This was later synthesised by Doug Laney, who in 2013 defined Big Data by a 3 {\it V}s model: Volume, Velocity and Variety \cite{laney20013d}. A different approach was proposed one year later by the National Institute of Standards and Technology (NIST), which focused on the technological aspects of Big Data: ``Big Data consists of extensive datasets primarily in the characteristics of volume, variety, velocity, and/or variability  that require a scalable architecture for efficient storage, manipulation, and analysis'' \cite{NIST2014}.

It is also worth analysing a recent trend in the study of Big Data, which can be synthesised through the HACE Theorem \cite{vilas2012big, wu2014data, tamhane2015big}: ``Big Data starts with large-volume, \emph{H}eterogeneous, \emph{A}utonomous sources with distributed and decentralised control, and seeks to explore \emph{C}omplex and \emph{E}volving relationship among data.'' Two of the characteristics underlying this definition have a clear relation with complex networks:

\begin{description}

	\item \emph{Autonomous Sources with Distributed and Decentralised Control}: Each data source is able to produce and collect information without involving any centralised control. Such distributed nature offers opportunities for understanding patterns of interconnections as complex networks. Examples include the World Wide Web \cite{pastor2001dynamical, ravasz2003hierarchical}, and distributed wireless sensor systems \cite{pottie2000wireless, akyildiz2002wireless}. The challenges are to control and optimise the dynamics of those systems, even in the absence of a central director.
	
	\item \emph{Complex and Evolving Correlations}: In standard data sets, {\it e.g.} such as those used in many biomedical problems, the environmental conditions are kept strictly under control in order to assume the stationarity of the data. In other words, starting records obtained at the beginning of the analysis are as valid as the final ones. This is not necessarily true in a Big Data environment. The real world is dynamic and so are the data acquired from it. Therefore, all relations should be studied taking into account their temporal evolution \cite{holme2012temporal,holme2013temporal}.
	   
\end{description}

Three main features characterise Big Data:

\begin{itemize}
    \item \emph{Fast Processing}: In order to extract new insights from vast volume of data, as close to real time processing as possible is desired.  	
	\item \emph{Prediction}: Accurate prediction is used to apply mathematical and statistical methods to these large quantities of data, in order to infer probabilities, with the aim of extracting useful insights.
	\item \emph{Relation}: The individual pieces of information are not independent and acquire a meaning when combined together.
\end{itemize}

Despite the variety of definitions and concepts, a common notion is recognised: Big Data is a reality that mankind must both face and exploit in a structured, aggressive and ambitious way, to create value for society, its citizens, and its businesses in all sectors. Data availability and the access to data sources is paramount. Furthermore, there is a broad range of data types and sources, including structured and unstructured data, multi-lingual data sources, data generated from machines and sensors, data-at-rest and-data-in-motion. The overall value is extracted by acquiring those data, combining data from different sources, and providing access to them minimising delays; all while ensuring data integrity and preserving privacy.

While complex network theory and Big Data have had very few contact points, this is expected to change in the near future, as applications combining both concepts are starting to appear - see Section \ref{sec:BDandCN} for an example.

\subsection{The limits of data mining in physics}

As previously discussed, there are many real-world problems, especially in the field of biophysics, which can be seen through the lens of data mining. As an example, whenever confronted with situations in which subjects need to be diagnosed into patients and healthy subjects, this boils down to a classification task. However, data mining has not been historically well received by the physics community. Before delving into how the two fields can (and should) be leveraged for one other, we analyse the intrinsic limitations of data mining in terms of its application to physical problems.

One of the distinguishing characteristics of physical science is its effort to achieve higher levels of abstraction. Consider gravity as an example: the aim is not just to explain the trajectory of a pencil falling through the air, but also to explain the dynamics of {\it any} free-falling object. The concrete nature of the pencil is thus disregarded, or rather, abstracted from the considered dynamics. The same is true in complex networks analysis. The details of each system element are disregarded, and the representation of their interactions is obtained in an abstract form. On the one hand, while data mining provides some level of abstraction, this is based on an inductive (sometimes, an abductive) reasoning over the available data, and may be far from what a physicist would like to achieve.
To further illustrate this point we can use a simple biological example. Since the seminal work of Gregor Johann Mendel in 1865, the laws of genetic inheritance are well understood \cite{bowler2000mendelian}. In his original experiments two types of pea plants were crossed: purple and white flowers. After seeing the characteristics of the offspring, he postulated some simple rules for inheritance traits. Mendel then concluded that each plant has two alleles of the same genes that are randomly inherited from the two parents. This defines the colour of the flowers, with each allele being of two variants ($p$ and $w$). Purple is the {\it dominant} form, and flowers will only be white if the pair $ww$ is inherited. Note the abstraction: the plant colour ({\it i.e.} the phenotype) is explained by something not visible, the genes ({\it i.e.} the genotype).

\begin{figure*}[!tb]
	\centering
		\includegraphics[width=0.99\textwidth]{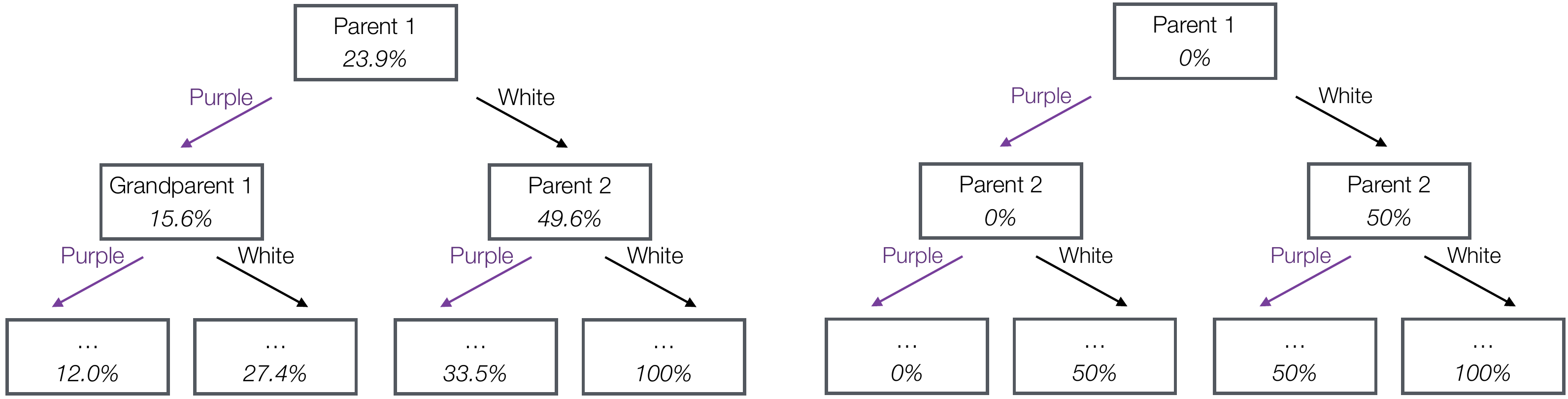}
	\caption{Formulating traits inheritance laws using data mining, using raw data analysis (Left) and by including heuristic about the statistical nature of results (Right). See main text for details.}
	\label{fig:GeneticExample}
\end{figure*}

If these laws were not known, one may attempt to understand inheritance traits through data mining. As the alleles ({\it i.e.} the genotype) are not known, the initial data set would contain a list of flower colours (the phenotype), for the target plant, its two parents, and four grandparents. A data mining task could then be executed to predict the colour of the target plant, just by knowing the colour of its predecessors.
When this is executed for a data set comprised of $1000$ different trials, it results in the Decision Tree model depicted in Fig. \ref{fig:GeneticExample} Left: each box is a decision node, where one feature (reported within the box) is evaluated, as represented by the arrows. The italic number inside the boxes represents the probability of obtaining a white flower offspring. In other words, if both parents are white then there is a $100\%$ probability of getting a white offspring (right-most sequence of arrows).
In terms of precision, the data mining model is not able to exactly forecast the colour of the offspring, and yields an error rate of $18.6\%$ (using a Leave-One-Out cross-validation). Although we know that inheritance traits are random, a preliminary reaction may be to assume that the model is wrong and that it not adequately captures the inheritance principles. In the end this is not completely true, as part of the model is correct; when two parents are white, the offspring will also be white.

To take this approach one step further, we can forecast offspring colour in statistical terms. Here, we introduce {\it heuristic} into the data mining problem \footnote{{\it Heuristic} in data mining refers to any approach to problem solving that employs a practical method not guaranteed to be optimal or perfect, but sufficient for the immediate goals. In other words, heuristic implies the inclusion of {\it a priori} information inside the problem.}. Instead of trying to predict the colour of the offspring, we may predict the probability of having white offspring. The resulting model is depicted in Fig. \ref{fig:GeneticExample} Right; note that numbers inside the box now indicate the predicted probability of having a white offspring. The model is still not accurate (for instance, the probability of having white offspring from two purple-coloured parents is not zero, as displayed in the left-most sequence), but it is now able to identify the two most important factors as the colours of each parent.

In this simple example, we observe that it is not sufficient to create a theory of inheritance traits. Data mining is simply unable to provide us with a sufficient level of abstraction from the overall information. Fig. \ref{fig:GeneticExample} Right does not yield concepts such as the presence of two alleles, or the existence of a dominant trait. Still this exemplifies how data mining can help detect the elements of the analysis that are important, how elements can be combined or simplified, and how data mining can provide insight into a large set of data in an automatic way.

Data mining is an abductive, rather than a deductive process: it simply induces, from the available data, what has happened in the past. If certain important variables are not present, or their description is faulty, then data mining comes to wrong conclusions. Therefore data mining cannot provide insight into what is omitted within the data, and in the case alleles existence, it does not yield higher levels of abstraction. If such data were available, or if all of the data a biologist would use to create an abstract theory were accessible, then such theory could be automatically inferred from the data.

Through this example we can observe a necessary condition. Although data mining creates abstract models from initial data, the data themselves should contain all the elements required to perform an accurate abstraction. Several examples can be found in the Literature in which such abstraction process is performed. For instance, Ref. \cite{lenat1977automated} proposes an algorithm for rediscovering concepts in number theory, given an initial set of concepts and heuristics.
Ref. \cite{langley1981data} describes a computer program for the discovery of physical laws, which is able to recover the law of ideal gas, using input values observed in experiments, along with number of moles, temperature, pressure and volume. With this the program is able to recover the law of ideal gas. This latter case exemplifies the previously discussed limitation: given all the parameters, it is possible to recover the true law. Nevertheless if some of these elements are missing, no data mining algorithm can suggest, for instance, that the temperature is a key element.

Therefore, this important limitation has to be taken into account by any researcher who wants to use data mining in a physical (or biological) scenario. However, to move from a simple prediction to a general model, another essential ingredient is required: the human ability for creating abstract models. Indeed, good data scientists are not only people who simply applies different algorithms; their {\it spark} and intelligence should also be included in each step described in Section \ref{sec:dataMining}, such as business understanding, data understanding and data preparation.

\subsection{Hands-on: analysing EEG data}
\label{sec:HandsOn_Intro}

\begin{figure*}[!tb]
	\centering
		\includegraphics[width=0.9\textwidth]{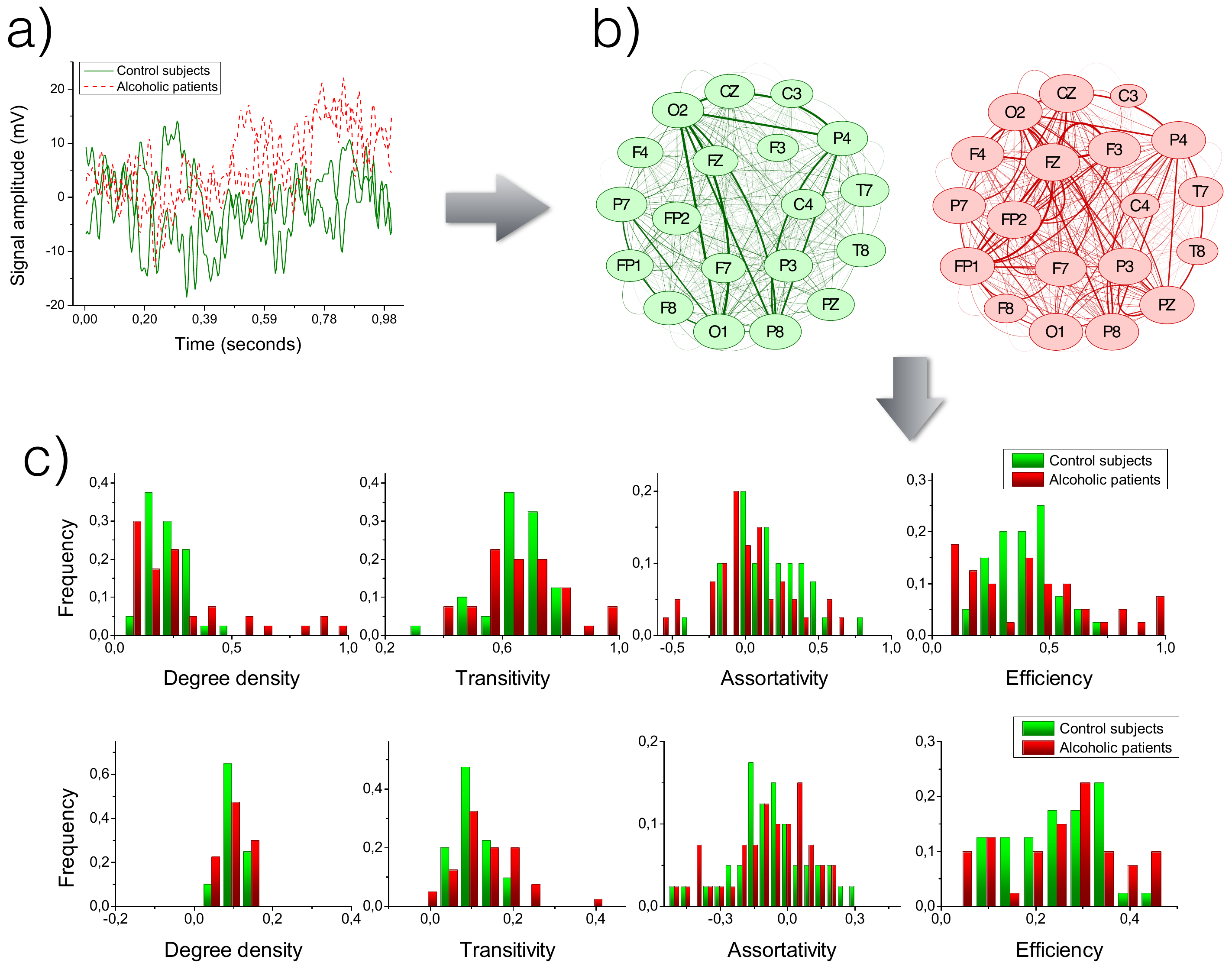}
	\caption{A standard functional network analysis. (Top Left) Example of four time series, two for control subjects and two for patients, representing the brain activity during a cognitive task. (Top Right) Two functional networks (green for the control subject, red for the patient) obtained through linear correlation. The links weight is represented by the width of the line; the size of nodes indicates their centrality. (Bottom) Histograms of four metrics, for networks obtained through correlation (top) and Granger Causality (bottom).}
	\label{fig:FigureHO01}
\end{figure*}

How would a network theory practitioner understand the structure encoded in EEG data? The analysis should, of course, start with a real data set, and in this case we consider a set of EEG recordings for healthy (control subjects) and alcoholic (patients) subjects; the aim will be to assess whether this specific health condition has an effect on the overall brain organisation.
The data set is freely accessible at \url{http://archive.ics.uci.edu/ml/}) \cite{Lichman2013}. It contains measurements from 19 electrodes placed on subject’s scalps which were sampled at 256 Hz (3.9-msec epoch) for 1 second \cite{zhang1995event}, with the electrode positions located at standard sites \cite{sharbrough1991american}. Each subject was exposed to a single stimulus, corresponding to pictures of objects chosen from the 1980 Snodgrass and Vanderwart picture set \cite{snodgrass1980standardized}.  Of the whole set, we randomly selected 40 recordings of control subjects, and 40 of patients. Our data set is then represented as a $80 \times 19 \times 256$ tensor. An example of four randomly chosen time series is depicted in Fig. \ref{fig:FigureHO01} Top Left.

To construct a (functional) network for each subject, it is necessary to define the meaning of nodes and links. The solution to the former problem is simple, as EEG electrodes provide a natural set of nodes. A link can then represent whether and the extent to which two nodes are connected, being synchronisation is a typical choice \footnote{The attentive reader would note that some intermediate steps may be required, {\it e.g.} filtering the data for spurious or erroneous samples, or selecting only some frequency bands. As our aim is not to yield a correct neurological analysis, but instead to exemplify the process to be followed, we have omitted these pre-processing tasks.}. We here consider two possibilities: a linear Pearson's correlation, and a Granger Causality \cite{granger1988some, granger1988causality}. As an illustration, in the case of the linear correlation, a pair of nodes $i$ and $j$ would be connected with a strength equal to $w_{i, j} = cov(x_i, x_j) / ( \sigma_{x_i} \sigma_{x_j} )$, where $x_i$ and $x_j$ are the time series recorded at nodes $i$ and $j$, and $cov$ is the covariance. For each subject, we thus obtain a {\it weighted adjacency matrix} $\mathcal{W}$, whose element $w_{i,j}$ indicates the strength of the connection between nodes $i$ and $j$ - see two examples in Fig. \ref{fig:FigureHO01} Top Right.

Finally, we calculated four common topological metrics from each network after applying an arbitrary threshold; the weighted adjacency matrix is thus converted to an unweighted one $\mathcal{A}$, such that $a_{i, j} = 1$ if $w_{i, j} > \tau$, and $a_{i, j} = 0$ otherwise. Here we consider the following topological metrics: the degree density, transitivity, assortativity and efficiency. The corresponding histograms, comparing control subjects and patients, can be found in Fig. \ref{fig:FigureHO01} Bottom, for networks created with correlation (top) and Granger Causality (bottom).

Some interesting conclusions can already been drawn. For instance, in the case of the linear correlation, patients seem to have higher degree density, transitivity and efficiency, thus indicating a more strongly connected network. Although less evident, similar results, particularly a higher efficiency for patients, are observed in the case of the Granger Causality.

One may stop here, add some statistical tests to confirm that the differences are indeed significant, and probably have a nice preliminary paper on the effects of alcoholism on brain activity. However, one may also start asking important questions. Is it better to use a linear correlation, or the Granger Causality? In other words, are these two metrics assessing different and significant aspects of brain dynamics? What is the best topological metric to describe the differences between both groups? Is the applied threshold the best possible one or can it be optimised? And finally, what is the discriminative power of this analysis? Could we diagnose the health condition of a new subject?

Answering these questions, or more specifically, understanding how we can answer them by using data mining is the objective of this review. We leave these questions unanswered till the next section where the use of classification algorithms is further discussed.

\section{Classification}
\label{sec:classification}

In this section, we start dealing with one of the most important tasks in data mining: {\it classification}, {\it i.e.} the creation of models to predict the class of an unseen instance. As discussed in the introduction, this has many applications in cross-disciplinary fields, as in medicine, with the creation of diagnostic tools. Before presenting examples of classification tasks involving complex networks (Sections \ref{sec:ClassifyNet} to \ref{sec:class_omics}), two additional issues should be discussed. First, the relevance of using classification algorithms, against resorting to simple statistical metrics (Section \ref{sec:pvalue}); second, the advantages associated with using network topological metrics, against the direct analysis of the network adjacency matrix (Section \ref{sec:NecessityMetrics}).

\subsection{Statistics vs. classification}
\label{sec:pvalue}

Suppose one is studying a given network metric, {\it e.g.} the clustering coefficient, to see if it indicates relevant differences between populations. Its relevance can be assessed in two different ways. The first relies on the use of some non-parametric statistical test, {\it e.g.} the Kolmogorov-Smirnov test, to assess if samples corresponding to both groups are drawn from the same distribution. If the hypothesis can be rejected, then the two groups present a statistically significant difference in their network structure. The final result of this analysis would be a $p$-value, which can then be compared against a significance level threshold (usually denoted by $\alpha$).
The second option involves performing a classification task, in which the subjects within the two groups are classified, using the network metric as the input feature. In this case, the output will be a classification score (or equivalently the classification error). The considered metric would be relevant if the score is sufficiently high (or conversely, the error is sufficiently low).

\begin{figure*}[!tb]
	\centering
		\includegraphics[width=0.9\textwidth]{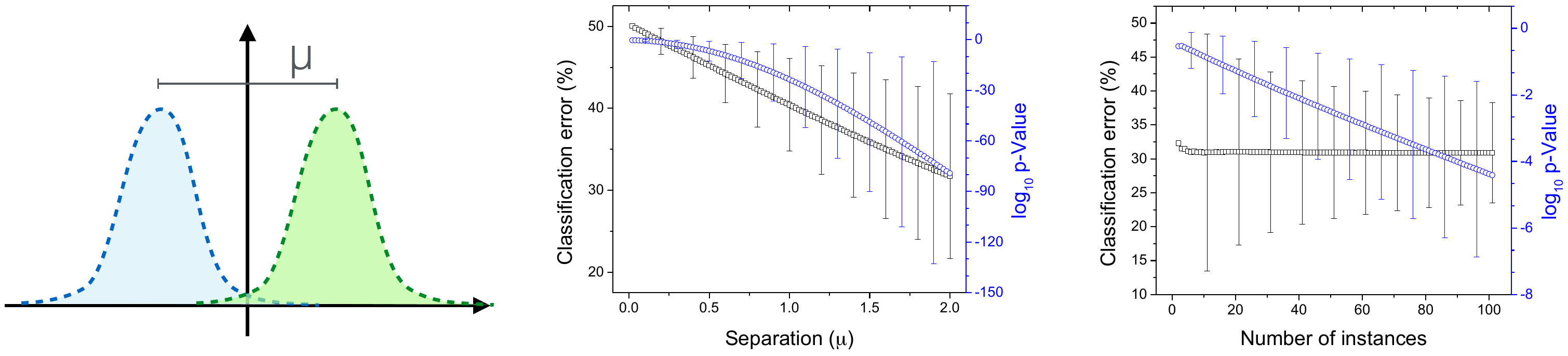}
	\caption{$p$-value {\it vs.} classification score. (Left) Data setup, with two normal distributions of $\sigma = 1.0$ and centres separated by a distance $\mu$. (Centre) $p$-value and classification error as a function of the separation $\mu$. (Right) $p$-value and classification error as a function of the number of samples drawn from each distribution.}
	\label{fig:PValue}
\end{figure*}

In principle, one may expect both methods to be equivalent. A low $p$-value signifies that the values of the considered metric are well separated between both groups, and thus a classification task can be easily performed. Nevertheless, there are situations in which this is not true.
Fig. \ref{fig:PValue} Left depicts a simple situation, with two probability distributions representing the topological values obtained for the two groups of subjects. For the sake of simplicity, both distributions are normal, with $\sigma_1 = \sigma_2 = 1.0$; their centres are separated according to a parameter $\mu$, such that $\mu_1 = - \frac{1}{2} \mu$ and $\mu_2 = \frac{1}{2} \mu$.
When a large number of samples ($n = 1000$) are drawn from both distributions, and the resulting $p$-values and classification errors are calculated as a function of $\mu$, the obtained results are as depicted in Fig. \ref{fig:PValue} Centre. While the $p$-value drops very fast (note the $\log$ scale), the error presents a more flat slope. As an example, in the separation of $\mu = 1.0$, the $p$-value is of the order of $10^{-30}$, while the classification error is still around a $40\%$. This illustrates an important point: when two distributions are different in a statistically significant way, the difference may not be sufficient to enable a successful classification task.
While uncommon, the opposite situation can also appear. Fig. \ref{fig:PValue} Right reports the evolution of both the $p$-value and the classification error as a function of the number of instances (samples) available in both groups (for $\mu = 1.0$). In the case of very small sets ({\it i.e.} below $20$ subjects), the reduced dimension prevents obtaining a $p$-value below the desired significance level, even though a classification can be successfully performed. Notice that this situation is quite common in biomedical problems, where the number of subjects available is seldom in the hundreds.

These results suggest that a researcher should err on the side of caution. The $p$-value provides valuable information about the statistical significance of an analysis, but it is not enough to assess its usefulness. This is especially relevant when the objective is not just theoretical, as when a diagnostic tool is constructed. Conversely, when the available data set is small, the classification score alone may be misleading, and it should be complemented by an analysis of its significance.

\subsection{Are network metrics really necessary?}
\label{sec:NecessityMetrics}

Data mining has previously been defined as the computational process of discovering patterns in sets of data. As such, one may question the necessity and advantage of using a complex network representation to extract structural patterns, as this would be completed by data mining analysis. Suppose we are studying a set of networks, each fully described by an adjacency matrix $\mathcal{A}_i$. The same information can be translated into a vector $\overline{A}_i = \{ a_{1,1}, a_{1,2}, \ldots, a_{n,n} \}$, which can then be used as the set of features for training a classification algorithm. If such algorithm were optimal, in that it could detect any class of patterns, it would describe the system as a network and the topological features extracted would be meaningless.

\begin{figure*}[!tb]
	\centering
		\includegraphics[width=0.9\textwidth]{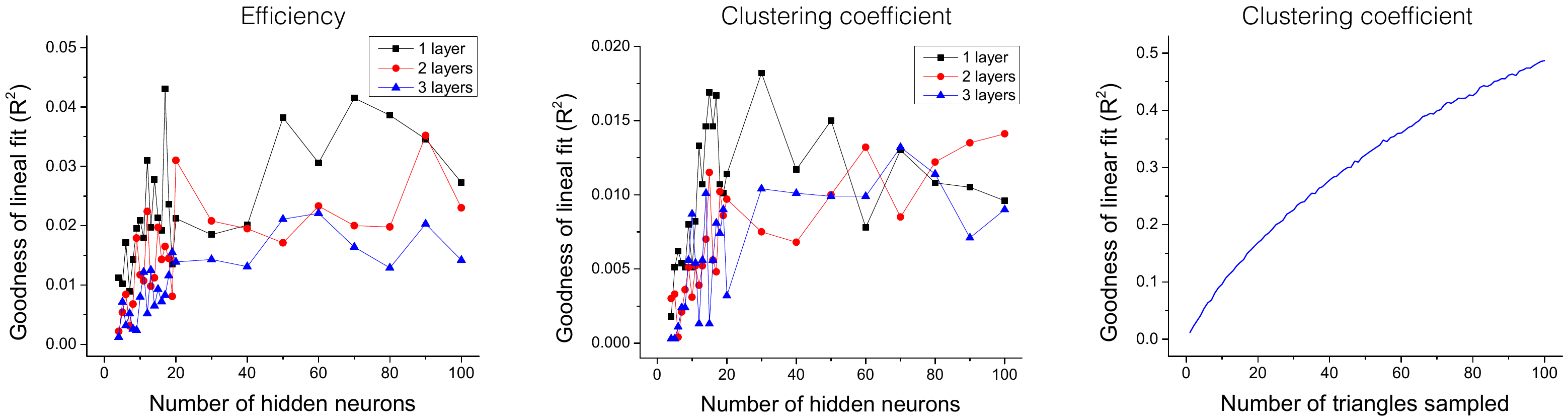}
	\caption{Network metrics {\it vs.} direct data mining. (Left) Coefficient of determination ($R^2$) for the linear fit between network efficiency and the prediction of an Artificial Neural Network - see main text for details. (Centre) $R^2$ for network clustering coefficient. (Right) $R^2$ for network clustering coefficient, and the approximation obtained sampling triangles of the network. }
	\label{fig:Metrics}
\end{figure*}

It is simple to disprove this hypothesis. Fig. \ref{fig:Metrics} presents the results of a process in which $20.000$ random networks of ten nodes have been generated with a fixed link density ($p = 0.3$). On one hand, two topological metrics (efficiency and clustering coefficient) were extracted from these networks. However, on the other hand, the adjacency matrices have been fed inside an Artificial Neural Network (ANN), trained to recover the obtained metric values. Both outputs, {\it i.e.} the true topological values and the estimation obtained by the ANN, are fit linearly, and the coefficient of determination $R^2$ represented in Fig. \ref{fig:Metrics} as a function of the number of nodes and hidden layers in the ANN. The small $R^2$ obtained in each case (below $0.04$) indicates that the data mining model is not able to recover the true topological indicator. This is especially noteworthy in the case of the clustering coefficient, as it is a local metric that can be estimated by just sampling triplets of nodes and evaluating the fraction of nodes that are connected. The $R^2$ obtained in this process, as a function of the number of triplets sampled, is shown in Fig. \ref{fig:Metrics} Right.

This simple numerical experiment demonstrates that  one single data mining algorithm is in general not sufficient to deal with the structure of a sufficiently complex system. Additional reasons for this can be identified:

\begin{description}
	\item \emph{Macro- and meso-scale perspectives.} Data mining algorithms are optimised to extract relationships between a small set of features, but are not optimised to analyse interactions on a global scale. For instance, calculating the efficiency of a network requires estimating each of the shortest paths within it; a task that cannot be solely accomplished by an ANN.
	\item \emph{Statistical perspective.} Most data mining algorithms hitherto designed do not handle a statistical approach well. On the contrary, they are optimised to detect patterns between fixed sets of features. Even in the case of local metrics, such as the clustering coefficient, the ANN is not able to statistically sample the data set.
\end{description}

It is then clear that a complex network approach provides an important added value to the study of complex systems, which cannot be obtained by data mining alone. In the following sections, we will review examples, mainly drawn from biomedical and engineering applications, in which data mining and complex networks are used to create classification models.

\subsection{Classifying with complex networks: Neuroscience}
\label{sec:ClassifyNet}

In biophysics it is often important to be able to {\it diagnose} a disease. Two or more states, {\it e.g.} normal (or control) and pathologic, typically need to be distinguished. The problem then consists in creating meaningful network representations, extracting topological metrics, and assessing the differences between the states. Within neuroscience, a natural way to build a network representation is the use of the functional network concept: synchronisation between time series recorded by means of EEG \footnote{{\it Electroencephalography} (EEG), an electrophysiological monitoring method to record the electrical activity of the brain, both in resting states and during a specific task.}, MEG \footnote{{\it Magnetoencephalography} (MEG): similar to EEG, but focuses on recording the magnetic signals generated by the brain activity.} or fMRI \footnote{{\it Magnetic Resonance Imaging} (MRI), and its {\it functional} equivalent (fMRI), is a medical imaging technique that uses magnetic fields and radio waves to form images of the brain.} can be assessed, and mapped in a functional network.

Historically, most of the classification tasks performed with functional networks and data mining models have been focused on MRI and fMRI. Within this, the most used classification model is by far the Support Vector Machine (SVM). The reasons for this choice are easily identified. SVM is conceptually simple, easy to represent, and is simple to extract biological rules from the model structure. In other words, SVM simplifies the translation of results to a biomedical language.

Numerous neurological pathologies have been studied by means of fMRI and SVM. 
Diagnosing Alzheimer's disease has been extensively discussed in the literature, both with linear \cite{li2013exploring, khazaee2014automatic, khazaee2015identifying} and non-linear kernels SVM \cite{li2012discriminant}.
The discrepancies between young and older adults, or in general children {\it vs.} adults, have also been addressed, due to its relevance in understanding how the brain develops and reaches a mature functional status \cite{dosenbach2010prediction, richiardi2011classifying, meier2012support}. Additional pathologies include:

\begin{itemize}
	\item Schizophrenia, a pathology whose main challenges are its complexity, in terms of multiple symptoms, and the lack of characterisation in terms of local abnormalities of brain activity. The fact that this affects the collective, emergent working activity of the brain suggests that a complex network approach can yield relevant results. This topic is discussed in Refs. \cite{rish2009discriminative, bassett2012altered}.
	\item Major Depressive Disorder (MDD) patients, and their discrimination from non-depressed controls, by means of whole-brain tractography \cite{sacchet2014elucidating}. Results indicate that the best network metric is the small-worldness, a measure of the balance between global integration and local specialisation \cite{humphries2008network, zanin2015alternative}.
	\item Social Anxiety Disorder (SAD), an extreme fear of being scrutinised and judged by others in social or performance situations \cite{liu2015multivariate}.
	\item Attention Deficit Hyperactivity Disorder (ADHD), in which symptoms include difficulty staying focused and paying attention, difficulty controlling behaviour, and hyperactivity \cite{colby2012insights}.
	\item Spinocerebellar ataxia type 7 (SCA7), a genetic disorder characterised by degeneration of the motor and visual systems \cite{hernandez2014whole}.
	\item Discrimination between shiverer (C3Fe.SWV $Mbp^{shi}$/$Mbp^{shi}$) and background control (C3HeB.FeJ) mice, a model for studying Parkinson's Disease \cite{iturria2011automated}.
\end{itemize}

Kernel learning, the concept underpinning SVM, which evolved to Multiple Kernel Learning, has received increasing attention in neuroscience and has yielded relevant results. As already discussed in Section \ref{sec:dataMining}, this presents two important advantages. First, it allows the merging of different types of information, for instance coming from different biological sensors. Second, when spatial information is embedded, it detects the regions of the brain which are relevant for the task. Examples of classification tasks using MKL include Ref. \cite{fekete2013combining}, where this technique has been applied in two tasks: a classification of patients with schizophrenia {\it vs.} controls and a classification of wake {\it vs.} sleep healthy subjects. It has also been used to diagnose amyotrophic lateral sclerosis (ALS) \cite{fekete2013multiple}, Alzheimer's patients \cite{wee2013prediction, dyrba2015multimodal}, Mild Cognitive Impairment (MCI) \cite{wee2013prediction, jie2014integration}, post-traumatic stress disorder (PTSD) \cite{liu2014characterization}, and image-stimuli data \cite{hardoon2010decomposing}.

Most of the examples above share a common feature: fMRI data are analysed by means of a single classification algorithm; specifically SVM (or its evolution MKL). This is an important limitation, as {\it one algorithm does apply to all scenarios}. Different classification models make various assumptions on the underlying data, which can then translate into important differences in classification scores. The problem of comparing classification algorithms is not new in data mining. For example, the interested reader may refer to Ref. \cite{lim2000comparison} for a comprehensive comparison, and to Ref. \cite{salzberg1997comparing} for a discussion of the associated methodological problems. Without delving into theoretical details, we stress once again that, to maximise the relevance of obtained results, scientists should try more than one algorithm. Of the works previously described, three papers have performed such comparison. Specifically, Ref. \cite{rish2009discriminative} compares Markov Random Field, Gaussian Naive Bayes and SVM; Ref. \cite{richiardi2011classifying} Na\"ive Bayes with kernel densities, SVMs with linear and 2nd order normalised polynomial kernels, a radial basis function network, a multi-layer perceptron, and tree-based algorithms (C4.5, functional tree and random forest); Ref. \cite{colby2012insights} linear SVM, SVM with a RBF-SVM, decision stumps as a base classifier in adaboost, random forests, and C4.5 decision trees. Results are heterogeneous; the Ref. \cite{colby2012insights} reports that the best classifier is a SVM with radial kernel, while Markov Random Field outperforms SVM in Ref. \cite{rish2009discriminative}. These results highlight the importance of comparing different classification algorithms, as even similar data (fMRI) may be better understood by different approaches under different conditions.

Examples of classification tasks on functional networks can also be found in studies using EEG and MEG data. In Ref. \cite{van2013improved}, a diagnosis in children with partial epilepsy is performed; resting EEG activity away from epileptic seizures is used to create networks, and then classified by means of Decision Trees (DT). The detection of epilepsy in newborn infants is also the topic of Ref. \cite{temko2012instantaneous}; this study aims at identifying important EEG channels, corresponding to nodes, using SMVs and Gaussian Mixture Models (GMMs) \cite{reynolds2009gaussian}. Ref. \cite{jamal2014classification} uses EEG recordings of children carrying out a face perception task to diagnose autism; the model used is a SVM with polynomial kernels, giving a notable $94.7\%$ accuracy result.

A similar goal is pursued in Ref. \cite{pollonini2010functional} but this time using MEG data. The functional networks associated to resting activity of eight autistic subjects reconstructed using Granger Causality is discriminated with an $87.5\%$ accuracy using SVMs. SVM and Granger causality are also used in Ref. \cite{zouridakis2012functional}, to discriminate mild traumatic brain injury (mTBI) patients. Finally, while the classification is not the main topic of the work, SVM are also used in Ref. \cite{zanin2012optimizing} to discriminate MCI and healthy subjects.

While functional networks obtained from EEG, MEG and fMRI data have yielded important results, Ref. \cite{antiqueira2010estimating} sounds a cautionary note. One of the problems of both EEG and MEG analyses is that the true network, or the network created by all neurons, cannot be recovered due to intrinsic limitations in the available technology. All previous reported studies thus deal with {\it downsampled} networks, whose number of nodes is defined by the number of sensors in the machine. The authors showed, using a large set of synthetic networks with standard topologies and corresponding down-sampled sub-networks, to then compare the resulting topologies using nearest neighbour classification (NNC), multi-layer perceptron (MLP) and support vector machine (SVM), that the topological measures may substantially deviate from that of an underlying, larger network, when sampled at a few surface sites. This suggests that the use of EEG and MEG recordings introduces an important error in the analysis. The problem of defining, or {\it sampling}, the nodes of a network will be further discussed in Section \ref{sec:FSelectionNodes}.

Finally, brain connectomics has also been addressed with data mining tools. For instance, Ref. \cite{fagerholm2015disconnection} studies the effect of traumatic brain injury, with white matter connections between 165 grey matter brain regions defined using tractography, and structural connectivity matrices calculated from skeletonised diffusion tensor imaging data. Support vector machines trained with graph metrics of white matter connectivity matrices from the microbleed group were able to identify patients with a history of traumatic brain injury with $93.4\%$ accuracy.

\subsection{Classifying with complex networks: Brain-Computer Interfaces}
\label{sec:Class_BCI}

While the field of Brain-Computer Interfaces (BCI) also deals with the analysis of the human brain dynamics, its focus is different from the previously reported examples of neuroscience classification problems. Instead of analysing differences between normal and pathological conditions, BCI aims at detecting patterns in brain activity to control external machines, such as computers. Examples of applications include the discrimination between movements involving the right hand and right foot, which, when performed by means of SVM, reaches a $99.1\%$ accuracy \cite{li2010eeg}; and the discrimination of left and right hand movements, also performed by means of SVM \cite{hassan2008classification}, with accuracy rates ranging from $97.77\%$ to $100\%$.

Beyond SVM, other classification algorithms have been used in BCI problems. An interesting example can be found in Ref. \cite{saa2012latent}, in which a classification of imaginary motor tasks from EEG data was performed by means of {\it hidden conditional random fields} (HCRFs) \cite{quattoni2007hidden}. An HCRF generates sequences of labels for sequences of input samples, and thus allows exploiting the temporal structure present in EEG data. In this study, the interest of this methodology goes beyond its improvement of classification accuracy, as it allowed extracting valuable information on phenomena underlying a motor imagery task.

Finally, brain activity associated with the execution of a given task may not be representative of a complete functional network. It may be interesting to reduce their complexity, for instance by extracting the corresponding minimum spanning tree (MST), and by performing the classification on its characteristic parameters. This was proposed in Ref. \cite{demuru2013brain}, in the task of discriminating between imagery hand movements (both right and left) and resting state conditions.

\subsection{Classifying with complex networks: -omics data}
\label{sec:class_omics}

{\it Omics} is a term that generically refers to research fields in biology ending in {\it -omics}: genomics, lipidomics, proteomics, foodomics, transcriptomics and metabolomics \cite{lederberg2001scientist}. Omics has received increased attention in the last decade due to its possibilities regarding {\it system biology}. The causes and development of a disease are seldom due to individual elements; {\it e.g.} single genes, proteins or metabolites. On the contrary, it is usually necessary to characterise their interactions, both within the same omics family, and between elements of a different kind. As such, omics studies naturally merge network representations with data mining analyses.

One of the most important problems is {\it lethality mining} (also called {\it dispensability}), which aims at detecting which omics elements are essential for the functions of the cell. Notably, in most organisms knocking out most genes has little or no affect on the cell viability, and only in few cases is lethal. The problem is then detecting such lethal elements, as understanding which biological mechanisms have been affected.

\begin{figure*}[!tb]
	\centering
		\includegraphics[width=0.7\textwidth]{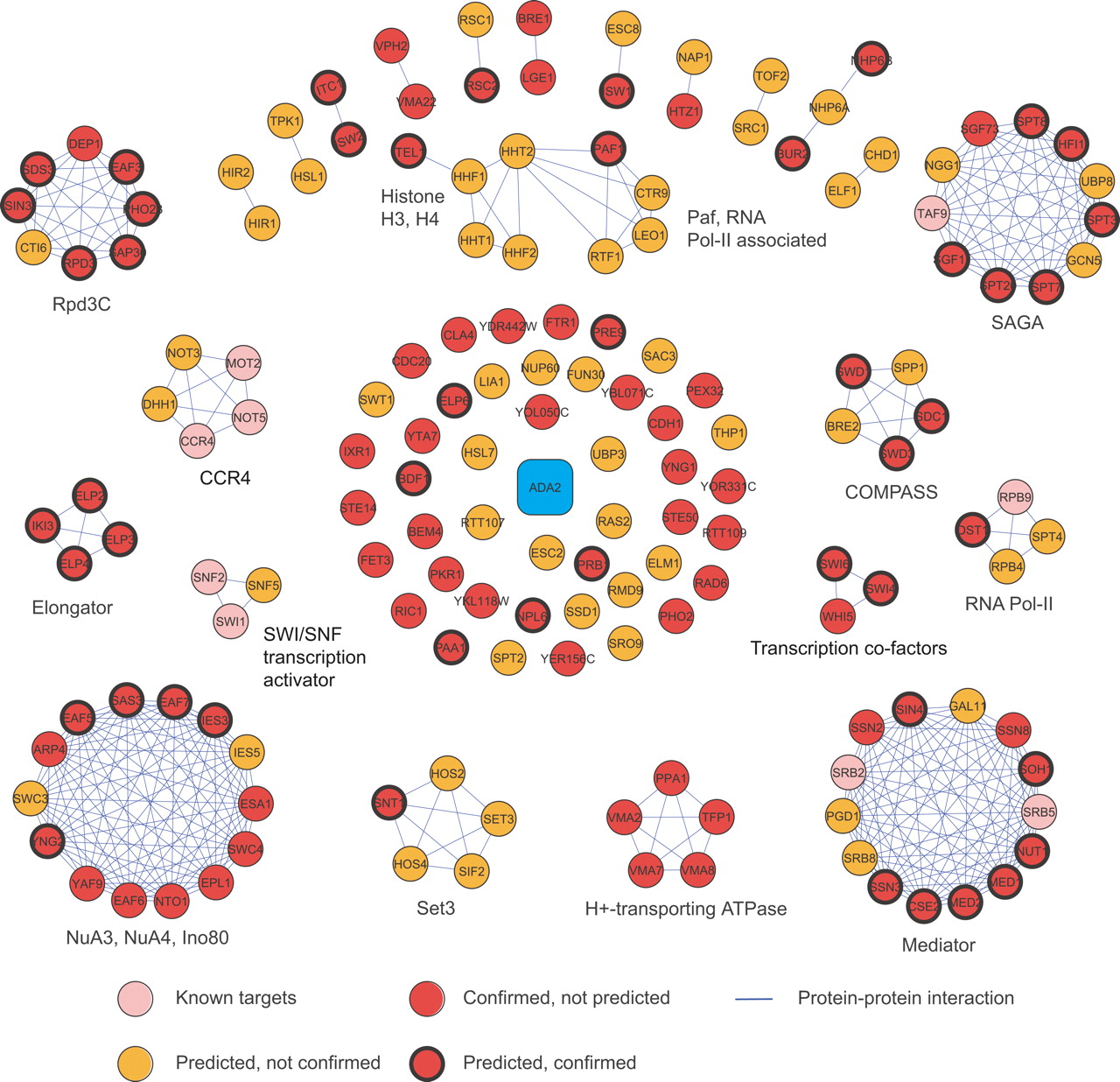}
	\caption{Friendly gene interactions identified by means of data mining techniques, organised according to the involved pathways. Reprinted figure with permission from Ref. \cite{qi2008finding}.}
	\label{fig:FriendConn}
\end{figure*}

Gene dispensability has been studied by combining network topological features with individual properties related to essentiality, such as flux deviations, centrality, codon frequencies of the sequences, co-regulation and phyletic retention. When this information is combined in a decision tree, it has been possible to identify lethal genes in {\it Salmonella typhimurium} with an AUC score of 75\% - 81\% \cite{plaimas2010identifying}. Expanding from single genes to gene pairs, synthetic lethal (SL) interactions have been studied and can be predicted from the topology of the interaction network. Ref. \cite{pandey2010integrative} presents results combining multiple classifiers, namely SVM, neural network, RIPPER (rule-based classifier), random forest, k-nearest neighbours (KNN) and decision trees. Random walks on the interactions networks, combined with a J48 decision tree, also yielded excellent results, which include: a true positive rate of 95\% against a false positive rate of 10\% for {\it Saccharomyces cerevisiae}, and a true positive rate of 95\% against a false positive rate of 7\% for {\it C. elegans} \cite{chipman2009predicting}. A classification task can also be performed to identify ``friendly'' connections, {\it i.e.} interactions which are beneficial to the cell. This has been studied in {\it S. cerevisiae} gene networks through graph diffusion kernels, with a 50\% precision and 20\% to 50\% recall \cite{qi2008finding} - see Fig. \ref{fig:FriendConn} for a graphical representation of the results.

Protein's dispensability has been relatively less studied, due to the challenges associated with modifying the presence of these molecules {\it in vivo}. Nevertheless, it has been proposed the use of neural networks and SVMs, combined with complex network features ({\it e.g.} the number of neighbours in the protein-protein interaction network and the gene expression cooperativity network), to predict essential and non-essential genes, for then using trained models to predict associated protein dispensability \cite{chen2005understanding}. 
SVM classifiers and network properties of protein interaction networks have also been used to predict synthetic sick or lethal (SSL) interactions between proteins, specifically in {\it S. cerevisiae} \cite{paladugu2008mining}.

Another relevant problem in omics analysis is the association of functions with specific genes and proteins, such as function annotation. Historically, this task has been performed by using various methods based on sequence or structural similarity \cite{huang2009bioinformatics, wang2010annovar}; more recently, the topology of protein interaction networks has been proposed as a viable alternative \cite{zhang2007discovering, zahiri2013computational}. It has further been proposed that lethal interactions and functional similarity are related problems. For example, kernel machines have been used to predict gene functions starting from lethality and interaction data in yeast \cite{lippert2010gene}; and a logical learning system called eXtended Hybrid Abductive Inductive Learning (XHAIL) for mapping genes of unknown function to enzymes in a known metabolic network \cite{ray2008inferring}.

Beyond the use of lethality information, protein function prediction has been performed by means of a large number of complex networks and data mining techniques. Among others, these include: Markov random fields, to infer a protein's functions using protein-protein interaction data and the functional annotations of its interaction protein partners \cite{deng2003prediction}; decision trees and na\"ive Bayesian networks for predicting protein interactions in yeast \cite{lu2005protein}; and random forest based methods, combined with bipartite network representations \cite{lee2013hidden}. Related problems are the identification and annotation of proteins that perform their jobs as part of a multi-protein unit, {\it i.e.} of a {\it complex}. Methods combining complex networks and data mining have been proposed to tackle this task. To illustrate, the analysis of level-2 neighbours, {\it i.e.} proteins that do not interact but share interaction partners \cite{chua2008using}; and SVM classifiers on yeast protein interaction networks, achieves a coverage of 89\% at a false discovery rate of 10\% \cite{qiu2008predicting}.

Omics data and omics interactions have also been used to understand the origin and development of different diseases. The study of endocrine disorders is of special importance. The endocrine system is a network of glands that produce and release hormones that help control many important body functions, including the body's ability to change calories into energy that powers cells and organs. Endocrine disorders are usually the result of the failure of multiple omics elements, and are therefore well suited to be studied by means of complex networks \cite{stevens2014network}. Studies include a protein interaction network-based analysis of Juvenile Idiopathic Arthritis (JIA), a common early onset disease encompassing all forms of arthritis, of unknown origin \cite{stevens2014network}. Additionally, Hepatocellular Carcinoma (HCC) has also been studied through a omics network-based analysis, by identifying clusters of co-expressed proteins between HCC and Metastatic Hepatic Carcinoma (MHCC) \cite{pan2014analysis}.

Finally, it is worth citing a few studies proposing different approaches to study relationships between omics networks and diseases. Specifically, text mining has been proposed as a way to predict gene-disease associations, by automatic literature mining based on dependency parsing and support vector machines \cite{ozgur2008identifying}; the trained system can then be used to classify the new sentences as describing an interaction between a gene pair or not \cite{erkan2007semi}. A new way of creating networks of omics interactions has also been recently proposed, based on the identification of expression deviations from a reference level detected in control subjects, by means of different data mining techniques. These networks, called {\it parenclitic}, have been applied to problems like the detection of genes responsible for abiotic responses in plants \cite{zanin2014parenclitic}, or the detection of key genes and metabolites in different diseases \cite{zanin2011complex, menasalvas2012preprocessing, zanin2013feature, zanin2013knowledge, karsakov2015parenclitic}.

\subsection{Hands-on: classification}
\label{sec:HandsOn_Class}

Now that we have a better view of how data mining algorithms can be used to classify complex networks, let us go back to and expand the results presented in Section \ref{sec:HandsOn_Intro}. Specifically, we will show how two commonly used algorithms, Decision Tree and SVM, would interpret the topological features resulting from those analyses.

\begin{figure*}[!tb]
	\centering
		\includegraphics[width=0.3\textwidth]{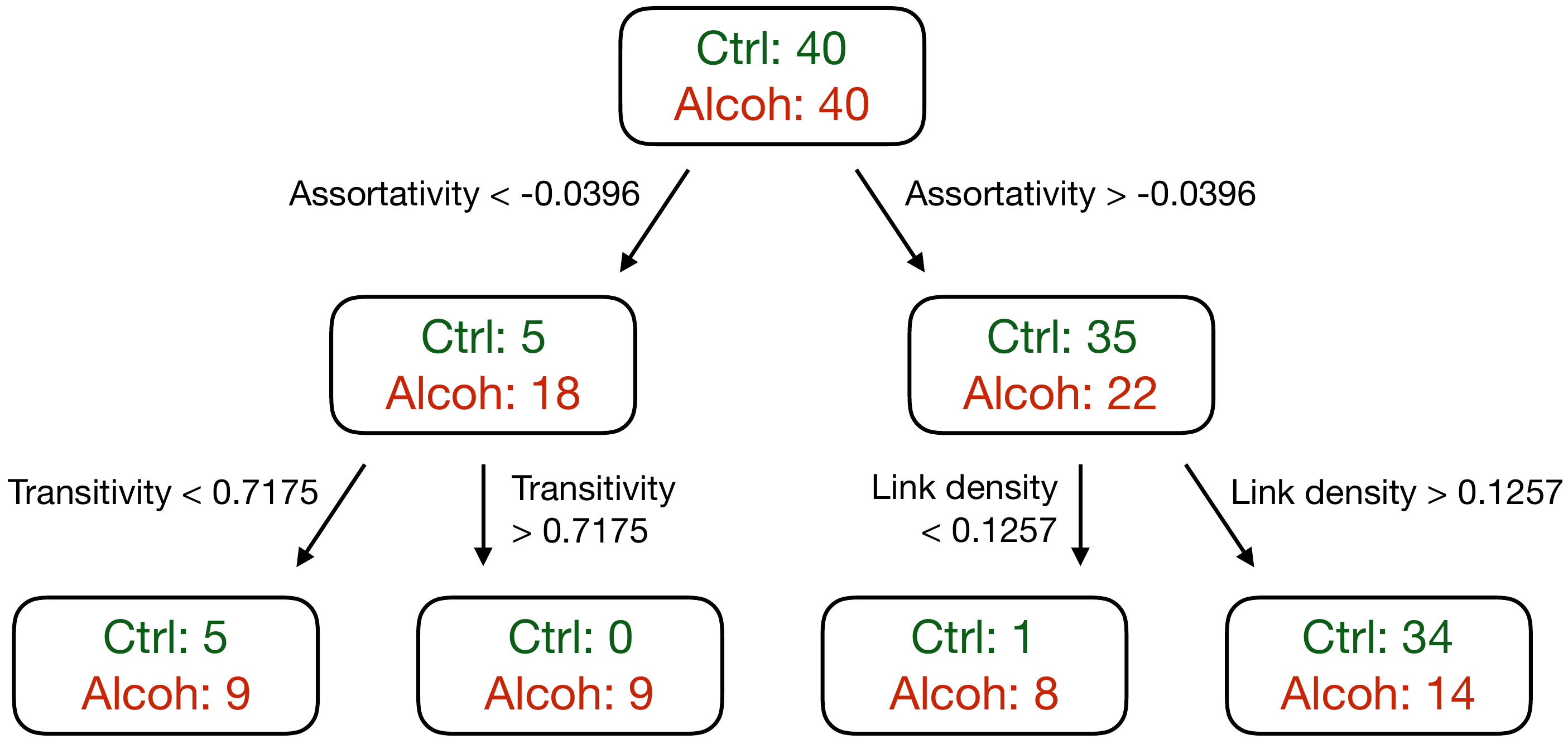}
		\includegraphics[width=0.3\textwidth]{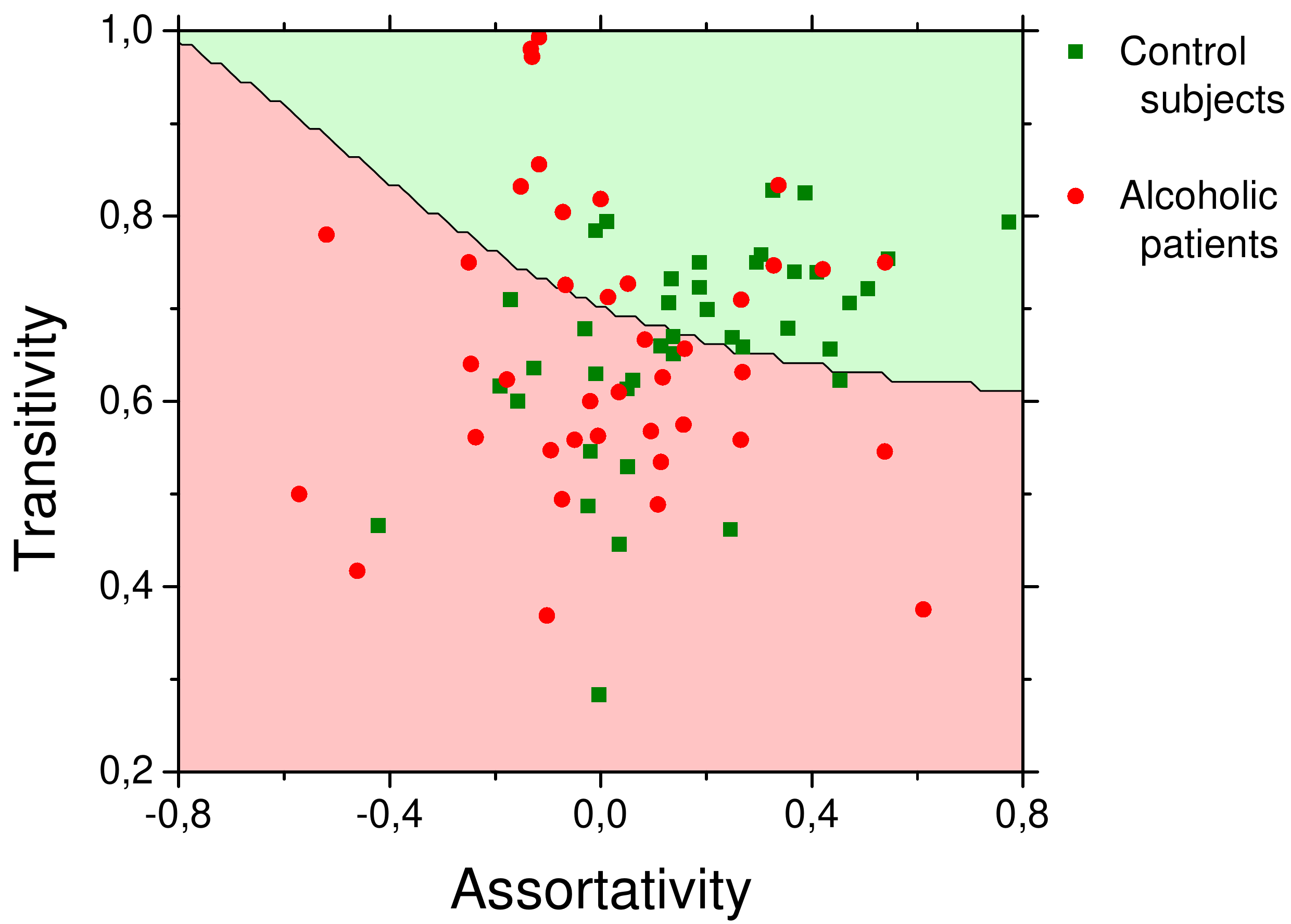}
		\includegraphics[width=0.3\textwidth]{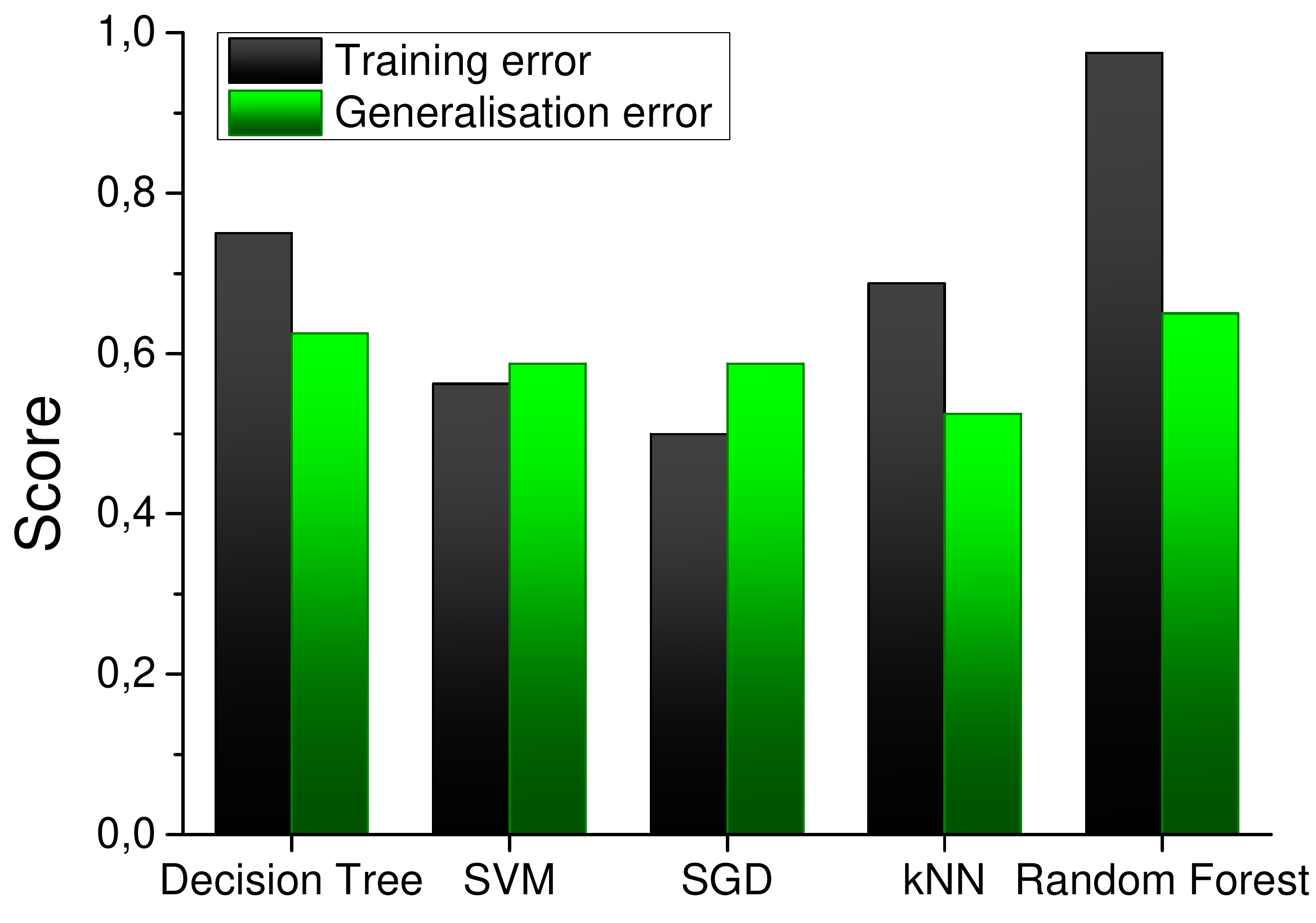}
	\caption{Classification of the data obtained in Section \ref{sec:HandsOn_Intro}, for networks obtained with a correlation metric. Left and Centre panels respectively depict the result of applying a Decision Tree and a SVM model. Right panel reports the training and generalisation scores obtained through five different algorithms.}
	\label{fig:FigureHO02}
\end{figure*}

First, let us consider a Decision Tree. As discussed in Section \ref{sec:dataMining}, this class of algorithms constructs a top-down (or reversed) tree structure, in which each node represents an attribute, and branches different values of those attributes. A graphical representation of the resulting classification model is depicted in Fig. \ref{fig:FigureHO02} Left. Note that the top (or main) node includes all $40+40$ subjects. A first selection is subsequently performed, according to the value of the assortativity; for values lower than $-0.0396$, we have $5+18$ subjects (respectively control and alcoholic), and thus more alcoholic than control subjects; on the other hand, assortativity values greater than $-0.0396$ give rise to the complementary situation. How can this model be used to classify? Quite simple: if we have a subject with an assortativity lower than $-0.0396$, and a transitivity higher than $0.7175$, we can conclude that he/she is alcoholic, as we have not seen an instance of a control subject with such values (second node in the lower row). Of course, this is not always that simple: the last node indicates that when the assortativity is higher than $-0.0396$ and the link density higher than $0.1257$, the subject is {\it possibly} healthy (although we are not certain).

Let us further consider the case of SVMs. Fig. \ref{fig:FigureHO02} Centre depicts the plane created by transitivity and assortativity \footnote{Note that the full classification model would require representing a four-dimensional space, quite beyond the possibilities of modern journals!}, with a black solid line representing the best separation between control subjects and patients (respectively represented by green squares and red circles). Above this line, there are more control subjects than patients, while the opposite occurs below; thus, a new subject could be classified by assessing the position in the plane.

In both cases, we can see that the classification is not perfect, although both algorithms do detect some patterns. Using these 80 subjects, we can calculate the effectiveness of the classification, {\it i.e.} the training score ({\it training} due to the fact that the same subjects are used in the training and in the validation). Results are presented in Fig. \ref{fig:FigureHO02} Right as black bars, both for the two considered algorithms, and for three additional ones (Stochastic Gradient Descent, kNN and Random Forest). The classification task seems quite successful, especially with the latter algorithm, which yields a score close to a $100\%$. Nevertheless, one should never forget the problem of validation, as discussed in Section \ref{sec:validation}: what would happen if the algorithm was tested with a brand new subject, not present in the training set? To answer this, Fig. \ref{fig:FigureHO02} Right also reports the {\it generalisation score} (green bars), calculated through a Leave One Out Cross Validation (LOOCV).

What conclusions can be drawn from this analysis? The differences in topological features that we observed, depicted in Fig. \ref{fig:FigureHO01}, can indeed be used to classify patients, and different algorithms seem to converge towards a score of $60\%$. This score is not very high, and essentially useless for clinical applications; but has been obtained without optimising any of the parameters (like the threshold for binarising networks). In the next section we will see how data mining can be used to optimise these parameters and ultimately improve this seemingly disappointing result.

\section{Improving network significance}
\label{sec:improve}

As previously introduced in Section \ref{sec:functional}, the analysis of {\it functional} networks requires tackling three interconnected problems: the selection of a synchronisation metric, of a threshold, and of a set of topological metrics.
This Section explores how data mining can help in performing a guided selection, {\it i.e.} how it can provide a series of rules for choosing the best combination of these three elements given a global objective.

\subsection{Selecting the synchronisation metric}
\label{sec:SynchMetric}

The presence of functional relationships between pairs of nodes ({\it e.g.} of sensors in brain networks) is usually assessed through statistical relations between the corresponding time series. Two problems emerge. First, interpreting the meaning of links, {\it i.e.} what aspect of the relationship each metric is assessing; and second, electing the most appropriate metric out of the great number of available ones.
In neuroscience, this is further complicated by the dynamical nature of brain functional links. Observed activity can be regarded as a sequence of waxing and waning synchronisation and desynchronisation episodes between distant brain regions. Synchronisation facilitates integrative functions, by transiently binding together spatially distributed neural populations, while desynchronisation may allow the brain to flexibly switch from one coherent state to another \cite{varela2001brainweb, papo2014reconstructing}. In brain functional networks, links are thus unstable objects, whose existence is bounded to specific time frames.

In the last decades, this problem has been tackled by means of different statistical approaches, generally based on assessing the differences between the networks obtained through different metrics. For instance, Directed Transfer Function (DTF), its modification known as direct DTF (dDTF) and the Partial Directed Coherence (PDC) have been compared using EEG recordings \cite{astolfi2007comparison}; unpartialised and partialised cross-correlations (the second trying to eliminate indirect relationships) \cite{jalili2011constructing}; and EEG recordings have been analysed through 9 bivariate and multivariate synchronisation measures, the latter directly ascribing a single value to the synchronisation in a group \cite{jalili2014synchronization}. In all these cases, differences between networks are studied through $p$-value analyses, and results indicate that different metrics lead to important differences in the final network structure. An isolated exception is represented by Ref. \cite{quiroga2002performance}, in which different metrics (non-linear interdependences, phase-synchronizations, Mutual Information, cross-correlation and the coherence function) appeared to detect similar phenomena in EEG recordings.

A step ahead in complexity is provided by those analyses that try to understand how some synchronisation metrics are able to discriminate between different groups of subjects. This has initially been performed through statistical tests. For instance, Ref. \cite{dimitriadis2009characterizing} characterises the differences between different stages of human sleep, based on scalp EEG and using Synchronisation Likelihood \cite{stam2002synchronization} and a non-linear interdependence measure \cite{quiroga2002performance}. Differences between sleep stages were assessed through the Variation of Information information criterion \cite{meilua2007comparing}.
Ref. \cite{lithari2012does} compared the magnitude square coherence (MSC), cross-correlation (CCOR), normalised Mutual Information (NMI) and normalised joint entropy (NJE) when different stimuli were presented to a group of subjects, ranging in pleasure (valence) and arousal.
Finally, Ref. \cite{bonita2014time} compares four metrics, {\it i.e.} Pearson product moment correlation, Spearman rank order correlation, Kendall rank order correlation and Mutual Information, in eyes open and closed EEG recordings. It is worth noticing that this last work explicitly studies the ability of each metric to discriminate between conditions, even if no data mining tools were considered.

In recent years, classification models have been included in this task, as for instance in Ref. \cite{hornero2009nonlinear}, to discriminate between control and Alzheimer's patients, both in EEG and MEG recordings. Five metrics were considered: approximate entropy, sample entropy, multiscale entropy, auto-Mutual Information and Lempel–Ziv complexity. The ability to discriminate AD patients from control subjects was evaluated using receiver operating characteristic (ROC) curves. The first two metrics resulted to be the most reliable for classification, being significantly lower in the EEG of AD patients. Furthermore, the sample entropy appeared to be less dependent on the signal length and more consistent across different parameters, confirming its usefulness in this classification task.
Additionally, it is worth citing Ref. \cite{zhang2013data}, in which a classification was performed between resting states and video watching tasks. Functional connectivities were measured via the Pearson correlation (PeCo), partial correlation (PaCo), Mutual Information (MI), and wavelet transform coherence (WTC). WTC yielded the best classification performance, suggesting that WTC is a preferable functional connectivity metric for functional brain network study.

\subsection{Selecting the threshold}
\label{sec:threshold}

\begin{figure*}[!tb]
	\centering
		\includegraphics[width=0.4\textwidth]{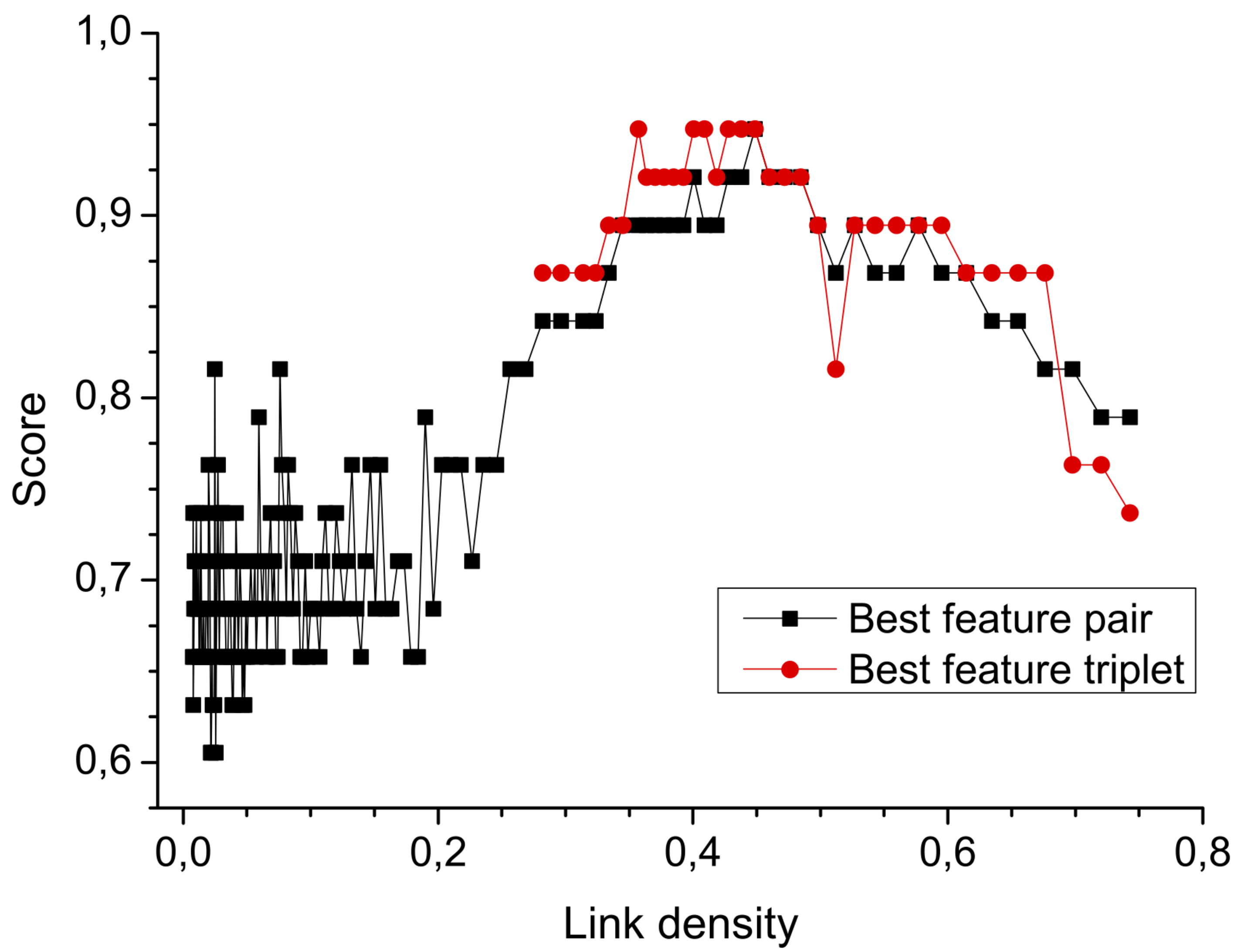}
		\hspace{1cm}
		\includegraphics[width=0.4\textwidth]{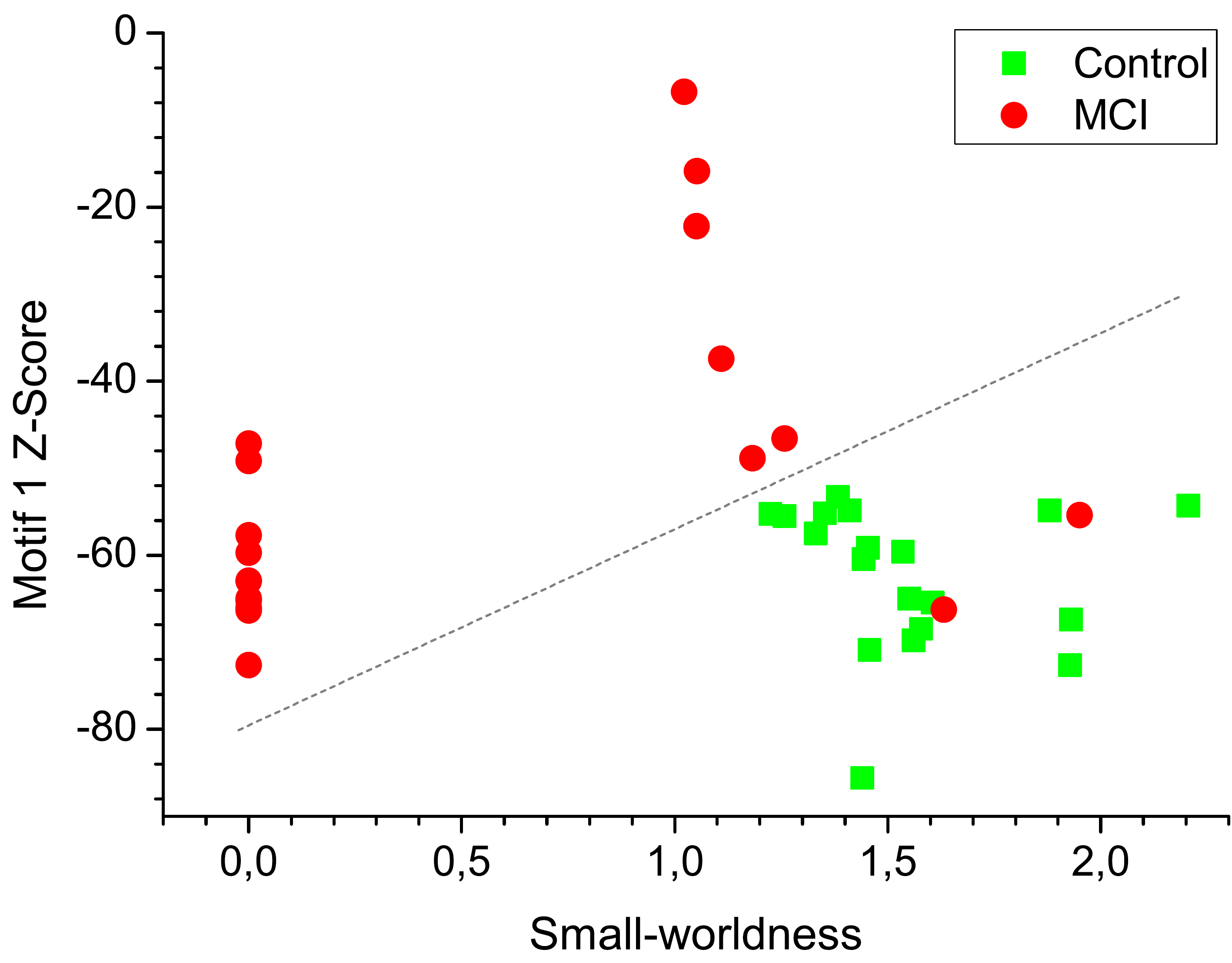}
	\caption{Optimising the network binarization process using classification scores, in a task of differentiating between control subjects and Mild Cognitive Impairment (MCI) patients. (Left) Best classification score as a function of the link density; black (red) points indicate the score obtained using pairs (triplets) of network features. (Right) Best classification of MCI and control subjects. Reprinted figure with permission from Ref. \cite{zanin2012optimizing}.}
	\label{fig:Threshold}
\end{figure*}

Once a synchronisation metric has been selected, a pairwise analysis of the node time series yields a set of weighted cliques, {\it i.e.} fully connected graphs whose weights correspond to the synchronisation strength detected between the corresponding time series.
The direct analysis of such weighted cliques would {\it prima facie} represent the best solution, as they codify all the available information about the system. Alternatively, one may binarise them, {\it i.e.} pruning the graphs according to some rule applied to link weights, and assigning unitary weights to surviving links.

While pruning always implies some information loss, it also yields important advantages. Let us consider the creation of brain functional networks - even though similar arguments would hold in other contexts.
First of all, brain networks are expected to be naturally sparse, as increasing the connectivity would imply a higher physiological cost. Most topological metrics available in network theory have originally been developed for unweighted graphs, for only subsequently been adapted to weighted ones; thus, in the former case the researcher has at his / her disposal a larger set of (better validated) tools. Furthermore, low synchronisation values may just be the result of statistical fluctuations, such that lowly weighted links may be present just because of noise; in such cases, deleting such noisy information can only improve the understanding of the system. Pruning can also help deleting indirect, second order correlations, which do not represent direct dynamical relationships. Finally, network analysis benefits from their graphical representation, which is meaningful only when depicting sparse structures.

The rules used to binarise networks can roughly be classified into two families: those that are based on statistical arguments, and those which rely on data mining tools. Of the first family, it is worth considering the deletion of all links whose weights are below a given fix threshold; maintains a given fraction of the strongest links; maintain a fixed node degree; or use a statistical test to select those that are significant - see Ref. \cite{van2010comparing} for a discussion on the pros and cons of each one of them.
Data mining provides two additional options, which we are going to review in the remainder of the section: use a classification problem as a criterion for the binarisation, and combine multiple binarisations into a single representation.

The first option does not aim at changing the way networks are binarised, {\it i.e.} the process is still performed by fixing either a threshold or a link density; it does use data mining to define a criterion for optimising the process. Ref. \cite{zanin2012optimizing} proposes the use of a classification task to define the best threshold, and specifically to consider the score obtained in the task as a proxy for the significance of the binarised networks. The process works the following way. First, binarised networks are created using different thresholds, covering the whole spectrum (from very dense networks, to sparse ones); second, these networks are used in a classification problem: the threshold corresponding to the highest classification score will be the optimal one, as it allows to extract the most information from the original system. The results of this process are represented in Fig. \ref{fig:Threshold}. While this method yields an optimal criterion, it has two important drawbacks: it is computationally costly, as many sets of networks have to be processed; and it yields a single threshold, so that information across different densities cannot be integrated.

This second problem has been tackled in Ref. \cite{jie2014topological}. Authors propose to split the analysis of a set of networks in different steps.
First, as in the method of Ref. \cite{zanin2012optimizing}, multiple thresholds are applied, to generate connectivity networks encoding different levels of details. Afterwards, each set of resulting networks are analysed, and their most relevant features selected by using a recursive feature elimination method based on graph kernel \cite{vishwanathan2010graph, shervashidze2011weisfeiler} and support vector machine. Lastly, a multikernel SVM is used to fuse all features from multiple thresholded networks, to obtain a global classification model. 
It is worth stressing the advantage of this method; by using a multiple-kernel learning technique, network properties from multiple thresholded networks are merged together. This allows, in principle, to simultaneously account for different topological metrics at different connectivity levels, {\it e.g.} motifs in dense networks and efficiency in sparse ones; and to describe the multiscale complexity observed in many real-world systems \cite{meunier2009hierarchical, ahn2010link}.

\subsection{Selecting the topological metrics}
\label{sec:TopMetrics}

Once a set of binary networks has been obtained, the next and final step involves the extraction of some topological metrics, which can ultimately be used {\it e.g.} to feed a classification task.
Most studies in which an automatic selection of topological metrics has been proposed, this was done by using a combination of two techniques: support vector machines (SVM) and Fisher’s criterion. The former allows to treat the selection process as a classification task: different sets of features can be tested, and the one yielding the best separation (in terms of classification score between two groups of subjects) is selected. The latter is similar in nature, although it does not explicitly rely on a classification task: if finds a linear combination of features that maximises the distance between the means of the two classes, while minimising the variance within each class.

\begin{figure*}[!tb]
	\centering
		\includegraphics[width=0.5\textwidth]{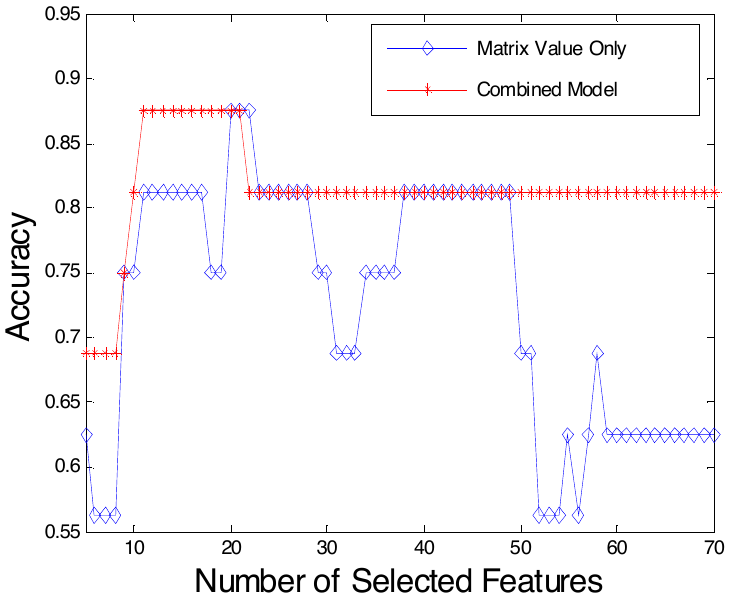}
	\caption{Classification scores obtained in a task of discriminating between autistic and control subjects, as a function of the number of features included in the analysis. The blue line represents the scores obtained by directly using the adjacency matrix as input feature, while the red one corresponds to the use of network topological metrics. Reprinted figure with permission from Ref. \cite{pollonini2010functional}.}
	\label{fig:Autism}
\end{figure*}

Both methods have been used to optimise the network analysis of different types of neurological signals. For instance, MEG time series were analysed in Ref. \cite{pollonini2010functional} through Granger Causality, to discriminate autistic and control subjects, reaching a $87.5\%$ accuracy. Two strategies were compared: in one case using the adjacency matrix alone as input feature; in the second, adding to the adjacency matrix different topological metrics. Results indicate that, while both strategies reach the same classification score, the latter is more stable with respect to the number of features included - see Fig. \ref{fig:Autism}. Additionally, Ref. \cite{zouridakis2012functional} dealt with mild traumatic brain injury (mTBI) patients, also through a Granger causality approach; and Ref. \cite{zanin2012optimizing} focused on the diagnosis of Mild Cognitive Impairment.
Structural brain connectivity has been tackled by means of functional magnetic resonance imaging (fMRI), in patients with Alzheimer's disease \cite{khazaee2014automatic}; whole-brain tractography, to identify graph metrics that best differentiate individuals with Major Depressive Disorder (MDD) from nondepressed controls \cite{sacchet2014elucidating}; and diffusion weighted magnetic resonance imaging (DWI), to classify brain connectivity networks based on sex and kinship \cite{duarte2012hierarchical}.

As we end this discussion it is once again useful to recall the work performed in Ref. \cite{jie2014topological}. By using a multikernel SVM, this paper demonstrates that the selection of topological features and that of thresholds can be carried out simultaneously. The importance of this solution resides in the fact that it tackles all the circle represented in Fig. \ref{fig:NetworkOptimisation} simultaneously.

\subsection{Hands-on: improving significance}
\label{sec:HandsOn_Improve}

\begin{figure*}[!tb]
	\centering
		\includegraphics[width=0.47\textwidth]{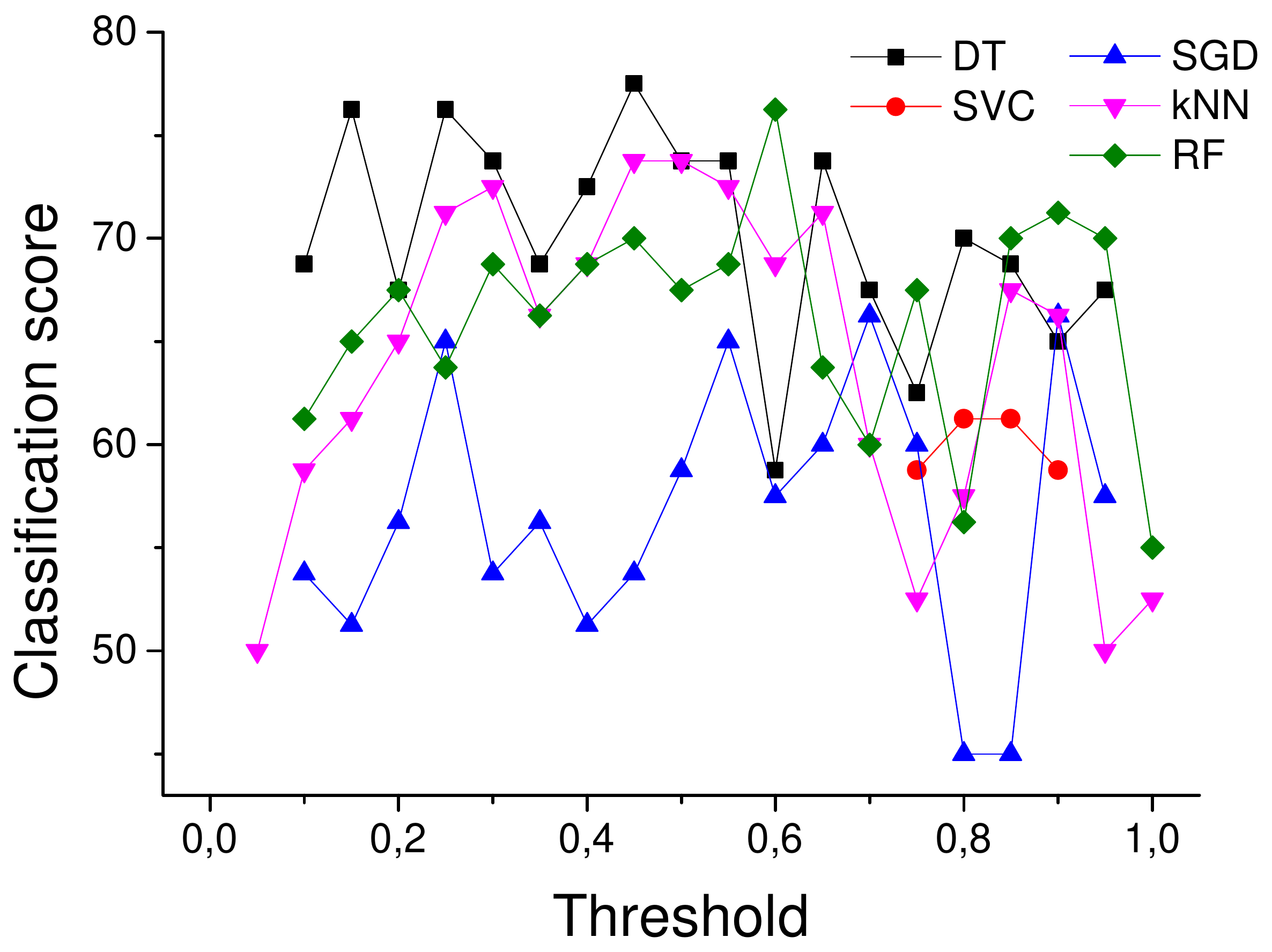}
		\includegraphics[width=0.47\textwidth]{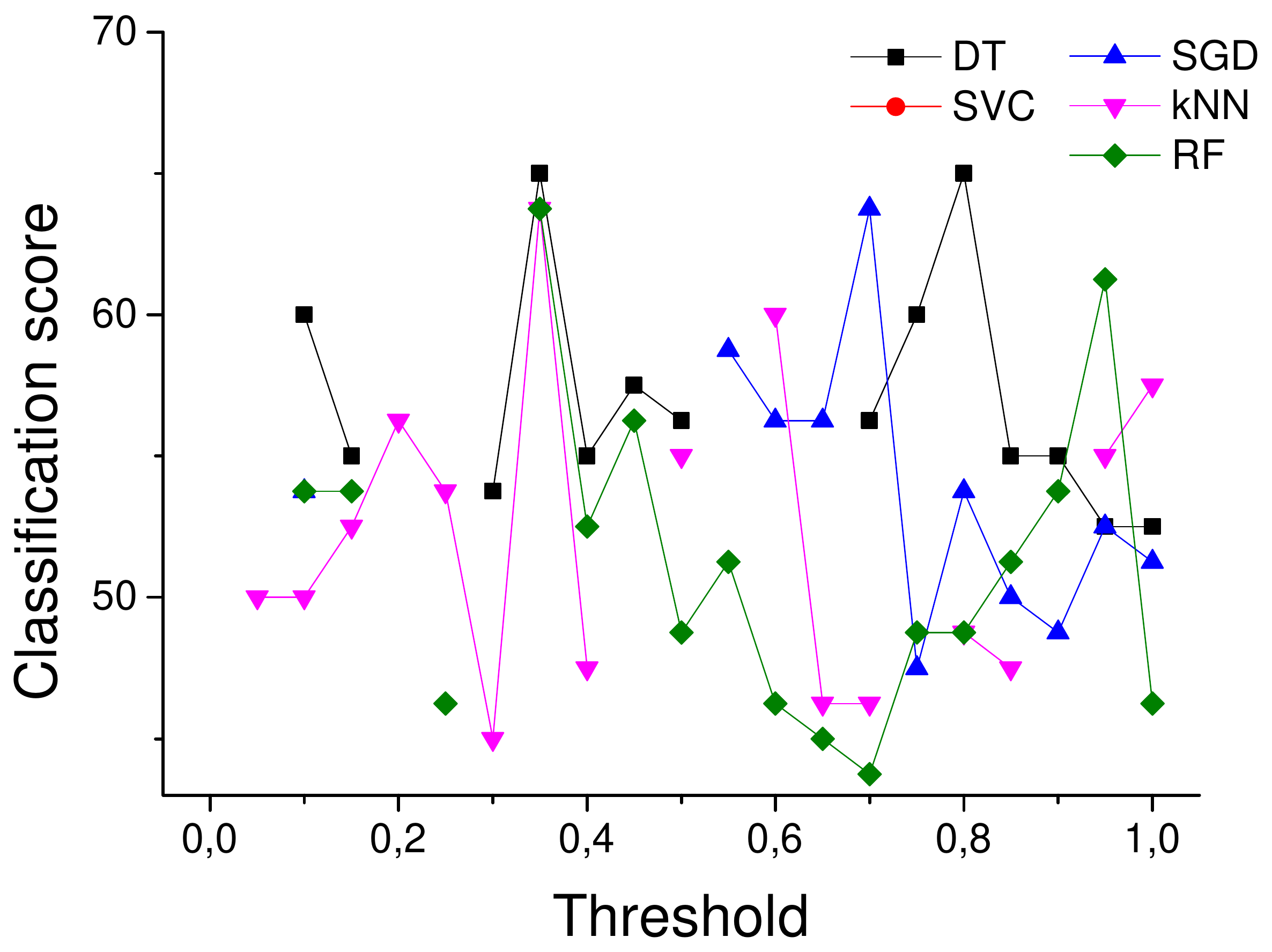}
	\caption{Classification of the data obtained in Section \ref{sec:HandsOn_Intro}, as a function of the threshold applied. Left and right panels respectively represent the results for correlation and Granger Causality networks.}
	\label{fig:FigureHO03}
\end{figure*}

Let us now go back to the hands-on exercise we started in Sections \ref{sec:HandsOn_Intro} and \ref{sec:HandsOn_Class}, and see how data mining can further be applied to increase our understanding of the system. We shall focus on two (critical) aspects: the selection of an optimal threshold and of the best synchronisation metric (in this case the choice is between correlation and Granger Causality).

As discussed in Section \ref{sec:threshold}, one of the simplest way to optimise the threshold involves extracting networks for different threshold values, and then compare the associated classification score. Fig. \ref{fig:FigureHO03} Left presents the results of this process, for the five classification models here considered (Decision Tree, SVM, Stochastic Gradient Descent, kNN and Random Forest). The maximum of the score is obtained with a Decision Tree, for a threshold of $0.45$.
Two conclusions can be drawn at this stage: {\it i}) Decision Tree is the most efficient classification model, for almost any value of the threshold; and {\it ii}) the best score ($77.5\%$) is much higher than the one previously obtained ($63\%$, threshold of $0.75$).

Moving to the results for the Granger Causality networks (Fig. \ref{fig:FigureHO03} Right), here too we observe a maximum score corresponding to Decision Tree, and for a threshold quite close to the previous one ($0.35$). On the other hand, it is clear that scores obtained with the Granger Causality are significantly lower than the ones obtained with linear correlation. For some reasons, the former is describing a type of information propagation that is not relevant for the problem at hand (distinguishing control {\it vs.} alcoholic patients).

In order to increase our understanding of the system, it would be necessary to understand the reasons behind such results. Why does a low threshold, and thus a high link density, better describe the two groups of people? What kind of information transfer is assessed by linear correlation, but not by Granger Causality? Why does Decision Tree systematically outperform other algorithms, which are in principle more advanced? Answering these questions is out of the scope of this review, but will certainly be of interest to all neuroscientists.


\section{Feature selection}
\label{sec:featureSel}

One of the typical obstacles in the learning phase of any data mining procedure is represented by the high dimensionality of the feature space, even when the model only depends on a reduced number of variables. Indeed, both theoretical and experimental studies indicate that many algorithms scale poorly when a large number of irrelevant features are included, even when many data mining algorithms attempt to automatically discriminate between important features and those that can be eliminated \cite{jain1997feature, blum1997selection}. The goal of feature selection procedures is threefold: reducing the sample size, and thus the computational cost of the analysis; focusing the analysis only on relevant data, thus improving its statistical relevance; and improving the quality of the data set.  Feature selection is a particularly important step in those domains that entail a large number of measured variables but a very low number of samples, {\it e.g.} gene and protein expressions, and MEG and EEG recordings.

This Section explores how data mining algorithms can be used to reduce the dimensionality of the problem, which, in the context of network analysis, is tantamount to reducing the set of nodes and links. Finally, Section \ref{sec:FSelectionByCN} will explore the other side of the coin, {\it i.e.} how complex networks can help performing a feature selection of an arbitrary data set.

\subsection{Selecting the nodes of a network}
\label{sec:FSelectionNodes}

A statistical physics approach to complex networks would suggest that the understanding of a network should improve with the number of nodes included in the analysis, as all nodes are {\it prima facie} equivalent; thus ideally no node should be discarded.
Nevertheless, there exist situations in which performing a selection can be useful. We explore three different applications: neuroscience, omics analysis, and other complex systems.

\subsubsection*{Neuroscience}
Probably the clearest example of the need for node selection is represented by the analysis of fMRI data. The output of an fMRI analysis is a set of images, representing the time evolution of the regional cerebral blood flow (and thus, indirectly, of neural activity). In modern equipment, the number of such regions (also called {\it voxels}) can be extremely large, of the order of $10^{5}$. Identifying each voxel with a network node supposes two main problems. First, the resulting networks would be extremely large, and thus difficult to analyse: $10^{5}$ nodes imply reconstructing adjacency matrices of $10^{10}$ elements; representing such networks would require in the order of 10 GB of memory. Second, assessing the synchronisation between each pair of nodes results in numerous statistical tests, which must be appropriately corrected for multiple comparisons in order to avoid a {\it curse of dimensionality} problem \cite{verleysen2005curse}.

Two different options have historically been explored to appropriately reduce the number of nodes \cite{richiardi2013machine}. The first uses anatomical knowledge to divide the brain into regions of contiguous voxels, which are known to mediate the same cognitive function \cite{salvador2005neurophysiological, richiardi2011decoding}.
Yet, it was shown that selecting regions can yield network properties different from the ones that would be obtained without grouping voxels. Grouping tends to underestimate node connectivity, but overestimates clustering degree; node centralities may also be substantially affected. Although many qualitative network properties are consistent, the voxel-level resolution reveals details that are not visible at the regional-level networks, providing extra information that can be useful in some applications \cite{tohka2012impact}.
When such information is not available, or one wants to avoid biases introduced by a parcellation, a data-driven procedure can be used. Voxels are then joined together according to their synchronous activity, by means of techniques like Spatial Independent Component analysis (ICA) \cite{calhoun2001method} or clustering \cite{cordes2002hierarchical, woo2014cluster}.

Another approach provided by data mining consists in the use of Recursive Feature Elimination (RFE), as described in Section \ref{sec:featureSelection}. This has been used, for instance, to classify functional imaging data using a Support Vector Machine classifier \cite{de2008combining}. Forward strategies have also been used - see for instance Ref. \cite{michel2008mutual}, in which a Mutual Information is used to assess the redundancy of features \cite{franccois2007resampling}. A recent study proposed the use of regularisation trees for fMRI node selection \cite{jenatton2012multiscale}, based on constructing trees penalising using a variable similar to the variables selected at previous tree nodes for splitting the current node \cite{deng2012feature}. This method has the advantage of providing a multi-scale representation of features, which allows selecting brain regions simultaneously at different scales.

Beyond the problems of computational cost and data dimensionality, applying a node selection procedure is expected to increase the reliability and precision of the subsequent classification tasks. While some studies claim an important increase in the classification score \cite{de2008combining}, the debate is still open regarding the usefulness of a data mining approach against the use of prior knowledge of where to expect signs of a given disease \cite{chu2012does}. In general, it seems that node selection is especially useful in cases of small sample sizes.

When reconstructing complex networks of brain activity using other signals, as for instance EEG and MEG, the problem of the large number of channels is less relevant - usually EEG and MEG are composed of between 60 and 250 channels. Some specific applications may nevertheless benefit from a reduction in the number of nodes. This is the case of Brain-Computer Interface (MCI), aimed at allowing severely motor-impaired patients to communicate and interact with a computer \cite{wolpaw2000brain, vallabhaneni2005brain} - see Section \ref{sec:Class_BCI}. If the BCI interface is to be embedded into a portable system, even creating and analysing networks of hundreds of nodes in real time can be a technical challenge, and would, as a minimum, increase the cost of the device.
To reduce the complexity of the analysis, Ref. \cite{santana2012regularized} proposed the use of an Evolutionary Algorithm (EA) for selecting relevant subsets of MEG channels, in a classification task aimed at assessing whether the subjects were paying attention to the left or the right.
Similarly to the fMRI case, such selection may lead to a distortion in the observed topological features. A full analysis of the effects derived from sampling a subset of nodes (and sensors) in EEG studies can be found in Ref. \cite{joudaki2012eeg}. It is based in the comparison of the topological metrics of EEG functional networks of 32 normal subjects, when the number of nodes is changed. Larger networks appear to have higher efficiency, higher assortativity and lower modularity, compared to those with same density but different size, highlighting the importance of taking into account the network size when comparing networks across studies.

\subsubsection*{Omics analysis}

The analysis of genomics, proteomics and metabolomics data has yielded promising results in the last years, especially in cancer biology; nevertheless, this has come at the cost of the analysis of high-dimensional data spaces \cite{clarke2008properties} - see Section \ref{sec:class_omics}. Starting from the hypothesis that the functional state of an organism is determined by its genes expression patterns, functional genetic networks can be created starting from the expression level of each gene and detected pairwise correlations or causalities. The same can also be performed with the other omics data. If no pre-processing step is performed, the result may be a very large network, with tens of thousands of nodes, only a fractions of them being relevant for the study at hand.

Examples of the application of a feature selection algorithm for reducing the number of nodes can be found in Ref. \cite{lopes2014feature}, in which a Sequential Backward Selection (SBS) is used: it starts with the full feature set and eliminates the least relevant feature according to the criterion function (top-down approach), repeating this process until a stop condition is satisfied \cite{pudil1994floating}. A similar approach has also been explored in Ref. \cite{menasalvas2012preprocessing}.

Furthermore, it has been proposed that omics data can be analysed in their raw format, {\it i.e.} starting from spectral information that can be extracted from the analysis of cell or blood samples, and reconstructing a complex network representing the relationships encountered between pairs of spectral frequencies. When compared with the standard binning procedure, in which the full spectrum is divided into a number of non-overlapping windows \cite{griffin2001metabolic}, the use of a Mutual Information criterion substantially improves the usefulness of resulting networks \cite{zanin2013feature}.

An alternative approach to feature selection consists in identifying clusters of omics elements that share a similar dynamics; and then map a few cluster-representative elements, or the mean expression level of all elements in a cluster, into a new node. Several algorithms can be used towards this aim, the one most used being $k$-means and fuzzy $c$-means \cite{granzow2001tumor, dougherty2002inference}. A special note should be devoted to hierarchical clustering techniques: in spite of yielding, in principle, some advantages (like the possibility of choosing the scale of the partition), and their wide use, they also impose artificial restrictions on the results (nodes joined in one cluster cannot be separated and assigned to different clusters in later steps) and perform comparatively poorly \cite{radke2004quantitative}.

\subsubsection*{Other complex systems}

In recent years, complex networks have found an important field of application in the analysis of climate data, {\it i.e.} in the identification of patterns in the global climate. This approach is important as it enables a deeper understanding of the complex processes underlying observed phenomena, with important consequences in prediction tasks. Thanks to the high resolution recordings of weather data, and to reanalysis models, it is possible to construct networks with tens of thousands of nodes, pairwise connected when some functional relationship is detected between the corresponding dynamics \cite{donges2009complex}. While it is technically feasible to analyse the whole network, it may be interesting to cluster regions, {\it i.e.} join together all nodes with a similar dynamics (even if they are not spatially contiguous), in order to simplify the final results representation. This has been performed by means of several algorithms, {\it e.g.} $k$-means \cite{fovell1993climate} and a Shared-Nearest Neighbor approach \cite{steinbach2003discovery}. It has also been proposed to use the complex network topology towards this clustering task \cite{steinhaeuser2011complex, tsonis2011community} - a topic that will be further discussed in Section \ref{sec:ClusteringAndCommDetection}.

When complex networks are applied to the study of transportation systems, as {\it e.g.} the study of the air transportation network \cite{zanin2013modelling}, a node selection problem arises. Even if we are studying a specific region, {\it e.g.} Europe or the USA, the airports (and the flights) in that region are connected to external elements: the problem is thus to select those elements that, even being outside the region of interest, are relevant to understand a given problem. An example of such analysis is presented in Refs. \cite{zanin2013synchronization, zanin2014network}, where networks of aircraft are reconstructed to predict the appearance of safety-related events, and relevant aircraft are selected through a wrapper strategy.

\subsection{Selecting the links of a network}
\label{sec:FSelectionLinks}

Once a network has been reconstructed, one may face the problem of selecting a subset of links that are representative of the network structure. There are several reasons for wanting to perform such sampling. First, networks reflect the complexity of the systems they represent, and it is thus easy to obtain intricate interweaved and densely connected structures. Ways of singling out key information by filtering irrelevant links are essential to obtain simpler relevant subgraphs. Second, in the case of functional networks, links may represent both direct and indirect paths: it is then necessary to isolate the link between $X \rightarrow Z$ when the true interaction is suspected to be $X \rightarrow Y \rightarrow Z$, as the former confounds our ability to identify true pairwise interactions. Notice that this is not equivalent to the problem of selecting a suitable threshold, as discussed in Sec. \ref{sec:threshold}; here we tackle the problem of extracting a more meaningful representation of an already optimised network, or of a network which has a clear topology - {\it e.g.} the traffic backbone of a large-scale transportation infrastructure. On the other hand, this is partly related to the problem of link prediction, which, as will be discussed in Sec. \ref{sec:LinkPrediction}, can be also seen as a problem of filtering non-relevant links.
In the last years, several techniques for link selection have been proposed. We review the most important ones, which we organise into two groups depending on their main assumptions: network- {\it vs.} data mining-based.

\subsubsection*{Network-based methods}

One of the simplest solutions to the link selection problem is provided by Minimum Spanning Trees (MST) \cite{graham1985history, gabow1986efficient}. A classical concept of graph theory, a MST is the shortest-length tree subgraph that contains all the nodes of the original network. In simple words, they allow to find the minimum set of links connecting a network, or its {\it superhighways}. Thanks to their simplicity, they have been largely used in biomedicine \cite{tewarie2014functional, stam2014trees} and finance \cite{micciche2003degree, tumminello2007spanning, bonanno2004networks}. They yet present one big limitation, {\it i.e.} that are by construction acyclic. This means that all local cycles are destroyed, thus distorting the clustering coefficient and the clustering hierarchies present in the starting network.
An evolution of the MST concept was discussed in Ref. \cite{tumminello2005tool}. This paper proposed the construction of graphs with different degrees of complexity by linking the most strongly connected nodes, as in MST; the difference is that the resulting graphs should be embeddable on a surface of a given genus, {\it i.e.} the topologically invariant property of a surface defined as the largest number of nonisotopic simple closed curves that can be drawn on the surface without separating it \cite{aste2005complex}. Resulting graphs have the same hierarchical tree structure of MSTs, but contain a larger amount of information.

If the network of interest has been created from time series, like in the case of functional networks, one may be interested in identifying those links that are the results of robust statistical properties of the time series, and not of simple fluctuations. This can be performed by calculating surrogate networks \cite{rheinwalt2012boundary}: shuffle the data such that they preserve all the statistical quantities, {\it i.e.} distribution of values and autocorrelation properties, and reconstruct different functional networks. These network properties are only due to the conserved statistical quantities, and therefore represent the spurious links in the original network.
A complementary approach was proposed in Ref. \cite{serrano2009extracting}, involving the local identification of the statistically relevant weight heterogeneities; this allows to filter out the backbone of dominant connections in weighted networks with strong disorder, preserving structural properties and scale hierarchies.

\begin{figure*}[!tb]
	\centering
		\includegraphics[width=0.7\textwidth]{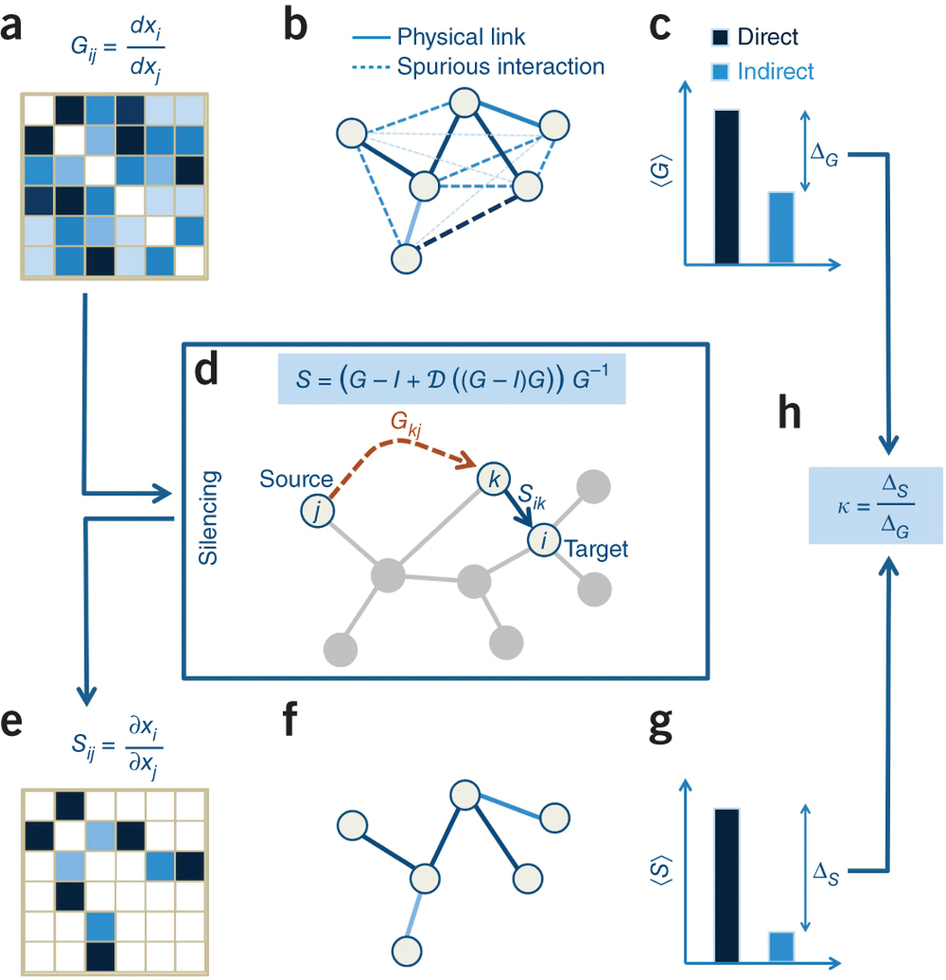}
	\caption{Graphical representation of the method for silencing indirect links in a network. It starts from the observed network $G$, which accounts for direct as well as indirect correlations (see panel {\it b}). Silencing is then achieved through the transformation reported in panel {\it d}, which results in a silenced matrix $S$ (panels {\it e} and {\it f}) whose nonzero elements correspond to direct links. Reprinted figure with permission from Ref. \cite{barzel2013network}.}
	\label{fig:Silencing}
\end{figure*}

Finally, it is worth citing the method proposed in Ref. \cite{barzel2013network}. By exploiting the properties of dynamical correlations in networks, this method takes a correlation matrix and uses a matrix transformation to turn it into a highly discriminative silenced matrix, which enhances only the terms associated with direct causal links. Mathematically, let us denote the original observed adjacency matrix by $G$; the silenced adjacency matrix is then obtained as:
\begin{equation}
	S = (G - I + D( (G - I) G )) G^{-1},
\end{equation} 
$D(M)$ being the result of setting the off-diagonal terms of $M$ to zero. This transformation thus detects the indirect paths between then nodes of the network and silences them, keeping only the contribution of the direct paths - see Fig. \ref{fig:Silencing} for a graphical representation of the method, and specifically panels {\it d - f} for the silencing part. When tested against the {\it Escherichia coli} regulatory interaction network, it yields a $50\%$ predictive improvement over traditional correlation measures.

\subsubsection*{Data mining-based methods}

Most of the data mining-based methods for link selection have been developed around two ideas: a recursive feature selection algorithms applied on links; and a metric for deciding the relevance of each link. For the latter, Mutual Information (MI) has demonstrated to be a successful option: while some implementations are reviewed in what follows, the reader may refer to Ref. \cite{lopes2009comparative} for a more detailed review and comparison of different algorithms.

One of the first MI-based algorithms to be developed, ARACNE (Algorithm for the Reconstruction of Accurate Cellular Networks) \cite{margolin2006aracne} aims at detecting and eliminating indirect interactions. It is based on the analysis of each triplet of nodes, in order to eliminate the edge with lowest MI (the indirect one) in cases where the difference between the lowest and the second lowest MI is greater than a given threshold.
MINET \cite{meyer2008minet} is based on the Maximum Relevance / Minimum Redundancy (MRMR) strategy for feature selections \cite{peng2005feature}. Starting with a partial set of possible links, new ones are iteratively added, chosen in order to simultaneously maximise their relevance (in terms of their MI with a target external feature) and minimise their redundancy (the MI with the other links on the partial set).
SFFS+MCE is a method based on the Sequential Floating Forward Selection feature selection strategy, a wrapper approach that includes features according to some criteria. In the case of SFFS+MCE, the criterion is a conditional entropy that penalises rarely observed links \cite{lopes2008feature}.
Finally, MIDER \cite{villaverde2014mider} is a method for distinguishing between direct and indirect interactions and to assign directionality; given a set of time-series describing quantitative features of the network nodes, this is accomplished by applying several definitions and normalisation of MI.

Many important link selection methods have been developed with Protein-Protein Interaction (PPIs) networks in mind. PPIs are of special interest for understanding most of the cellular functions of the cell; nevertheless, experimentally detected PPIs have high levels of false positive rate, requiring computational tools for pruning low-confidence networks. Ref. \cite{zhu2014two} proposes a new geometric approach, called Leave-One-Out Logistic Metric Embedding (LOO-LME), for assessing the reliability of interactions. It first transforms the learning task into an equivalent discriminant form, for then directly dealing with the uncertainty in the networks using a leave-one-out-style approach. Experimental results show that LOO-LME substantially outperforms previous methods on PPI assessment problems. 
On the similar line, Ref. \cite{lei2013increasing} proposes the use of a  fast isometric feature mapping (fast-ISOMAP), transforming a PPI network into a low dimensional metric space, thus reinterpreting the problem of assessing protein interactions into the one of measuring similarity between points of its metric space. A link reliability index is then created, based on the similarity between the corresponding nodes (points in the embedding space). Validation results for yeast data demonstrate that the interactions ranked top by this method have high functional homogeneity and localisation coherence.

A rather different approach is presented in Ref. \cite{lezon2006using}, by leveraging on the Boltzmann's concept of entropy maximisation as a way of performing statistical inference with minimal reliance on the form of missing information. The underlying rationale is that each macroscopically observable state of a complex system corresponds to a number of microscopic states. Because the number of ways of realising a given macroscopic state can vary, the most likely state of the system as a whole is the one that corresponds to the largest number of microscopic states. Translated to the problem of filtering network links, entropy maximisation involves the identification of those links that increase the probability of giving rise to the macrostate captured in the observed data. As an illustration, the method is applied to {\it Saccharomyces cerevisiae} genetic microarray data exhibiting energy metabolic oscillations, for the reconstruction of intracellular communication pathways associated with metabolic oscillations.

\subsection{Feature selection through complex networks}
\label{sec:FSelectionByCN}

In this last subsection, we reverse the approach to the problem: instead of analysing how feature selection techniques can be used in studying complex networks, in this section we sketch how complex networks could be used to improve current feature selection techniques.

One of the most recent trends in feature selection is the so-called Structure Learning \cite{aliferis2010localA, aliferis2010localB}, which is based on the idea of finding all possible relationships between pairs of variables, and expressing them by mean of a graph. The attentive reader will notice the parallelism between this approach, and the reconstruction of functional brain networks discussed in previous sections. 
To simplify the analysis, it is usually assumed that data are generated by a Bayesian Network, so that the resulting structure is a directed graphical model. Following this hypothesis, the optimal solution is the Markov blanket of the target node, which is unique in a Bayesian Network. A Markov blanket of a given node is composed of its parents, its children, and its children's other parents; in terms of data relationships, this is equivalent to saying that, in order to understand a given feature, it is necessary to include its direct causes, its direct effects, and the direct causes of its direct effects (also called the ``spouses'' of the target feature) \cite{pearl2014probabilistic}.
It is clear that an essential issue is the correct detection of the relationships between the features; using linear methods, such as SVMs, does not provide reliable results, as linear SVM may give zero weight to important features \cite{hardin2004theoretical, statnikov2006using}, and it is thus necessary to resort to causality analyses - see for instance Refs. \cite{tsamardinos2003towards, tsamardinos2003time}.

Once the graph of relationships has been established, complex networks can help identifying the most relevant nodes ({\it i.e.} features) in an efficient way, without any prior hypotheses on the structure of the network. This can be accomplished in at least two ways. First, by ranking nodes according to their importance, using one of the numerous available centrality metrics \cite{costa2007characterization}; second, by considering that identifying the causal nodes in a network is tantamount to establishing which nodes are controlling its dynamics, thus converting the feature selection into a controllability problem \cite{liu2011controllability, cowan2012nodal}. It is worth noticing two advantages that this approach would yield with respect to the Markov blanket one. First, there is no need to define {\it a priori} the target node(s), as complex network analysis would provide a global ranking of nodes importance. Second, any complex network analysis takes into account the global structure of the network, thus including structural information that may be missing in a local Markov analysis.

\subsection{Hands-on: feature selection}
\label{sec:HandsOn_FS}

\begin{figure*}[!tb]
	\centering
		\includegraphics[width=0.47\textwidth]{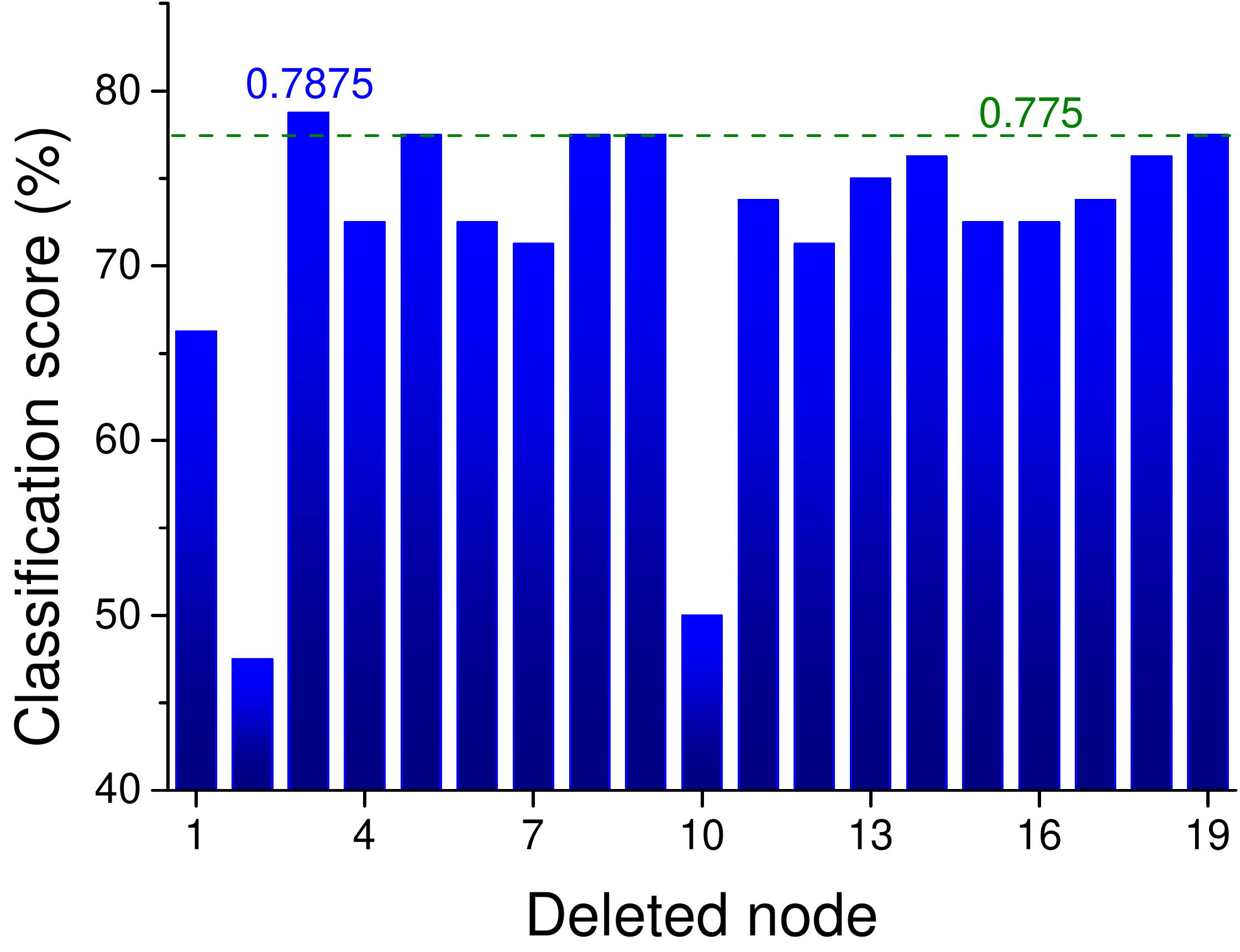}
		\includegraphics[width=0.47\textwidth]{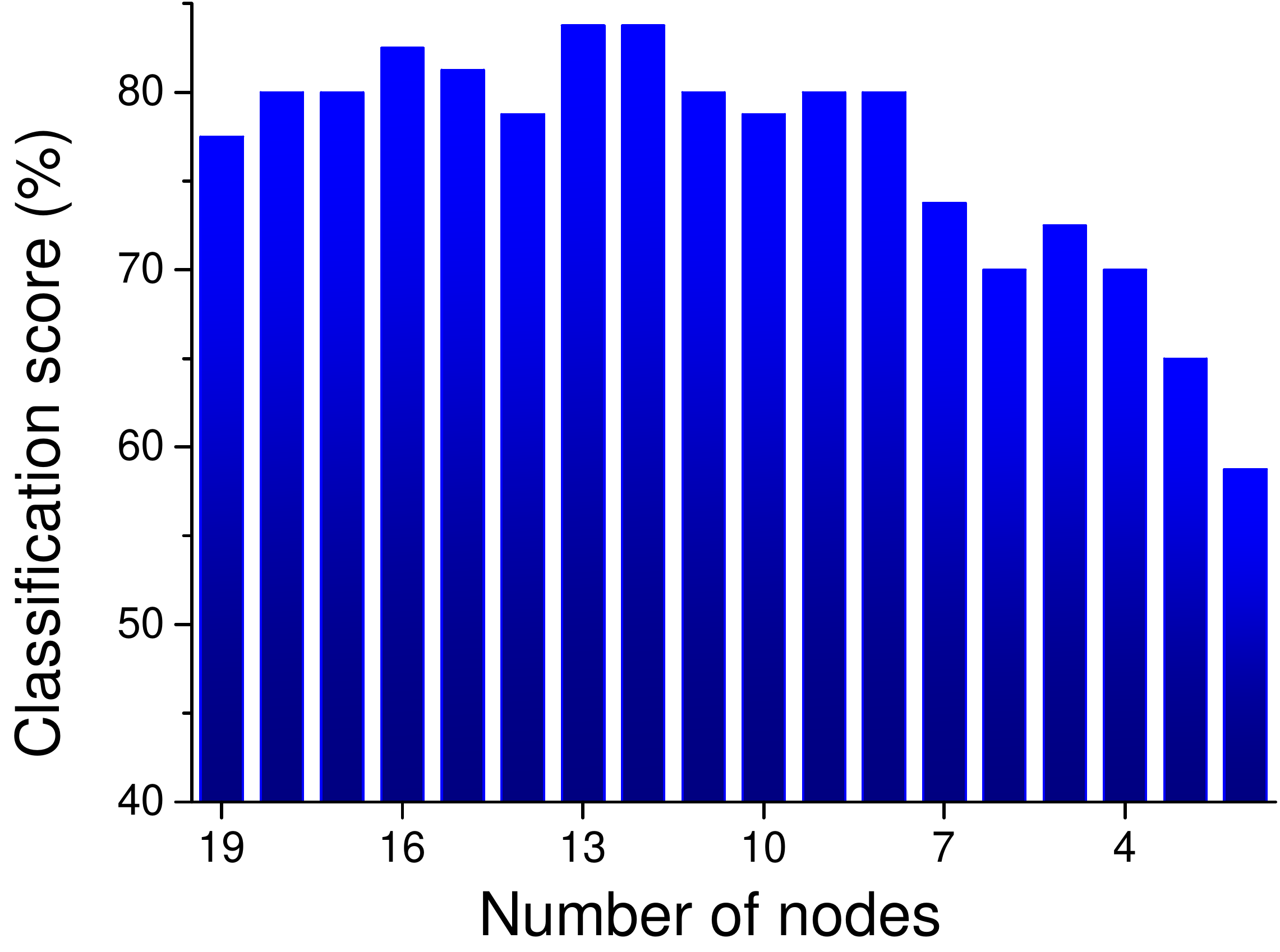}
	\centering
		\includegraphics[width=0.4\textwidth]{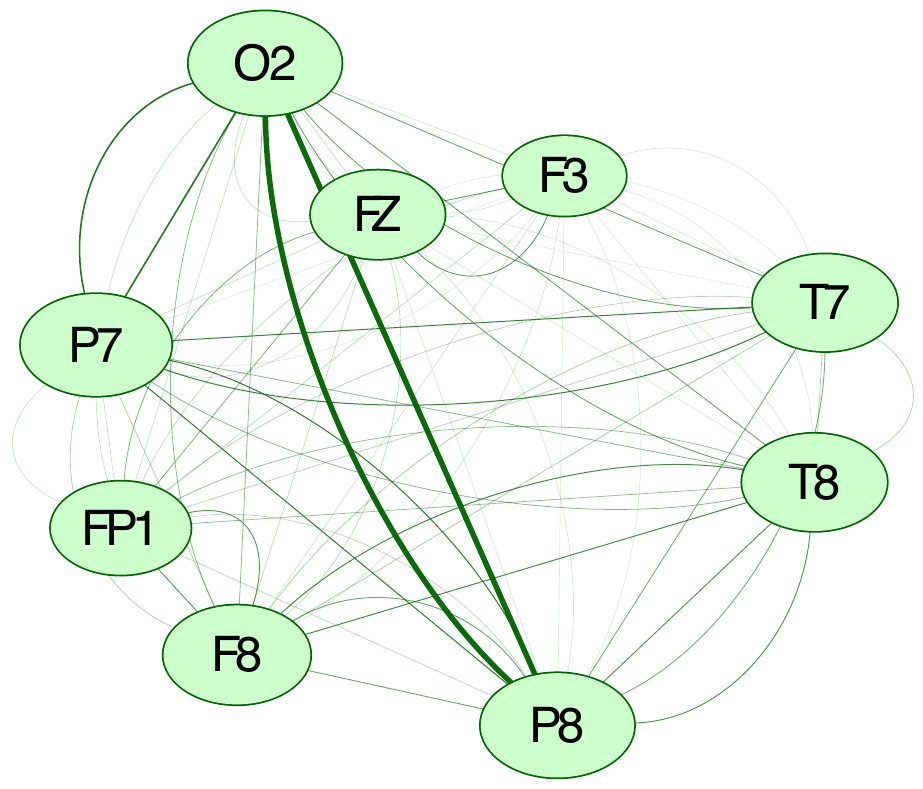}
		\includegraphics[width=0.4\textwidth]{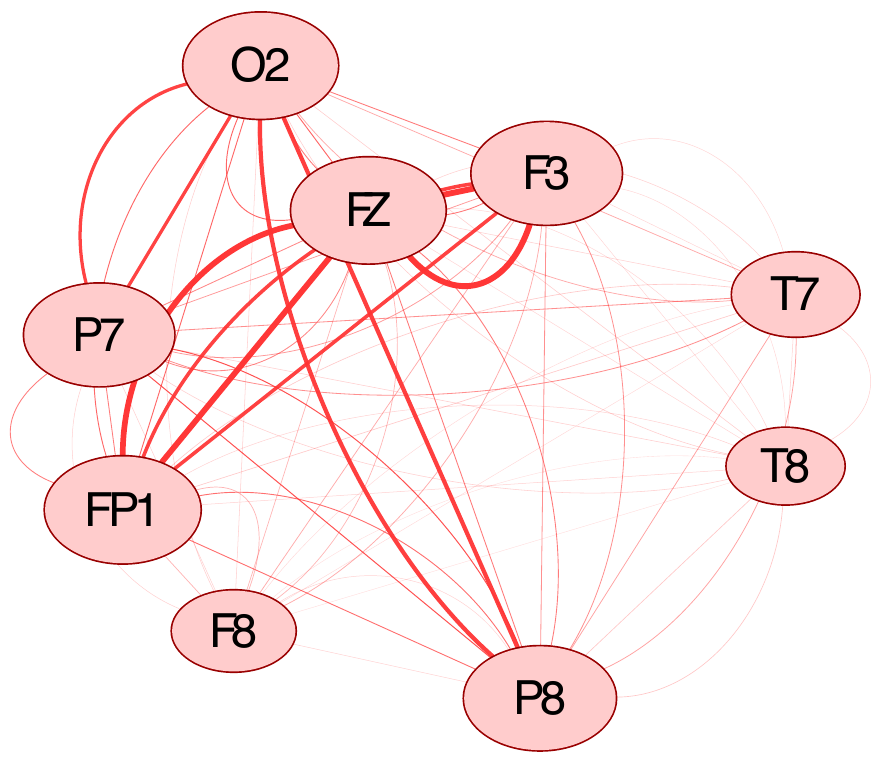}
	\caption{Discarding nodes from the networks. (Top Left) Classification score as a function of the node discarded. The dashed horizontal line represents the best classification score with the complete network ($77.5\%$). (Top Right) Classification score as a function of the number of surviving nodes. (Bottom) The networks of Fig. \ref{fig:FigureHO01} when only the 9 most important nodes are included.}
	\label{fig:FigureHO04}
\end{figure*}

As a final task in this hands-on exercise, we explore how data mining can be used to select a sub-set of significant nodes within the network. While EEG data set is relatively low-dimensional, stemming from only $19$ electrodes, further reducing dimensionality may be useful to gain more knowledge on the disease under study.

To do so, we use a {\it greedy algorithm}. The full network is considered, and nodes are deleted one at the time - thus yielding 19 sets of networks, each one with 18 nodes. A classification score is calculated for each set, and the one yielding the best classification is passed to the following phase. Next, nodes are deleted from this reduced networks, that is, nodes and all associated links are excluded from further analyses; and a new set of scores is calculated - and so forth, until further reduction would make the network too small to be of interest.

Fig. \ref{fig:FigureHO04} Top Left depicts the results, in terms of classification scores, for the first phase - {\it i.e.} when only one node is deleted. It can be seen that, in general, the score does not significantly decrease, and even increases when the third node is deleted. This may be surprising, but just confirms what has been discussed throughout this section: some nodes and links may just represent noise, for not being related with the pathology under study, and their deletion only improves the analysis by focusing it to the important elements. Fig. \ref{fig:FigureHO04} Top Right further depicts the results of the whole process: note how deleting up to ten nodes does not affect our classification capabilities (the score is still above $80\%$), which in turns illustrates the fact that the nine surviving nodes are the truly important ones to understand the pathology (Fig. \ref{fig:FigureHO04} Bottom).

In summary, a complex network analysis with arbitrary parameters seems to indicate some interesting differences between control subjects and patients of our EEG data set. Nevertheless, a classification task of these topological features yielded a rather low score ($\approx 60\%$). This suggests that these differences were not as important as initially thought, possibly because the analysis (and specifically, its parameters) was not properly tuned. The same data mining techniques were the instrument to increase the significance of the networks, and to point us towards the synchronisation metrics and brain regions most relevant. The upshot is that we end up with higher prognostic capabilities and better understanding of the pathology at hand, when compared with what would be obtained with an unguided complex network analysis.


\section{Additional bridges}
\label{sec:other}

In the three previous sections we have seen numerous examples of how data mining and complex network theory can be used in a synergistic way. The discussion has been organised around three of the most important tasks in data mining, namely classification, validation and feature selection. Nevertheless, this is clearly not the whole picture. This section presents a selection of four additional problems, in which bridges between the two communities have been successfully created.

\subsection{Clustering and community detection}
\label{sec:ClusteringAndCommDetection}

Clustering in data mining and community detection in complex networks analysis are two tasks that present many similarities. Clustering is defined as the task of grouping a set of objects in such a way that objects in the same group (or cluster) are more similar to each other than to those in other groups \cite{kaufman2009finding, duran2013cluster}. If we take a set of objects, map them into nodes of a network, and pairwise connect nodes whose corresponding objects are similar, clustering is tantamount to finding communities within the network, {\it i.e.} groups of nodes for which internal connections are much denser than those with the rest of the nodes in the network \cite{girvan2002community, fortunato2010community}. 
The network approach presents some advantages.

\begin{figure*}[!tb]
	\centering
		\includegraphics[width=0.6\textwidth]{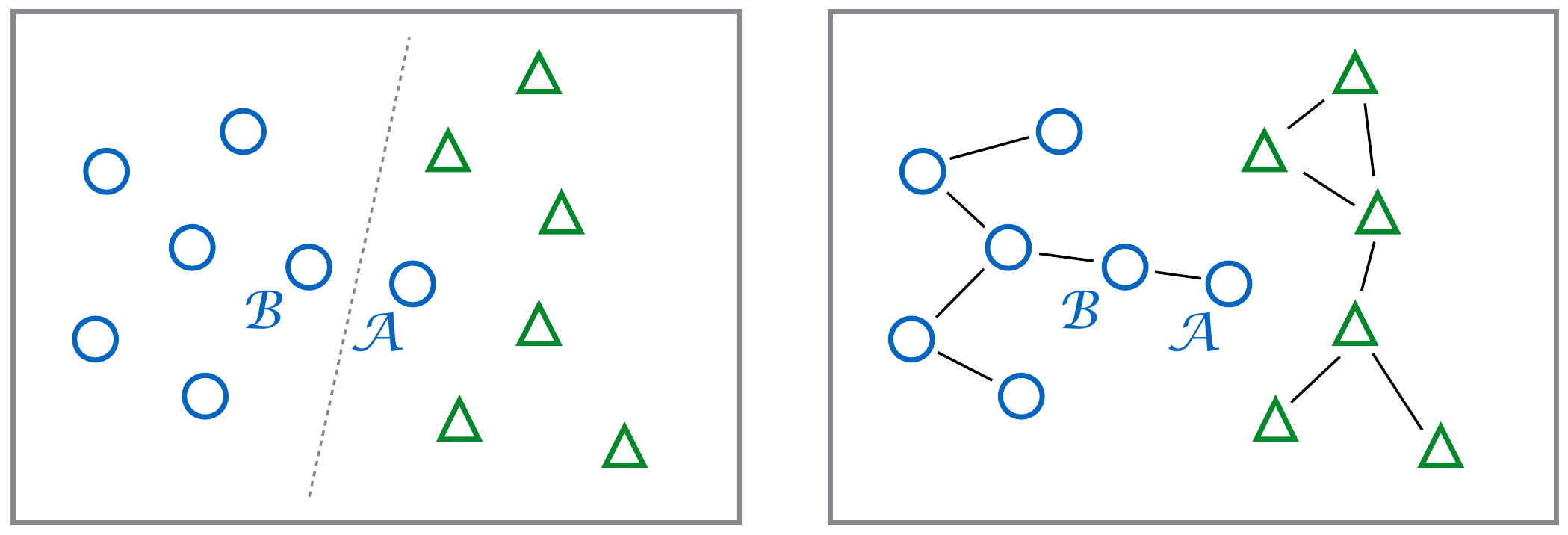}
	\caption{The clustering / community detection problem (Left) Clustering by means of a centroid model; object $A$ is wrongly assign to the right cluster, due to the non-convex nature of such cluster (Right) Result of a community detection process.}
	\label{fig:Clustering}
\end{figure*}

Most clustering algorithms exploit local ({\it i.e.} pairwise) information about similarity. For instance, connectivity models are based on a measure of distance between groups of items; centroid models create clusters based on the distance between objects and a single mean vector; and graph-based models (an intermediate version of both problems) define clusters in which at least a fraction of the nodes are connected by an edge. While the low computational cost represents one of the main advantages of these algorithms, there are some situations in which their detection ability is severely affected. Consider, for instance, the case of an object $A$ which is similar to another object $B$ of its true cluster, but not to the other objects of the same cluster - in other words, $B$ is a mediatory item between $A$ and the cluster itself, see Fig. \ref{fig:Clustering} Left. In this case, a graph-based model fails, as $A$ has not enough connections with its true cluster; a centroid model can also yield wrong results, especially in the case of non-convex clusters \cite{estivill2002so}.

A complex network approach, on the other hand, allows including the full topology of the network into the analysis - provided, of course, that a suitable similarity metrics has been defined, a problem which may be far from trivial. As depicted in Fig. \ref{fig:Clustering} Right, the object $A$ is correctly assigned to the left cluster, although it only shares one link with it, and is closer to the centre of the right cluster (note that, in general, spatial information is eliminated in a network representation). Complex networks also facilitate the detection and characterisation of communities in sets of heterogeneous elements or among which a distance measure cannot easily be assessed \cite{tumminello2011community}; as well as of networks that evolve with time \cite{mucha2010community, hopcroft2004tracking}. On the negative side, community detection algorithms can fail in some scenarios, for instance in the identification of very small communities, or when one community is much smaller than the other ones \cite{danon2006effect, fortunato2010community}.

\begin{figure*}[!tb]
	\centering
		\includegraphics[width=0.9\textwidth]{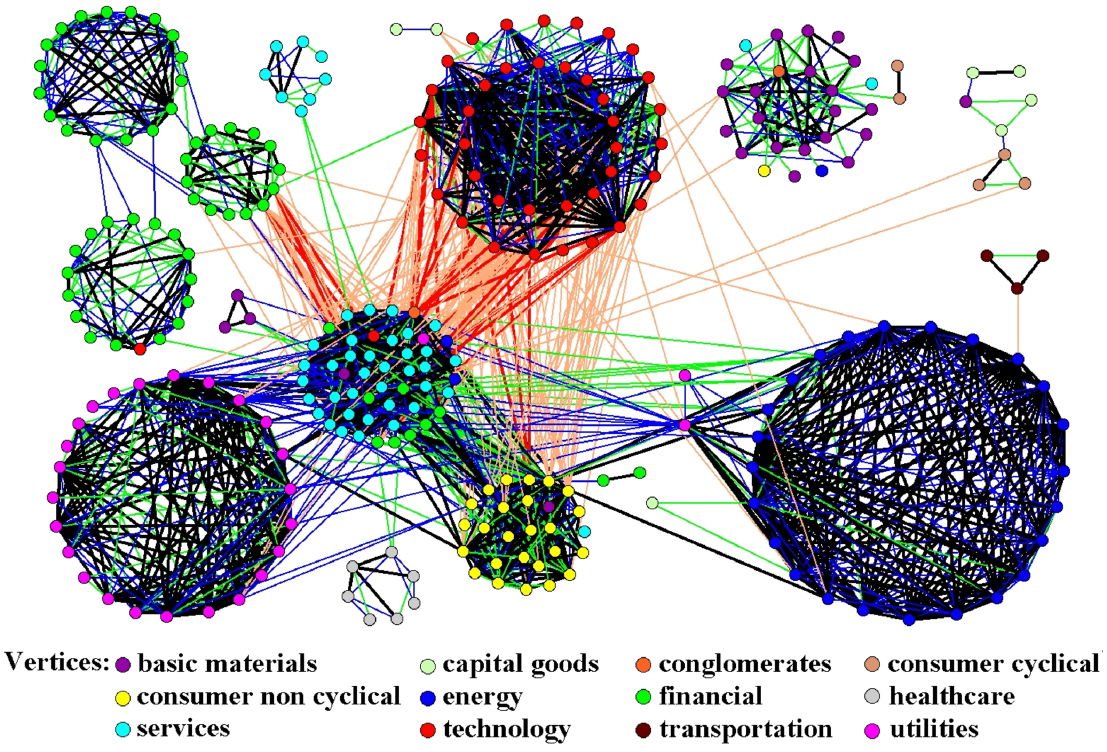}
	\caption{The largest connected component of the Bonferroni network associated with the system of 500 stocks, and its community structure. The nodes represent stocks, and links connecting different stocks correspond to statistically validated relationships. The node color identifies the economic sector of the corresponding stock. Reprinted figure with permission from Ref. \cite{tumminello2011statistically}.}
	\label{fig:Stocks}
\end{figure*}

Overall, a complex network approach to the clustering problem has historically yielded better results than the ones afforded by classical data mining approaches. Beside those reported in Section \ref{sec:FSelectionNodes} \cite{steinhaeuser2011complex, tsonis2011community}, a few examples are worth mentioning. For instance, in music technology a community detection algorithm was used to detect groups of {\it covers} of the same song, with a result outperforming traditional clustering algorithms \cite{serra2012characterization}. In econophysics \cite{tumminello2012identification}, clusters of investors trading in a financial market are identified as communities in statistically validated networks \cite{tumminello2011statistically} - see Fig. \ref{fig:Stocks} for a graphical representation; or stocks are clustered \cite{micciche2003degree} according to the correlation detected between return and volatility time series \cite{mantegna1999hierarchical}.

It is also worth noticing that clustering (and other data mining) algorithms have recently been used to perform a community detection task. These include fuzzy $c$-means \cite{zhang2007identification}, fuzzy clustering \cite{sun2011identification}, adaptive clustering \cite{ye2008adaptive}, or genetic algorithms \cite{liu2007effective}.

\subsection{Link prediction}
\label{sec:LinkPrediction}

Another problem of great interest in complex networks, especially for its biological implications, is the one of {\it link prediction}. All previous examples covertly assumed that the network is completely known - either because relationships are explicit, or because they could be reconstructed from functional analyses. However, in biological networks such as food webs, protein-protein or metabolic networks, assessing the presence of a link between two elements requires performing costly field and/or laboratory experiments; as a consequence, only a fraction of the total number of links is typically tested, {\it e.g.} $20\%$ of the molecular interactions in yeast cells \cite{yu2008high} and $0.3\%$ in human ones \cite{amaral2008truer} have been described. A solution to this problem involves using some kind of {\it link prediction} algorithm, {\it i.e.} an algorithm to estimate the likelihood for a link (yet undiscovered) to be present.

The problem of link prediction has a clear counterpart in the wide family of {\it recommendation system} algorithms developed in data mining. In its original formulation, a recommender system (also called {\it information collaborative filtering} system) takes as input information about past purchases (or selections) of goods by a set of users, and suggests the following best buy for each of them. If one considers a bipartite network, where nodes represent both users and goods, and links past interactions between them, a recommendation system is completely equivalent to a link prediction system acting on such network.

In spite of their substantial equivalence, link prediction in complex network theory and recommendation system in data mining have mostly walked independent paths. In the former, network theory has relied more on random walk processes and maximum likelihood methods - see Ref. \cite{lu2011link}. In the latter, the most used approaches are based on concepts such as Markov chains and statistical models - see Refs. \cite{adomavicius2005toward, ricci2011introduction} for a full overview on the topic. Only recently have both communities started joining forces and mixing both approaches \cite{lu2012recommender}.

What advantages does the complex network approach offer to data mining in this particular context? As in other cases, like in the clustering problem (see Section \ref{sec:ClusteringAndCommDetection}), the answer lies in the global topological perspective of complex networks. For instance, if one were to try to construct a link prediction model for the Internet, a network analysis would be useful to unveil a hierarchical (or {\it medusa-like}) structure, with a strongly connected core, a peer-connected periphery, and a set of isolated nodes \cite{carmi2007model}. In this case, better performance could be obtained by taking into account the hierarchy of the network, and favouring those links that do not disrupt such a structure \cite{clauset2008hierarchical}.

\subsection{Evaluating and validating generative models}

Suppose that one is analysing the human brain under different conditions, {\it e.g.} control subjects {\it vs.} Alzheimer's patients, and that the topologies of the corresponding functional networks have been described. A further question may be asked: can such topologies, and the evolution from the former to the latter, be explained by a set of simple rules? Knowing these rules would allow a better understanding of the disease, possibly shedding light of the underlying causes of the observed phenomena. This can be seen as a ``genotype to phenotype problem'', that is, the process of identifying the relationships between hidden variables (the genotype) and measured observables (the phenotype) of a system.

{\it Generative models} are simple models designed to create synthetic networks with specific topological features. Proposed models include theoretical ones, {\it i.e.} independent of real-world networks \cite{albert2002statistical, klemm2002highly, schmeltzer2014percolation, rad2012topological}; and models describing brain networks: from cortical connectivity \cite{kaiser2004modelling, chen2013trade, nicosia2013phase, klimm2014resolving}, neural ensembles \cite{kwok2007robust, stam2010emergence}, to fMRI \cite{fraiman2009ising, simpson2011exponential, vertes2012simple, li2013exploring, vertes2014generative} and MEG \cite{buldu2011reorganization} functional networks. 

Proposed generative models need to be optimised, that is, one ought to obtain the best set of parameters, such that the networks yielded by the model are topologically equivalent to the real ones according to some criteria. Moreover, even if the model does indeed recover some topological characteristics, this does not guarantee that the generative model itself is representative of what is really happening in the brain, {\it i.e.} that the model yields new knowledge of the neurological processes involved in the disease. The model must therefore be validated. The attentive reader will notice the similarities between this problem, and the one of improving network significance as tackled in Section \ref{sec:improve}.

Evaluation and generation of generative models can be achieved by using a classification task in which two conditions are compared \cite{zanin2015probabilistic, zanin2016phenotype}. An increase in the knowledge about the conditions implies a higher capability of discriminating between both states; this, in turn, implies that the generative model should increase our capacity of discriminating between both conditions, in order to ensure an increase in knowledge.
It has further been demonstrated that, to obtain such an increase in discriminatory capabilities, the relationship between the phenospace ({\it i.e.} the observed topological metrics) and the genospace (the model parameters) should at least be a non-monotonic function \cite{zanin2016phenotype}.

\subsection{Complex networks in Big Data: semantic representations by graphs}
\label{sec:BDandCN}

\begin{figure}
	\includegraphics[width=0.8\textwidth]{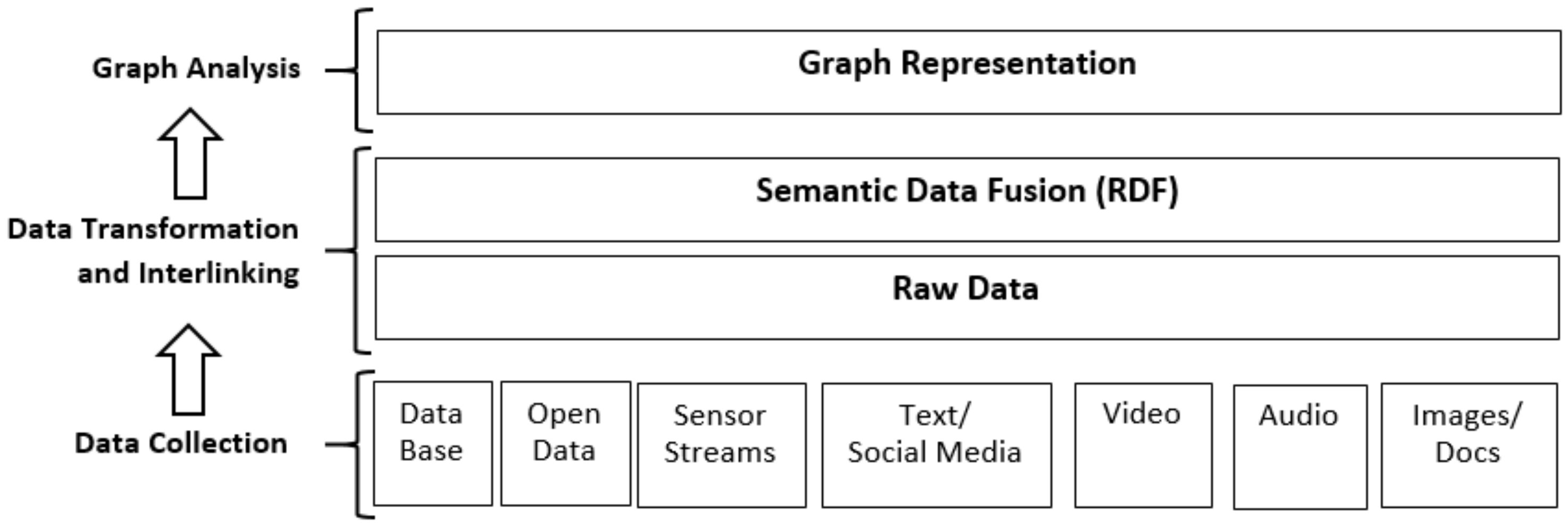}
	\centering
	\caption{Conceptual representation of a graphical Big Data analysis framework. See main text for details.}
	\label{fig:Framework}
\end{figure}

To illustrate the possibilities offered by complex networks to simplify the analysis of Big Data, we discuss a standard approach involving the use of non-planar graphs \cite{gupta2014graphical}. The approach consists of a methodology for transforming raw data into linked graphical representations of them (see Fig. \ref{fig:Framework}), and is composed of three main steps: Data Collection, Data Transformation and Interlinking and Graph Analysis.

In the initial phase, the user only has access to huge quantities of structured, semi-structured and unstructured data, originating from many independent sources. All these data are heterogeneous and raw, and thus require some processing to be understandable. This is achieved in a second phase, termed \emph{Data Transformation and Interlinking phase}, in which the raw data are transformed into \textit{smarted data}. The same data are expressed in a standard format using RDF (\textit{Resource Description Framework}, a standard defined for the web of semantic data \cite{pan2009resource}). In RDF every statement is expressed by three components ({\it i.e.} a {\it triple}): the \textit{subject}, {\it i.e.} the identifier of the resource; the \textit{predicate}, that indicates the property or attribute of the subject to be described; and the \textit{object}, the value of the predicate related to the subject. Using this semantic data model, it is possible to capture the semantic heterogeneity and the differences in the logical representation of data.
Finally, the last phase of the methodology, the \textit{Graph Analysis phase}, deals with the creation of a graphical representation using semantic data fusion. At the end of the process, the original data are mapped into a network structure, on which all network metrics and principles can be applied to discover hidden relationships.

\begin{figure}
	\includegraphics[width=0.8\textwidth]{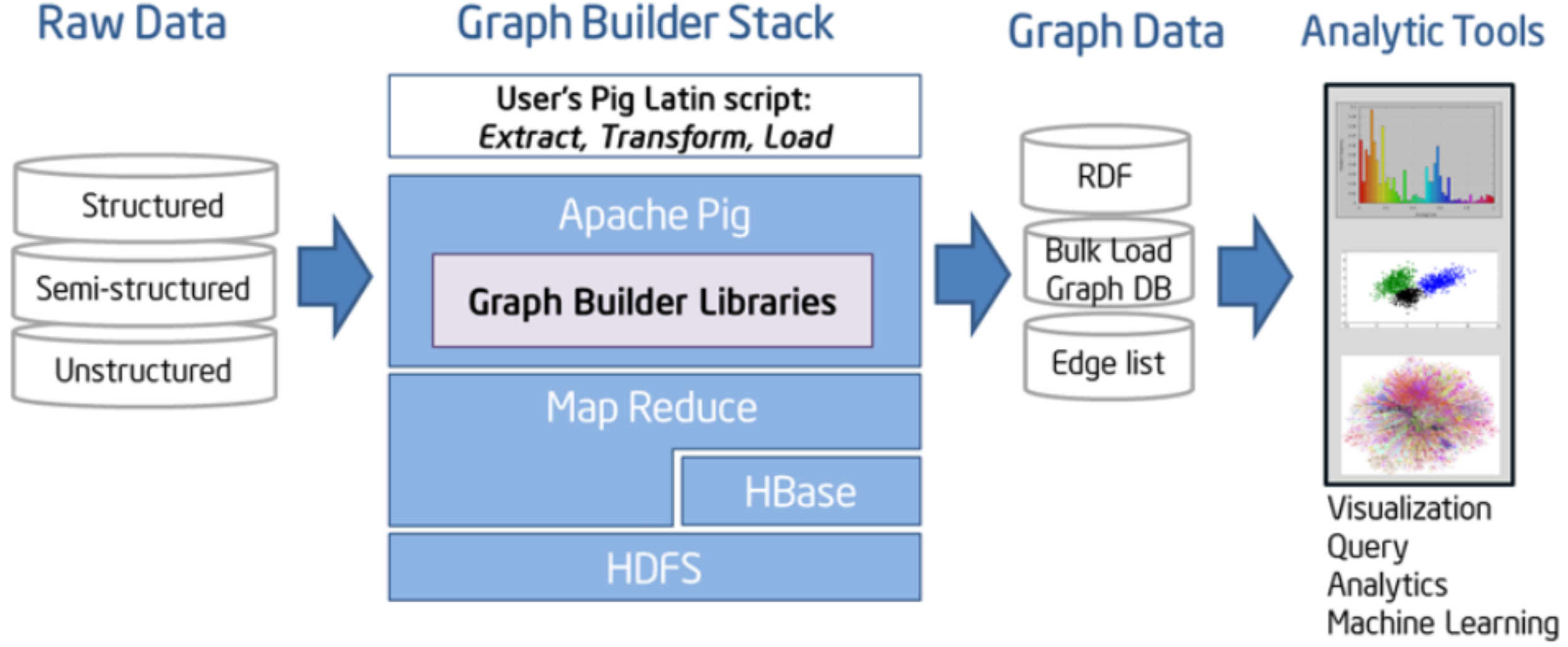}
	\centering
	\caption{Representation of the software ecosystem supporting a graphical Big Data analysis framework.}
	\label{fig:GraphBuilderSchema}
\end{figure}  
 
A more concrete example of the framework proposed in Fig. \ref{fig:Framework} is provided by the Intel Open Source Community \cite{Intel01}. Fig. \ref{fig:GraphBuilderSchema} presents the schema proposed to uncover hidden correlations in large volumes of data, by means of graph analytics. Starting from structured, semi-structured, or unstructured data, the Apache Pig library is used to form RDF triples for graph elements. Afterwards, some graph visualisation tools, such as Gephi (see Section \ref{sec:softwareNets}), are applied to this triples list.
 
\begin{figure}
	\includegraphics[width=0.7\textwidth]{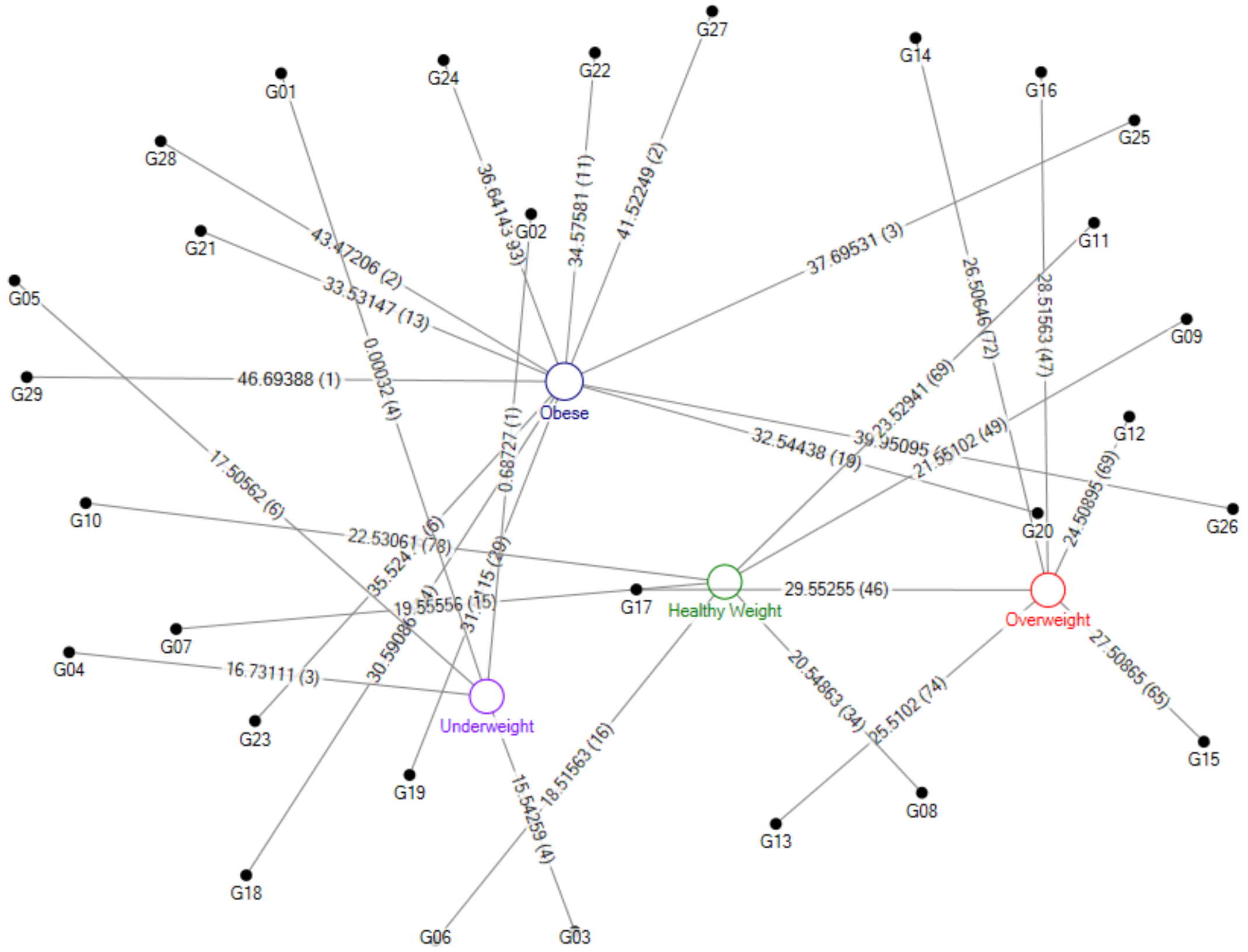}
	\centering
	\caption{An example of a directed labelled graph of a graphical Big Data analysis framework. See main text for details.}
	\label{fig:GraphBMI}
\end{figure}  

A simple example of an analysis performed following the framework clarifies this issue, as depicted in Fig. \ref{fig:Framework}. The example involves the representation of an anonymised real dataset including subjects with some kind of cardiovascular disease. In particular we consider subjects' BMI (Body Mass Index), following the formulas and the parameters of  the Centre for Disease Control and Prevention (CDCP) \cite{CDCP}. Following Fig. \ref{fig:Framework}, the first phase is Data Collection; note that BMI data are only part of the data set, but the only one relevant to our study. In the Data Transformation phase, raw data are transformed to an RDF triple list of the type (``Patient Group'' ``with BMI Value'' ``has Weight Status''); according to the previous definition, “Patient Group” is the object, “with BMI Value” the predicate/property and “has Weight Status” the object/value. Two examples of the elements contained in such lists are the following:

{\centering
\vspace{0.3cm}
<G02 \hspace{0.5cm} 00,68727 \hspace{0.5cm} Underweight>

<G06 \hspace{0.5cm} 18,51563 \hspace{0.5cm} Healthy Weight>

\vspace{0.3cm}
}

From this triple list, and by using the in-memory graph tools like NodeXL \cite{RefNodeXL}, one can produce a directed labelled graph as the one shown in Fig. \ref{fig:GraphBMI}. A simple visual inspection makes it clear that most people with cardiovascular disease have some overweight problems - notice the larger degrees of the ``Obese'' and ``Overweight'' nodes. Further analyses would require to connect this with other graphs, like for instance with one related to lifestyle, to understand the incidence of overweight in cardiovascular disease.

This simple example shows how RDF, as the basic data representation language for linked information, can be used to control the data variety and the semantic heterogeneity one usually encounters in Big Data, and to produce a graph representation of the available information. Beyond that, there is a clear technical challenge in the system scalability, {\it i.e.} in the ability of handling and representing very large data sets; nevertheless, this is continuously tackled, and the availability of tools like NodeXL \cite{RefNodeXL} make this problem less and less relevant in most applications.


\section{Conclusions}
\label{sec:conclusions}

The application of data mining concepts to the study of complex networks is generating great enthusiasm and increasing interest from several scientific communities. Data mining and network theory share the common objective of finding ways to represent complex systems, and ultimately provide a better understanding of their structure and dynamics. At the same time, they provide different view-points. Data mining focuses on the comparative analysis of different instances, to detect similarities and differences between them; complex networks allow representing the internal structure of individual instances in a simple and elegant way. That these two approaches have not yet merged is mainly to put down to the different background of researchers working in statistical physics and artificial intelligence. This problem notwithstanding, in this review we have shown that, when used in synergy these two approaches broaden the scope of complex systems that can be studied in a quantitative way, particularly in interdisciplinary biophysical contexts.

Data mining provides the complex network practitioner with numerous tools for improving the significance of representations. It can for instance be used to identify the important nodes and connections of the system, disregarding less important ones that can be thought to contribute noise. Data mining also allows creating models for discriminating between groups of instances, thus enabling diagnostic and prognostic applications. Additionally, such models can be used to objectively estimate the significance of the reconstructed networks. On the other hand, data scientists can find in network representations a new way of structuring information, providing a wealth of quantitative variables through which systems can be described. When both fields address the same basic goals, as in the case of data clustering and recommendation systems, a cross fertilisation allows coming up with innovative solutions. Finally, we foresee an increasing interest for complex network theory in the raising field of Big Data, in which relationships between data elements are of fundamental importance.

In spite of all the achievements outlined in this review, many issues still need to be tackled by the scientific community. First, joining data mining and complex network theory is a path often plagued with numerous pitfalls, particularly when some of their respective assumptions and limitations are not properly taken into account. Second, there is a lack of a specific theoretical and numerical foundation for the combined use of concepts from both fields. In this last section, we discuss some of these topics, with the hope of fostering debate in the scientific community.

\subsection{Problems to overcome}

After reading this review and some of the papers cited within, the attentive reader may notice some recurring issues. The complex network practitioner with insufficient experience with data mining, may be tempted to use it as a black box: results of a data mining task are reported, without a full appraisal of the limitations they present. In what follows, we want to briefly discuss some caveats of merging complex networks and data mining, which we hope will be a guide for future studies.

\begin{description}
	\item \emph{Method specification.} In most studies, data mining tools have by and large been considered as ``utilities'', used to achieve a specific goal, but not as an end in itself. As a consequence, many papers' abstract omits  a description of the classification algorithm or the validation procedure used. This is especially true of biomedical and neuroscience studies, in which the focus is on the biological and clinical aspects of the analysis. It is nevertheless important to understand that data mining techniques have a significant impact in the results. For instance, different classification algorithms make different hypotheses on the patterns to be detected in data; and different validation strategies may highlight limitations in the obtained models. The choice of the algorithm is a fundamental step in the overall methodology, and should be handled as such.

	\item \emph{One method does not fit all.} Using one single data mining technique in a specific task is bound to return a partial picture, however well-established and widely used the technique. As illustrated in Section \ref{sec:ClassifyNet}, different classification models may yield important differences in accuracy, due to their underlying hypotheses \cite{rish2009discriminative, richiardi2011classifying, colby2012insights}, even for similar data sets, as different problems (pathologies) may require different approximations. It is thus important to always compare different algorithms, and select the one yielding best results - possibly understanding why such differences arise.
	
	\item \emph{Avoiding overfitting.} In some cases, the full power of data mining is deployed in different phases of the study: for instance an initial feature selection can be performed through SVM, and the same technique used in a classification task. This may result in an {\it overfitting} problem: the classification is performed over data that have already been filtered using information about the real class of each instance; in other words, the classification uses data that have been selected using information about the final result. The solution may come from splitting the available data into training, validation and test sets, although a large number of instances is required (a condition not easy to fulfil in biomedical problems). In general, a special attention must be devoted to the problem of model validation - see Section \ref{sec:dataMining}.
	
\end{description}

\subsection{Open lines of research}

Merging data mining and complex network theory is an ongoing process, which the research community has just started dealing with. If such integration is to continue evolving, and especially if concepts such as Big Data are to be included, several technical and theoretical aspects will need to be further developed.

\begin{description}

	\item \emph{Merging complex networks and data mining.} As we have seen throughout this review, complex networks metrics and data mining algorithms have developed and are being applied in an independent fashion. In other words, a standard analysis requires extracting some topological metrics and use them in data mining tasks, see Section \ref{sec:HandsOn_Class}: two steps that have little cross-interactions between them. This is due to the fact that data mining algorithms are not able to extract a pattern in the data that develops in the meso- or macro-scale. An important step ahead would consist in developing models including the network concept in their definition; for instance, a classification model able to internally evaluate topological metrics, like the clustering coefficient, without the need of {\it a priori} knowledge.

	\item \emph{Optimising network metrics evaluation.} As data sets become larger and larger, the evaluation of complex network metrics on them is getting more complicated, especially from a computational point of view. In parallel with the advent of Big Data, the analysis of large social networks, like the one generated by on-line social systems, may be in the future called ``Big Complex Networks''. If this trend consolidates, complex network metrics will have to be designed with a special attention on their computational cost, in order to ensure their scalability to hundreds of million of nodes.

	\item \emph{Software infrastructures for complex network analysis.} The challenges in terms of data complexity and size created by the Big Data movement, have forced the IT community to design new software tools for managing and analysing such a large amount of information. In the same vein, important benefits would be obtained from a software infrastructure specifically designed for the analysis of large scale complex networks. While some steps have already been taken in this direction, for instance with the creation of {\it graph databases} \cite{robinson2013graph}, most solutions are limited to the traditional graph concepts, disregarding the complexity of real-world networks.

\end{description}

\section*{Acknowledgements}

At the end of this review, we would like to acknowledge all scientists with whom we had, have, and surely will have interactions on this topic. Discussions and dialogues have been both stimulating and fruitful, giving birth to many of the ideas included in this work. In this spirit, the authors would like to thank the following colleagues: Emilio \'Alvarez Pereira (Team\&Cloud, Spain), Salvatore Micciche' (University of Palermo, Italy), Fabrizio Lillo (Scuola Normale Superiore di Pisa, Italy), Andrew Cook (University of Westminster, UK), Alexey Zaikin (University College London, UK), Jos\'e Luis Vallejo (Sngular, Spain), and the ComplexWorld team.

At the same time, we also would like to gratefully acknowledge all the discussions we had (in meetings, conferences, workshops, and personal visits) with other colleagues who are not mentioned in the above list (and, for that, we deeply apologise), which equally inspired our efforts, and opened up our minds in a way that contributed, eventually and substantially, to the realisation of the present survey.

\section*{Final note}

After editing the present Report, we learned about the publication of other recent and important contributions which relate to subjects and arguments treated in this work. In the following, we provide a list of these additional References that the reader may find useful to consult.

\vspace{0.5cm}

Traxl, D., Boers, N., \& Kurths, J. (2016). Deep Graphs-a general framework to represent and analyze heterogeneous complex systems across scales. arXiv preprint arXiv:1604.00971.

Thomas, J. M., Muscoloni, A., Ciucci, S., Bianconi, G., \& Cannistraci, C. V. (2016). Machine learning meets network science: dimensionality reduction for fast and efficient embedding of networks in the hyperbolic space. arXiv preprint arXiv:1602.06522.

Armano, G., \& Javarone, M. A. (2013). Clustering datasets by complex networks analysis. Complex Adaptive Systems Modeling, 1(1), 1-10.

Akoglu, L., Tong, H., \& Koutra, D. (2015). Graph based anomaly detection and description: a survey. Data Mining and Knowledge Discovery, 29(3), 626-688.

Petri, G., Scolamiero, M., Donato, I., \& Vaccarino, F. (2013). Topological strata of weighted complex networks. PloS one, 8(6), e66506.

Gauvin, L., Panisson, A., \& Cattuto, C. (2014). Detecting the community structure and activity patterns of temporal networks: a non-negative tensor factorization approach. PloS one, 9(1), e86028.

Massara, G. P., Di Matteo, T., \& Aste, T. (2015). Network Filtering for Big Data: Triangulated Maximally Filtered Graph. arXiv preprint arXiv:1505.02445.

Barfuss, W., Massara, G. P., Di Matteo, T., \& Aste, T. (2016). Parsimonious modeling with Information Filtering Networks. arXiv preprint arXiv:1602.07349.

\section*{References}

\bibliography{References_Intro,References_CNDef,References_DM,References_LimitsDM,References_Handson,References_FSelection,References_Others,References_Classification,References_Improve,References_BigData,References_Class_Omics,References_Conclusions}

\appendix

\section{List of acronyms}

\begin{center}
    \begin{tabular}{ | l | p{10cm} |}
    \hline
    Acronym & Explanation \\ \hline

ANN & Artificial Neural Network \cite{hagan1996neural, Zurada92a}, see Section \ref{sec:dataMining} for definition. \\

AUC & Area Under the Curve. In Receiver Operating Characteristic (ROC) curves, AUC measures the probability that a classifier will rank a randomly chosen positive instance higher than a randomly chosen negative one. \\

BCI & Brain-Computer Interfaces, a computer-based system to acquire, and respond upon, brain signals. \\

EEG & ElectroEncephaloGram, an electrophysiological monitoring method to record electrical activity of the brain. \\

fMRI & Functional Magnetic Resonance Imaging, an MRI recorded during a cognitive task. \\

HCRF & Hidden Conditional Random Fields \cite{quattoni2007hidden}, see Section \ref{sec:dataMining} for a definition. \\

KDD	& Knowledge Discovery in Databases, the overall process of discovering useful knowledge from data \cite{fayyad1996advances}. \\

KNN & k-Nearest Neighbors algorithm, a classification algorithm based on the class of neighbour training examples. See Section \ref{sec:dataMining} for a definition. \\

LOOCV & Leave One Out Cross Validation, a cross-validation procedure in which one sample at the time is excluded from the training. See Section \ref{sec:featureSelection} for details. \\

MEG & MagnetoEncephaloGram, an electrophysiological monitoring method to record the magnetic field generated by the activity of the brain. \\

MKL & Multiple Kernel Learning \cite{lanckriet2004learning}, see Section \ref{sec:dataMining} for a definition. \\

MRF & Markov Random Field \cite{kindermann80mrf}, see Section \ref{sec:dataMining} for a definition. \\

MRI & Magnetic Resonance Imaging, a medical imaging technique that uses magnetic fields and radio waves to form images of the brain. \\

MST & Minimum Spanning Tree, tree that connects all the vertices of a network together with the minimal total weighting for its edges. \\

NNC & Nearest Neighbour Classifiers, classification method based on copying the class of the closest training examples in the feature space. \\

PMML 	& Predictive Model Markup Language, XML-based standard for interchanging predictive models \cite{pechter2009s}. \\

RDF & Resource Description Framework, general standard for conceptual description or modelling of information \cite{pan2009resource}. \\ 

RFE & Recursive Feature Elimination (RFE), a feature selection method based on repeatedly removing features with low relevance. See Section \ref{sec:featureSelection} for details. \\

ROC & Receiver Operating Characteristic, graphical plot illustrating the performance of a binary classifier system as its discrimination threshold is varied. It is created by plotting the true positive rate as a function of the false positive rate, for different thresholds. \\

SVM & Support Vector Machine \cite{cortes1995support}, see Section \ref{sec:dataMining} for a definition. \\

    \hline
    \end{tabular}
\end{center}

\end{document}